\documentclass[11pt]{elsarticle}
\usepackage{verbatim,amsmath,amssymb}
\usepackage{epsfig,float,color}
\usepackage{geometry}
\usepackage{setspace}
\usepackage[numbers]{natbib}
\usepackage{amsthm,mathrsfs,amsfonts,dsfont}
\usepackage{hyperref}
\usepackage[yyyymmdd,hhmmss]{datetime}
\usepackage{caption}
\usepackage{subcaption}
\usepackage{tablefootnote}
\usepackage{tabularx}
\usepackage{footmisc}  
\usepackage{placeins}
\usepackage{cmap}
\usepackage[toc,page]{appendix}
\usepackage[right,pagewise,displaymath, mathlines]{lineno}

\definecolor{cgn}{rgb}{0,0,0}
\definecolor{cgm}{rgb}{0,0,0}

\definecolor{cgn2}{rgb}{0,0,0}

\definecolor{cgr2}{rgb}{0,0,0}

\definecolor{darkblue}{rgb}{0,0,1}
\hypersetup{pdftex=true, colorlinks=true, breaklinks=true, linkcolor=darkblue, menucolor=darkblue, pagecolor=darkblue, citecolor=darkblue, urlcolor=darkblue}

\newcommand{\bitm}{\begin{itemize}}
\newcommand{\eitm}{\end{itemize}}
\newcommand{\bnumr}{\begin{enumerate}}
\newcommand{\enumr}{\end{enumerate}}
\newcommand{\bolds}[1]{\boldsymbol{#1}}

\newcommand {\auab}{a_{\alpha\beta}}
\newcommand {\agd}{a^{\gamma\delta}}

\newcommand {\Mab}{M^{\alpha\beta}}

\newcommand {\buab}{b_{\alpha\beta}}

\newcommand {\tauab}{\tau^{\alpha\beta}}

\newcommand {\eqb}[1]{\begin{equation}\begin{array}{#1}}
\newcommand {\eqe}{\end{array}\end{equation}}

\newcommand {\esb}[1]{\begin{equation*}\begin{array}{#1}}
\newcommand {\ese}{\end{array}\end{equation*}}
\newcommand {\ds}{\displaystyle}

\newcommand {\pa}[2]{\frac{\partial{#1}}{\partial{#2}}}
\newcommand {\paq}[2]{\frac{\partial^2{#1}}{\partial{#2}^2}}

\newcommand {\back}{\! \! \!}
\newcommand {\is}{\back &=& \back}
\newcommand {\dis}{\back &:=& \back}

\newcommand {\dif}{\mathrm{d}}

\newcommand {\II}{{I\kern-.3em I}}
\newcommand {\III}{{I\kern-.3em I\kern-.3em I}}

\newcommand {\mra}{\mathrm{a}}

\newcommand {\mf}{\mathbf{f}}

\newcommand {\mk}{\mathbf{k}}

\newcommand {\muu}{\mathbf{u}}

\newcommand {\ba}{\boldsymbol{a}}

\newcommand {\bn}{\boldsymbol{n}}

\newcommand {\bv}{\boldsymbol{v}}

\newcommand {\bx}{\boldsymbol{x}}
\newcommand {\by}{\boldsymbol{y}}

\newcommand {\mK}{\mathbf{K}}

\newcommand {\mM}{\mathbf{M}}
\newcommand {\mN}{\mathbf{N}}

\newcommand {\bE}{\boldsymbol{E}}

\newcommand {\bM}{\boldsymbol{M}}
\newcommand {\bN}{\boldsymbol{N}}

\newcommand {\bT}{\boldsymbol{T}}

\newcommand {\bX}{\boldsymbol{X}}
\newcommand {\bY}{\boldsymbol{Y}}

\newcommand {\bsig}{\mbox{\boldmath$\sigma$}}

\newcommand {\bone}{\mathbf{1}}

\newcommand {\bbR}{\mathbb{R}}

\newcommand {\IR}{{\rm\kern.24em
   \vrule width.02em height1.53ex depth-.05ex
   \kern-.3em R}}
\newcommand {\ic}{{\rm\kern.20em
   \vrule width.02em height1.0ex depth-.05ex
   \kern-.22em c}}
\newcommand {\ia}{{\rm\kern.20em
   \vrule width.02em height1.05ex depth-.0ex
   \kern-.25em a}}
\newcommand {\IC}{{\rm\kern.24em
   \vrule width.02em height1.4ex depth-.05ex
   \kern-.26em C}}
\newcommand {\ID}{{\rm\kern.34em
   \vrule width.02em height1.5ex depth-.05ex
   \kern-.36em D}}
\newcommand {\IS}{{\rm\kern.24em
   \vrule width.02em height1.6ex depth.05ex
   \kern-.26em S}}
\newcommand {\IT}{{\rm\kern.50em
   \vrule width.02em height1.55ex depth-.05ex
   \kern-.52em T}}

\newcommand {\IE}{{\rm\kern.24em
   \vrule width.02em height1.55ex depth-.05ex
   \kern-.33em E}}
\newcommand {\IEa}{{\rm\kern.24em
   \vrule width.02em height1.55ex depth-.05ex
   \kern-.33em E}^{1}_{ijkl}}
\newcommand {\IEb}{{\rm\kern.24em
   \vrule width.02em height1.55ex depth-.05ex
   \kern-.33em E}^{2}_{ijkl}}

\newcommand {\sJ}{\mathcal{J}}

\newcommand {\Ass}[2]{\kern 0.9ex \vrule width0.45em height0.2ex depth0ex \kern -2.1ex \bigwedge_{#1}^{#2}}
\newcommand {\ASS}[2]{\kern 1.45ex \vrule width0.5em height0.2ex depth0ex \kern -2.65ex \bigwedge_{#1}^{#2}}

\newcommand {\cabgd}{{c}^{\alpha\beta\gamma\delta}}
\newcommand {\dabgd}{{d}^{\alpha\beta\gamma\delta}}
\newcommand {\eabgd}{{e}^{\alpha\beta\gamma\delta}}
\newcommand {\fabgd}{{f}^{\alpha\beta\gamma\delta}}

\graphicspath{ {/} }

\begin{document}



\begin{center}
\Large{\bf{Modal analysis of graphene-based structures for large deformations, contact and material nonlinearities}}\\

\end{center}

\begin{center}

\large{Reza Ghaffari\footnote{email: ghaffari@aices.rwth-aachen.de} and Roger A. Sauer\footnote{Corresponding author, email: sauer@aices.rwth-aachen.de}}\\
\vspace{4mm}

\small{\textit{
Aachen Institute for Advanced Study in Computational Engineering Science (AICES), \\
RWTH Aachen University, Templergraben 55, 52056 Aachen, Germany \\[1.1mm]}}

\vspace{4mm}

Published\footnote{This pdf is the personal version of an article whose final publication is available at \href{https://doi.org/10.1016/j.jsv.2018.02.051}{http:/\!/sciencedirect.com}} in \textit{Journal of Sound and Vibration}, \href{https://doi.org/10.1016/j.jsv.2018.02.051}{DOI: 10.1016/j.jsv.2018.02.051}\\
Submitted on 7.~July 2017, Revised on 15.~February 2018, Accepted on 21.~February 2018

\end{center}

\vspace{3mm}

\rule{\linewidth}{.15mm}
{\bf Abstract: }
The nonlinear frequencies of pre-stressed graphene-based structures\textcolor{cgr2}{, such as flat graphene sheets and carbon nanotubes,} are calculated. These structures are modeled with a nonlinear hyperelastic shell model. The model is calibrated with quantum mechanics data and is valid for high strains. Analytical solutions of the natural frequencies of various plates are obtained for the Canham bending model by assuming infinitesimal strains. These solutions are used for the verification of the numerical results. The performance of the model is illustrated by means of several examples. Modal analysis is performed for square plates under pure dilatation or uniaxial stretch, circular plates under pure dilatation or under the effects of an adhesive substrate, and carbon nanotubes under uniaxial \textcolor{cgn}{compression} or stretch. The adhesive substrate is modeled with van der Waals interaction (based on the Lennard-Jones potential) and a coarse grained contact model. It is shown that the analytical natural frequencies underestimate the real ones, and this should be considered in the design of devices based on graphene structures.

{{\bf Keywords}: Carbon nanotube (CNT); circular and square graphene plates; hyperelastic shell model; nonlinear finite elements; nonlinear frequencies.}

\vspace{-4mm}
\rule{\linewidth}{.15mm}

\vspace{7mm}
{\Large\bf List of important symbols}

\begin{tabbing}
$\bone$ \qquad~~~~~~~ \=  identity tensor in $\bbR^3$  \\
$\ba_\alpha$  \> co-variant tangent vectors of $\mathcal{S}$; $\alpha=1,2$ \\
$a_{\alpha\beta}$ \> co-variant components of the metric tensor of $\mathcal{S}$ \\
$a^{\alpha\beta}$ \> contra-variant components of the metric tensor of $\mathcal{S}$ \\
$\ba^\alpha$ \> contra-variant tangent vectors of $\mathcal{S}$; $\alpha=1,2$ \\
$\mathcal{S}_{0}$ \>  reference configuration of the manifold\\
$\mathcal{S}$ \> current configuration of the manifold\\
$b_{\alpha\beta}$ \> co-variant components curvature tensor of $\mathcal{S}$ \\
$\bE^{(0)}$ \> logarithmic surface strain \\
$\bE^{(0)}_{\text{dev}}$ \> deviatoric part of the logarithmic strain \\
$\Gamma^\gamma_{\alpha\beta}$ \> Christoffel symbols of the second kind \\
$H$ \> mean curvature of $\mathcal{S}$\\
$J$ \> surface area change of $\mathcal{S}$ \\
$\kappa$ \> Gaussian curvature of $\mathcal{S}$ \\
$\kappa_1$, $\kappa_2$ \> principal curvatures of $\mathcal{S}$ \\
$k_{\mathrm{p}}$ \> penalty parameter \\
$\lambda$  \> square root of the stretch ratio, i.e. $\sqrt{\lambda_1/\lambda_2}$ \\
$\lambda_1$, $\lambda_2$ \> principal surface stretches of $\mathcal{S}$ \\
$\bn$ \> surface normal of $\mathcal{S}$ \\
$\xi^{\alpha}$ \> parametric coordinates; $\alpha=1,2$ \\
$\bX$ \> reference position of the manifold \\
$\bx$ \> current position of the manifold \\
\end{tabbing}
\section{Introduction}
The high mechanical strength \citep{Akinwande2016}, thermal conductivity \citep{Balandin2011,Renteria2014} and electrical conductivity \citep{Lemme2010,Schwierz2010,Berashevich2010} of graphene have received much interest in recent years. The vibrational properties \textcolor{cgn}{(i.e. frequencies and mode shapes)} of graphene play an important role in analysis and design of graphene-based sensors and resonators. There are several studies on the development of new sensors using graphene-based structures \citep{Pumera2011,Fazelzadeh2014}. For example, graphene can be used in oscillators and electro-mechanical resonators \citep{Chen2013,natsuki2015,Kim2016}. \\
The effect of pre-stressing on the vibrational properties of graphene have been investigated by \citet{Gupta2010} and \citet{mustapha2015vibration}, while foundation effects have been studied by  \citet{Murmu2009}, \citet{Lee2014}, \citet{lee2014b} and \citet{mustapha2015vibration}. \textcolor{cgr2}{\citet{Sadeghi2010_01} use an atomistic method at the temperature of 19.3 K to calculated the nonlinear frequencies of graphene sheets.} The nonlinear vibration of sandwiches with graphene and piezoelectric layers have been modeled by \citet{Li2015}. \textcolor{cgr2}{\citet{Favata2012_01} propose an orthotropic shell model for CNTs. \citet{Ansari2014_01} use a non-local shell theory to include the size-effects in the calculation of the frequencies of single and double walled carbon nanotubes.} \textcolor{cgr2}{\citet{Hussain2017_01} obtain the natural freqencies of single walled carbon nanotubes by using Donnell thin shell theory.} \citet{Li2016} used a nonlinear finite element (FE) method to analyze large deformations and obtain the nonlinear frequencies of graphene membranes for nanomechanical applications. \textcolor{cgn}{A linear material model works well for infinitesimal strains. But, the mechanical properties of graphene vary in large strains. Hence, nonlinear hyperelastic material models should be used to model the material behavior in large strains \citep{Kumar2014_01,Ghaffari2017_01}}. Thermal vibration of rectangular, circular and annual graphene sheets are studied by \citet{PrasannaKumar2013}, \citet{Wang2015}, \citet{MOHAMMADI2014} and \citet{biswal2015}. The vibrational properties of multi-layer circular and rectangular graphene sheets are obtained by \citet{Kitipornchai2005} and \citet{Allahyari2016}. \citet{KE2012} has modeled the size-effects on vibrational properties of rectangular plates. The vibrational properties of a graphene sheet can be calculated by molecular mechanics \citep{Chowdhury2011} and molecular dynamics \citep{Arash2011}. \citet{STROZZI2014} have calculated the natural frequencies and mode shapes of CNTs by analytical approaches and validated them by experimental, atomistic and FE results. \citet{ARGHAVAN2011} have computed the natural frequencies, mode shapes and force vibration of CNTs.\\
All mentioned continuum models are limited by linear elastic material behavior. However, graphene shows nonlinear and anisotropic behavior under large deformations \citep{Kumar2014_01}. \textcolor{cgn}{\citet{Kumar2014_01} develop a hyperelastic material model for graphene that is based on  three strain invariants and several unknown
material constants. Those constants need to be determined from appropriate tests. A suitable approach for this are ab-initio calculations.
They are more accurate than molecular dynamics simulations, and they do not have difficulties with applying homogeneous strain states as is the case in experiments}. In addition, atomistic potentials \citep{Brenner1990_01,Brenner2002_01} underestimate elastic modulus \citep{Arroyo2004_01}. A wide range for the elastic modulus for graphene have been reported by \citet{Cao2014_01} that under or overestimate experimental and ab-initio results \citep{Kudin2001_01,Lee2008_01}. It should be mentioned that \citet{Gupta2010} used the MM3 potential and obtained a very close results to experimental and ab-initio results, but further investigations should be considered for large deformations. The nonlinear material model of \citet{Ghaffari2017_01} is used here to remedy these deficiencies and the consistency of the model with experimental and ab-initio results is verified analytically. \textcolor{cgn2}{Neglecting the bending stiffness can result in large frequency errors for low pre-tension and/or small sheets. However, the bending stiffness can be neglected for a large graphene sheet under significant pre-tension \citep{Sajadi2017_01,Ghaffari2017_01}.} \\
Isogeometric analysis (IGA) is a new computational technique that connects CAD and FE analysis \citep{Hughes2005}.
Recently, an isogeometric FE formulation has been developed by \citet{Sauer2014_01} for the analysis of liquid and solid membranes based on inherent curvilinear coordinates. It has been extended to anisotropic membranes by \citet{Roohbakhshan2016_01} and rotation-free shells by \citet{Sauer2017_01} and \citet{Duong2016_01}. This shell formulation has been applied to biomaterials and composites by \citet{Roohbakhshan2016_02,Roohbakhshan2017_01} and to graphene by \citet{Ghaffari2017_01}. The latter work uses the anisotropic membrane model of \citet{Kumar2014_01} and extends it to a shell formulation by including the Canham model \citep{CANHAM1970_01}. This new model can simulate the anisotropic behavior of graphene-based structures under large deformation and it has been used to simulate indentation and peeling of graphene sheets and torsion and bending of carbon nanotubes (CNT). \textcolor{cgn2}{Thermal fluctuation are not considered in the current study. But the proposed model does allow for an extension to those. Thermal fluctuations can result in structural softening, but they can be suppressed with small pre-strains above 1\% \citep{Roldan2011_01,Gornyi2017_01}. Therefore \citet{Kumar2016_01} and \citet{Ghaffari2017_01} obtain similar material properties as the experimental results of \citet{Lee2008_01} by using a hyperelastic constitutive law that disregards thermal fluctuations.}\\
This work reports new data of the effect of stretching and contact on the vibrational frequencies. The major novelties of this work are
\begin{itemize}
  \item The frequencies of square and circular graphene plates, and CNTs are obtained under nonlinear deformations.
  \item The effects of substrate adhesion on the frequencies of a graphene plate are investigated and instabilities are found for certain adhesion energies.
  \item The analytical solutions of the natural frequencies are obtained for the Canham bending model.
  \item The hyperelastic material model is valid under large deformations. So, the limitation of a linear elastic material model in previous studies is surpassed.
  \item The current formulation can be extended to capture finite size effects, such as those reported in \citep{Polizzotto2001_01,Wang2011_01}.
\end{itemize}
The remainder of this paper is organized as follows: In Sec.~\ref{s:material_model}, the hyperelastic material model of \citet{Ghaffari2017_01} is summarized. Its FE formulation is described in Sec.~\ref{s:finite_element_formulation}. In Sec.~\ref{s:numerical_examples}, the proposed numerical formulation is verified with analytical solutions. The nonlinear modal analysis is benchmarked with several numerical examples. The nonlinear modal analysis is conducted for a square sheet under pure dilatation and uniaxial stretch, a circular sheet under pure dilatation, a circular plate under adhesive effects of a substrate and CNTs under axial compression or stretch. The paper is concluded in Sec.~\ref{s:conclusion}.

\section{Material model}\label{s:material_model}
\begin{figure}
    \begin{subfigure}[t]{1\linewidth}
        \centering
     \includegraphics[height=55mm]{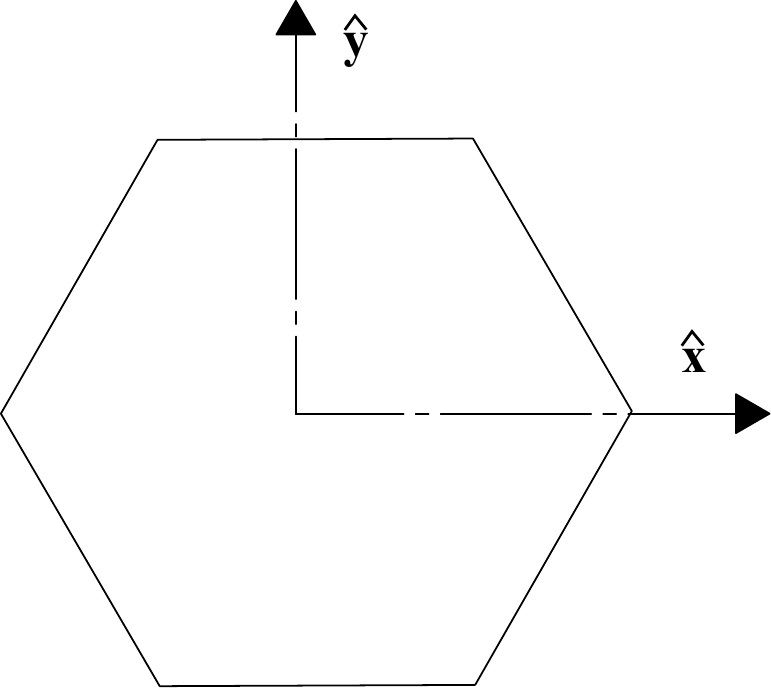}
    \end{subfigure}
    \caption{Anisotropy of the material: Orthonormal vectors characterize the graphene lattice. \textcolor{cgn}{$\hat{\bx}$ and $\hat{\by}$ are the armchair and zigzag directions.}}
    \label{f:lattice}
\end{figure}

\textcolor{cgn}{Measured strains up to $12.5\%$ \citep{Hugo2014_01}, $20\%$ \citep{Tomori2011_01} and even $25\%$ \citep{Lee2008_01,Kim2009_01} have been reported.} Therefore, a nonlinear hyperelastic constitutive law for graphene is considered in this work. The membrane part of the strain energy is based on logarithmic strain and calibrated by density function theory (DFT) data \citep{Kumar2014_01}. It is adapted to a curvilinear formulation by \citet{Ghaffari2017_01}. The membrane model is extended to a shell formulation by including the Canham bending strain energy. The required bending parameter is calibrated by quantum mechanics data \citep{Ghaffari2017_01}.\\
An anisotropic functional can be based on an isotropic functional by including structural tensors \citep{Zheng_Theory_of_representations_for_tensor_functions, itskov2015}. Based on the symmetry group of the graphene lattice and its structural tensor, a set of invariants can be introduced as \citep{Kumar2014_01}
\eqb{lll}
\sJ_1 \dis \epsilon_\mra = \ln J~;\quad J:=\lambda_1\,\lambda_2~, \\[3mm]
\sJ_2 \dis \ds\ \frac{1}{4}\,\gamma_i^2 = \frac{1}{2}\,\bE^{(0)}_{\text{dev}}:\bE^{(0)}_{\text{dev}} =(\ln\lambda)^2; \lambda:=\sqrt{\frac{\lambda_1}{\lambda_2}}\quad ;~\lambda_1 \ge \lambda_2~,\\[3mm]
\sJ_3 \dis \ds\frac{1}{8}\,\gamma_\theta^3 =  \ds\frac{1}{8}\left[ \left(\hat{\bM}:\bE^{(0)}_{\text{dev}}\right)^3 - 3\,\left(\hat{\bM}:\bE^{(0)}_{\text{dev}}\right)\left(\hat{\bN}:\bE^{(0)}_{\text{dev}}\right)^2\right]=(\ln\lambda)^3\cos(6\theta)~.
\label{e:defsJ}
\eqe
$\sJ_1$ and $\sJ_2$ capture the isotropic response of the material and $\sJ_3$ is related to the anisotropic features. $\bE^{(0)}_{\text{dev}}$ is the deviatoric part of the logarithmic strain.
In addition, $\hat{\bM}$ and $\hat{\bN}$ are defined as
\eqb{lll}
\hat{\bM} := \ds \hat{\bx} \otimes \hat{\bx} - \hat{\by} \otimes \hat{\by}~,
\eqe
\eqb{lll}
\hat{\bN} := \ds \hat{\bx} \otimes \hat{\by} + \hat{\by} \otimes \hat{\bx}~,
\eqe
where $\hat{\bx}$ and $\hat{\by}$ are the two orthonormal vectors (see Fig.~\ref{f:lattice}). $\theta$ is the maximum stretch angle relative to the armchair direction and defined as
\eqb{lll}
\theta \is \arccos{(\bY_{\!1}\cdot\hat{\bx})}~,
\eqe
\textcolor{cgn}{where $\bY_{\!1}$ is the direction of the maximum stretch.}
The strain energy density per unit area of the initial configuration is decomposed into the membrane and bending parts $W_{\mathrm{m}}$ and $W_{\mathrm{b}}$ parts. The membrane part includes the pure dilatation and deviatoric parts $W^{\mathrm{dil}}_{\mathrm{m}}$ and $W^{\mathrm{dev}}_{\mathrm{m}}$. The strain energy density can thus be written as
\eqb{l}
W(\sJ_1,\sJ_2,\sJ_3,\kappa_1,\kappa_2) = W^{\mathrm{dil}}_{\mathrm{m}}(\sJ_1) + W^{\mathrm{dev}}_{\mathrm{m}}(\sJ_2,\,\sJ_3;\sJ_1)+W_{\mathrm{b}}(\kappa_1,\kappa_2)~.
\eqe
These terms are defined as
\eqb{lll}
 W^{\mathrm{dil}}_{\mathrm{m}} \dis \varepsilon\big[1 - (1+\hat{\alpha}\,\epsilon_\mra)\,\exp(-\hat{\alpha}\,\epsilon_\mra)\big]~,\\[2mm]
 W^{\mathrm{dev}}_{\mathrm{m}} \dis 2\,\mu(\epsilon_\mra)\,\sJ_2 + \eta(\epsilon_\mra)\,\sJ_3~,
\eqe
\eqb{lll}
\ds W_{\mathrm{b}} \dis \ds J\frac{c}{2}\left(\kappa_1^2+\kappa_2^2\right)~.
\eqe
where $\mu$ and $\eta$ are defined as
\eqb{lll}
\mu(\epsilon_\mra) \dis \mu_0 - \mu_1\,e^{\hat{\beta}\,\epsilon_\mra}~,\\[2mm]
\eta(\epsilon_\mra) \dis \eta_0 - \eta_1\,\epsilon_\mra^2~.
\eqe
The material constants $\varepsilon$, $\hat{\alpha}$, $\mu_0$, $\mu_1$, $\hat{\beta}$, $\eta_0$, $\eta_1$ and $c$ are defined in Tabs.~\ref{t:Graphene_cons} and \ref{t:graphene_bending_material_cons}. The parameters in Tab.~\ref{t:Graphene_cons} are computed from local density approximation (LDA) and generalized gradient approximation (GGA). In Tab.~\ref{t:graphene_bending_material_cons}, FGBP and SGBP indicate the first and second generation Brenner potential respectively, while QM indicates quantum mechanics.\\
The first derivative of the strain energy density, relative to the metric and bending tensors, gives the co-variant components of the Kirchhoff and moment tensor as
\eqb{l}
\tauab = \ds 2\pa{W}{\auab}~,
\eqe
\eqb{l}
\Mab_{0} = \ds \pa{W}{\buab}~.
\eqe
$\tauab$ and $\Mab_{0}$ are given in \citet{Ghaffari2017_01} and \citet{Sauer2017_01}.
\begin{table}
  \centering
    \begin{tabular}{c c c c c c c c }
      \hline
        & $\hat{\alpha}$ & $\varepsilon~[\textnormal{N/m}]$  &  $\mu_0~[\textnormal{N/m}]$ & $\mu_1~ [\textnormal{N/m}]$ & $\hat{\beta}$ & $\eta_0~[\textnormal{N/m}]$ & $\eta_1~[\textnormal{N/m}]$ \\
      \hline
      GGA  & 1.53 & 93.84  & 172.18 & 27.03 & 5.16 & 94.65 & 4393.26 \\
      LDA  & 1.38 & 116.43 & 164.17 & 17.31 & 6.22 & 86.9${}^{\text{a}}$ & 3611.5${}^{\text{a}}$ \\
      \hline
    \end{tabular}
  \caption{Material constants of graphene \citep{Kumar2014_01}. ${}^{\text{a}}$ See correction of \citet{Kumar2016_02}. (\citet{Kumar2014_01} contains errors)}\label{t:Graphene_cons}
\end{table}

\begin{table}
  \centering
     \begin{tabular}{c c c c}
     \hline
     & FGBP & SGBP & QM \\
     \hline
    $c$~[nN$\cdot$nm] & 0.133 & 0.225 & 0.238 \\
    \hline
  \end{tabular}
  \caption{Bending stiffness according to various atomistic models \citep{Kudin2001_01,Lu2009_01}. \textcolor{cgn}{FGBP = first generation Brenner potential; SGBP = second generation Brenner potential; QM = quantum mechanics.}}
  \label{t:graphene_bending_material_cons}
\end{table}
\section{Finite element formulation}\label{s:finite_element_formulation}
In this section, the nonlinear modal analysis based on the FE method is presented. Based on the Galerkin method, the weak form of the equations of motion is obtained \textcolor{cgn}{by using principle of virtual work \citep{Bonet2008_01}}. The standard linearization is utilized and discretization is conducted based on NURBS shape functions \citep{Sauer2017_01}.
Using the Kirchhoff-Love shell theory, the equations of motion for a two-dimensional (2D) manifold can be written as
\eqb{l}
\bT_{;\alpha}^{\alpha} + \bolds{f}=\rho\,\dot{\bv}~~~\forall~\bx~\in~\mathcal{S}~,
\eqe
where $\bolds{f}$, $\dot{\bv}$ and $\rho$ are body force, acceleration and mass density, respectively.  $\bT^{\alpha}$ is defined based on the Cauchy stress tensor $\bsig$ as
\eqb{lll}
\bT^{\alpha} \dis \bsig^{\mathrm{T}}\cdot\ba^{\alpha}=N^{\alpha\beta}\,\ba_{\beta}+S^{\alpha}\,\bn~,
\eqe
where $\bsig$ is defined as
\eqb{lll}
\bsig \dis N^{\alpha\beta}\,\ba_{\alpha}\otimes \ba_{\beta}+S^{\alpha}\,\ba_{\alpha}\otimes\bn~,
\eqe
where
\eqb{lll}
N^{\alpha\beta} \is \ds \sigma^{\alpha\beta}+b_{\gamma}^{\alpha}\,M^{\gamma\beta}~,
\eqe
\eqb{lll}
S^{\alpha} \is \ds -M_{;\beta}^{\beta\alpha}~,
\eqe
and $b_{\gamma}^{\alpha}=a^{\alpha\,\eta}\,b_{\eta\,\gamma}$ , $\sigma^{\alpha\beta} = \tauab/J$ and $M^{\alpha\beta} = M_{0}^{\alpha\beta}/J~.$
The mixed in-plane components of $\bsig$ are defined as
\eqb{lll}
N^{\alpha}_{\beta} \dis N^{\alpha\gamma}\,a_{\gamma\beta}~.
\label{e:mix_N}
\eqe
The discretized weak form can be written as
\eqb{lll}
\mM\,\ddot{\muu} +\mf_{\text{int}}=\mf_{\text{ext}}~,
\eqe
where $\mM$ is the mass matrix and $\mf_{\text{int}}$ and $\mf_{\text{ext}}$ are the internal and external force vectors. \textcolor{cgn}{Using Taylor expansion around $\hat{\muu}$ such that $\muu=\hat{\muu}+\dif\muu$ ,} the linearized relation can be written as
\eqb{lll}
\mM\,\dif\ddot{\muu} +\mK\,\dif\muu =-(\mf_{\text{int}}+\mM\,\ddot{\hat{\muu}}-\mf_{\text{ext}})~,
\label{e:weak_form_dyn}
\eqe
where $\mK:=\partial(\mf_{\text{int}}-\mf_{\text{ext}})/\partial\muu$ is the stiffness matrix. $\mK$ includes the material and initial stress (geometric) stiffness matrices. The mass and stiffness matrices are given in \ref{s:stiffness_mass_matrix}. The detailed derivation of the mass and stiffness matrices and a efficient formulation for their numerical implementation can be found in \citet{Sauer2017_01} and \citet{Duong2016_01}.\\
Graphene has a single atom thickness and its frequencies change severely under external loads. If the structure is excited by a dynamic load which has the same frequency and mode shape as the specimen, the structure may be unstable and fail. The vibrational properties of structures can be manipulated to avoid instabilities due to resonance. A pre-stretch can be used to change the natural frequencies and shift them away from the frequency of the applied external loads. \textcolor{cgn}{The influence of damping on the vibrational properties of graphene-based resonators is important and should be considered to improve the modeling of these sensors. Experimental results suggest a nonlinear damping formulation for graphene-based the sensors \citep{Eichler2011,Moser2013}}. In addition, sensors can be used to measure the mass of a nano particle based on the vibrational properties of sensors \citep{natsuki2015}. The vibrational properties of these \textcolor{cgn}{sensors} can be tuned by specific pre-stretching to increase sensor precision for a suitable ranges of mass. Furthermore, the substrate influence on the vibrational properties of the graphene sheet should be modeled for an accurate design to reduce the cost calibration.\\
Eq.~(\ref{e:weak_form_dyn}) is linearized to obtained $\mK$ and calculate the frequencies \citep{kerschen2014}. In modal analysis, it is assumed that structures have a periodic response that can be defined as
\eqb{lll}
\dif \muu \dis \dif \bar{\muu}\,e^{-i\omega t}~,
\eqe
where $\bar{\muu}$ are $\omega$ are the mode shape and frequency of the structure. So, the generalized \textcolor{cgn}{eigenvalue problem of the nonlinear system} can be written as
\eqb{lll}
\mK\,\dif\bar{\muu}=\omega^2\,\mM\,\dif\bar{\muu}~.
\eqe

\section{Numerical examples}\label{s:numerical_examples}
In this section, three numerical examples are shown: a square plate, a circular plate and a carbon nanotube. For the first two examples the FE formulation is verified in the linear elastic regime by comparison with analytical solutions. The analytical solution for the natural frequencies\footnote{The natural frequencies are the frequencies in the linear elastic regime without pre-deformation or pre-loading. The natural frequencies are indicated by ``$\hat{\omega}$''.} and mode shapes are provided in \ref{s:analytical_sol_modal}. The load cases of pure dilatation, uniaxial stretch and interaction with an adhesive substrate are considered. The investigation is considered up to the instability point. The elasticity tensor loses its ellipticity \citep{Kumar2014_01} at the instability point and the frequencies of the membrane mode shapes become zero.

\subsection{Vibrating square plates}
\begin{figure}
\begin{center} \unitlength1cm
\begin{picture}(18,9.5)
\put(2.,7.5){\includegraphics[width=35mm,trim=1.5cm 0cm 1.5cm 3cm,clip]{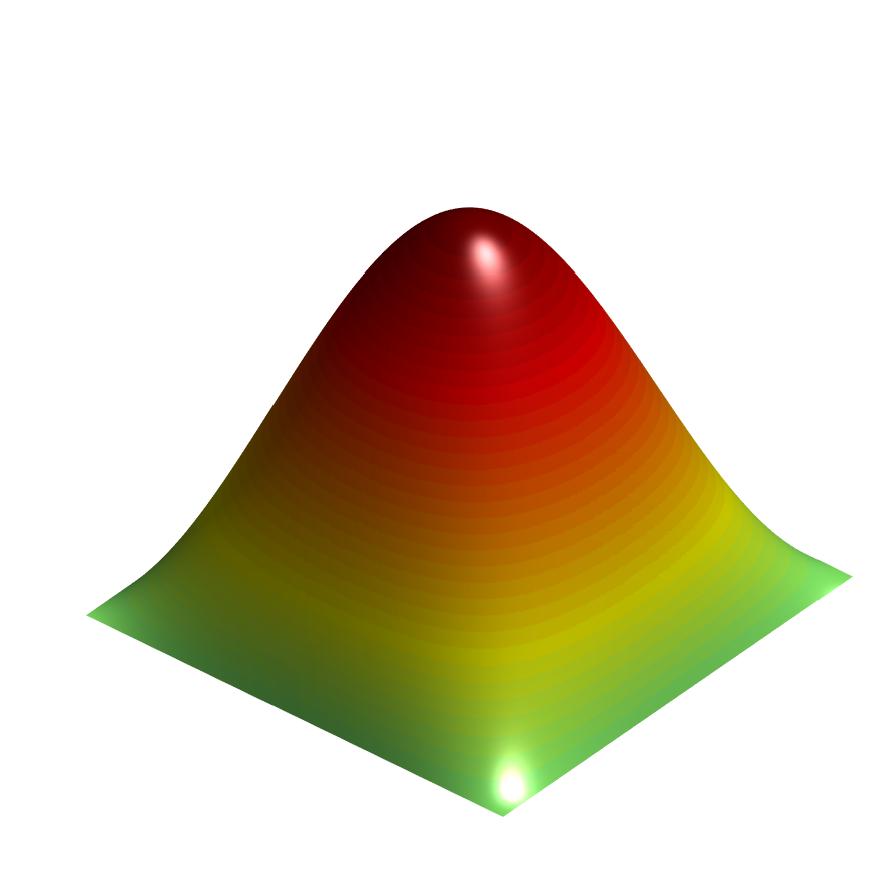}}
\put(5.5,7.5){\includegraphics[width=35mm,trim=1.5cm 3cm 1.5cm 3cm,clip]{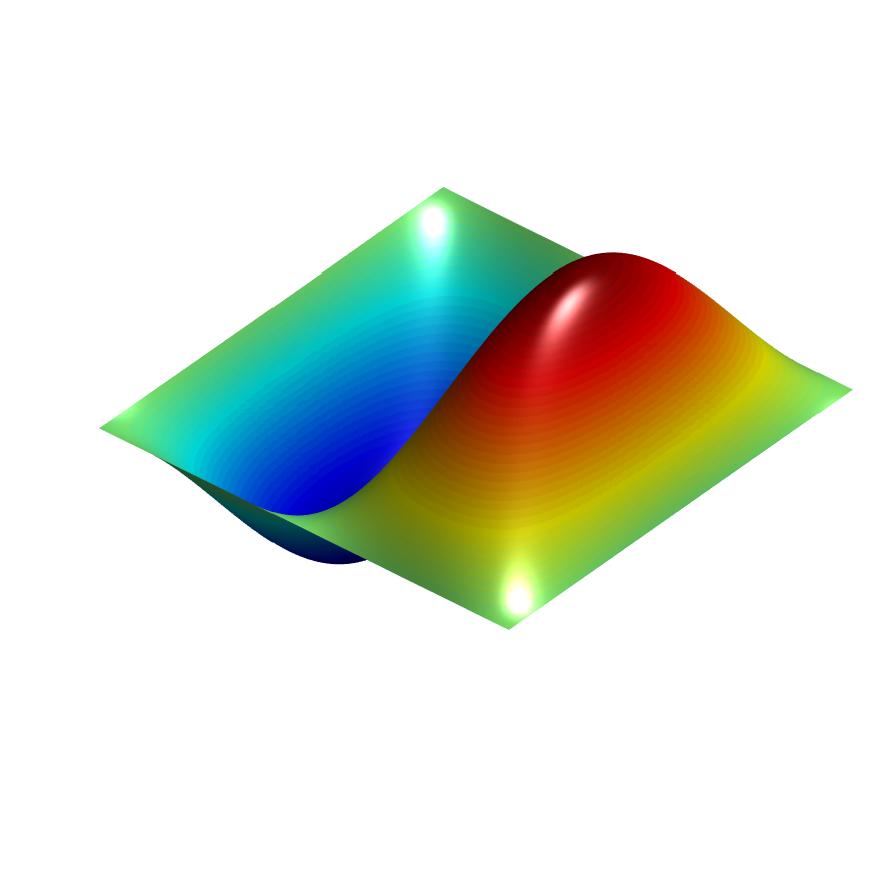}}
\put(9,7.5){\includegraphics[width=35mm,trim=1.5cm 3cm 1.5cm 3cm,clip]{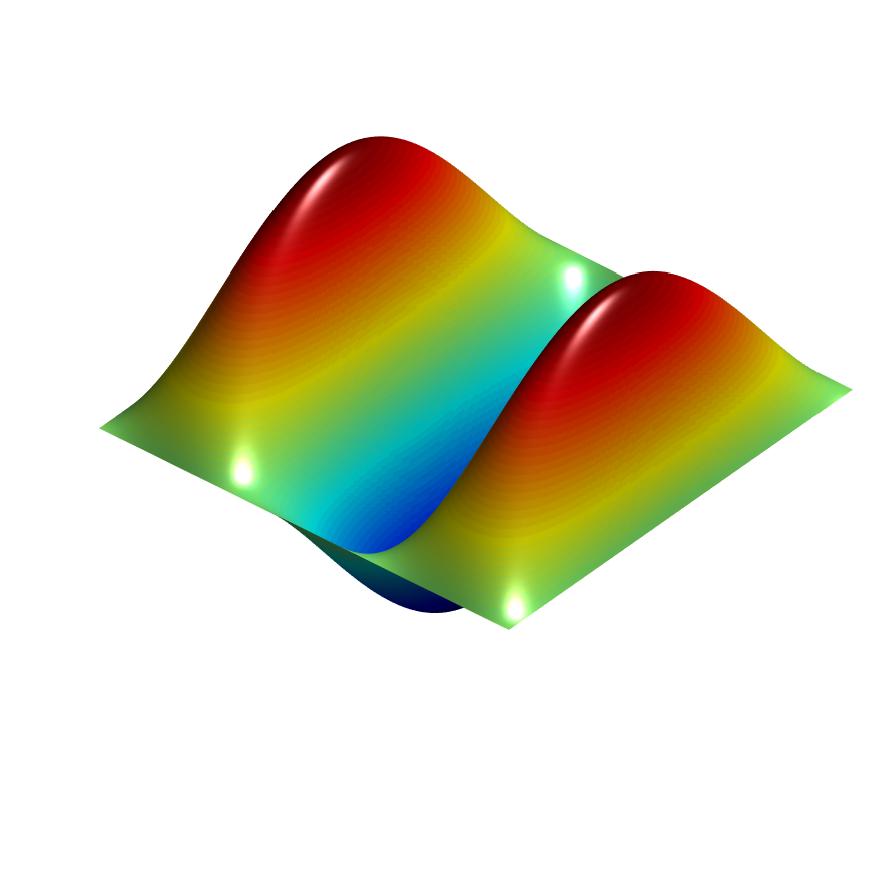}}
\put(2,4){\includegraphics[width=35mm,trim=1.5cm 1.5cm 1.5cm 3cm,clip]{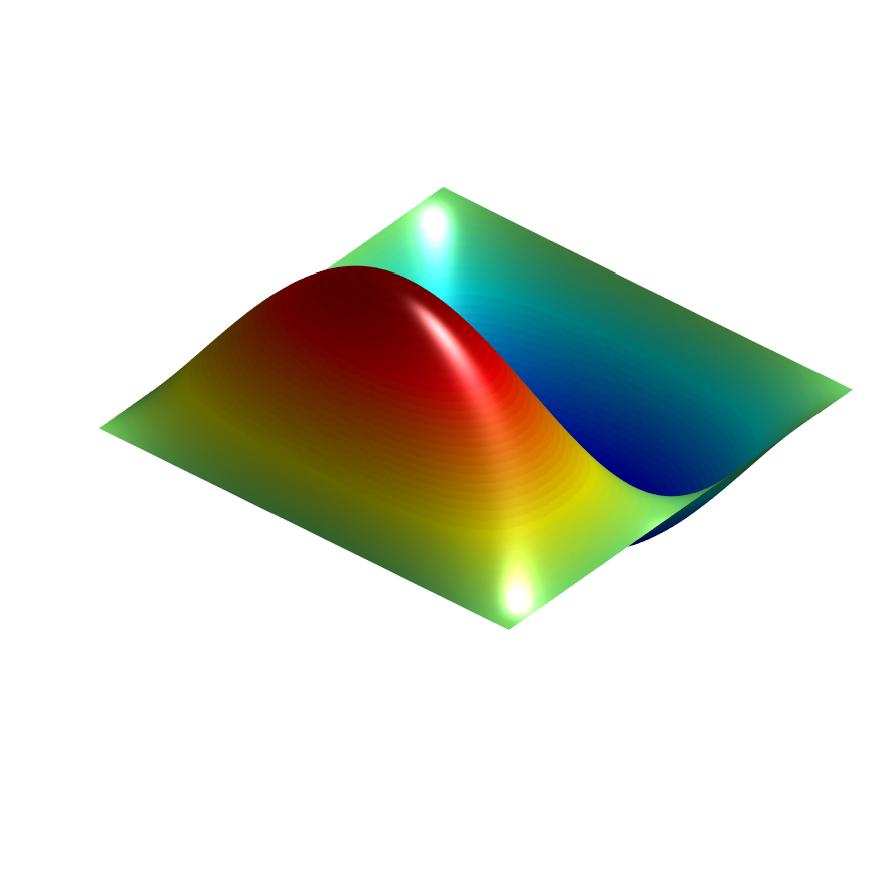}}

\put(5.5,4){\includegraphics[width=35mm,trim=1.5cm 1.5cm 1.5cm 1.5cm,clip]{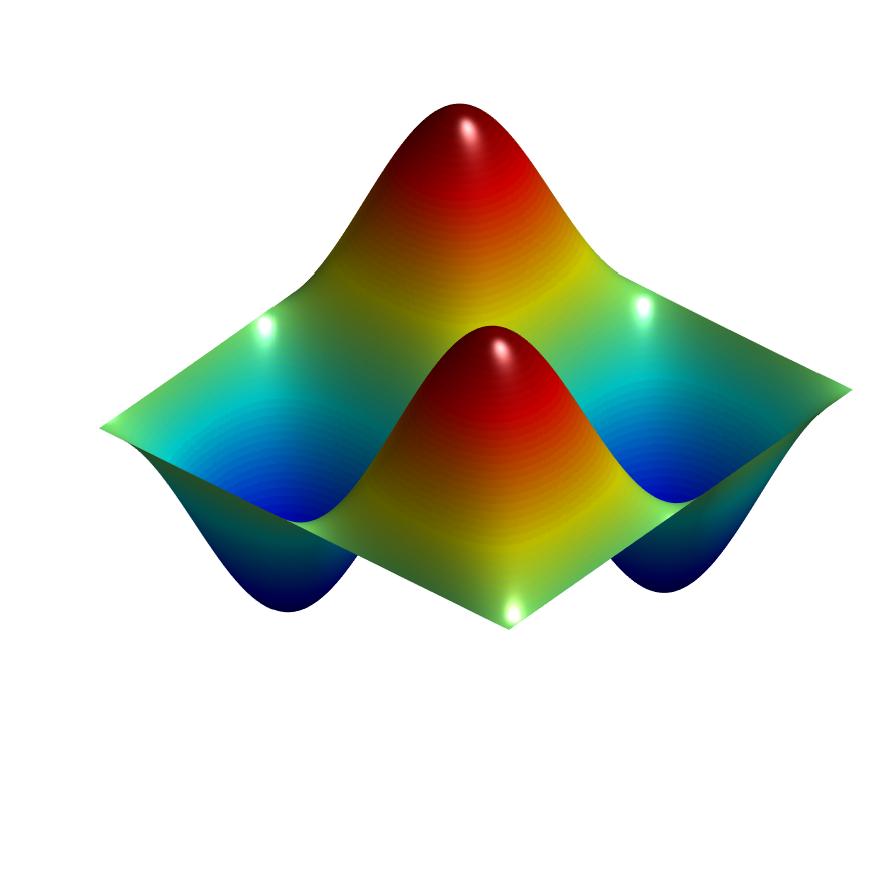}}
\put(9,4){\includegraphics[width=35mm,trim=1.5cm 1.5cm 1.5cm 1.5cm,clip]{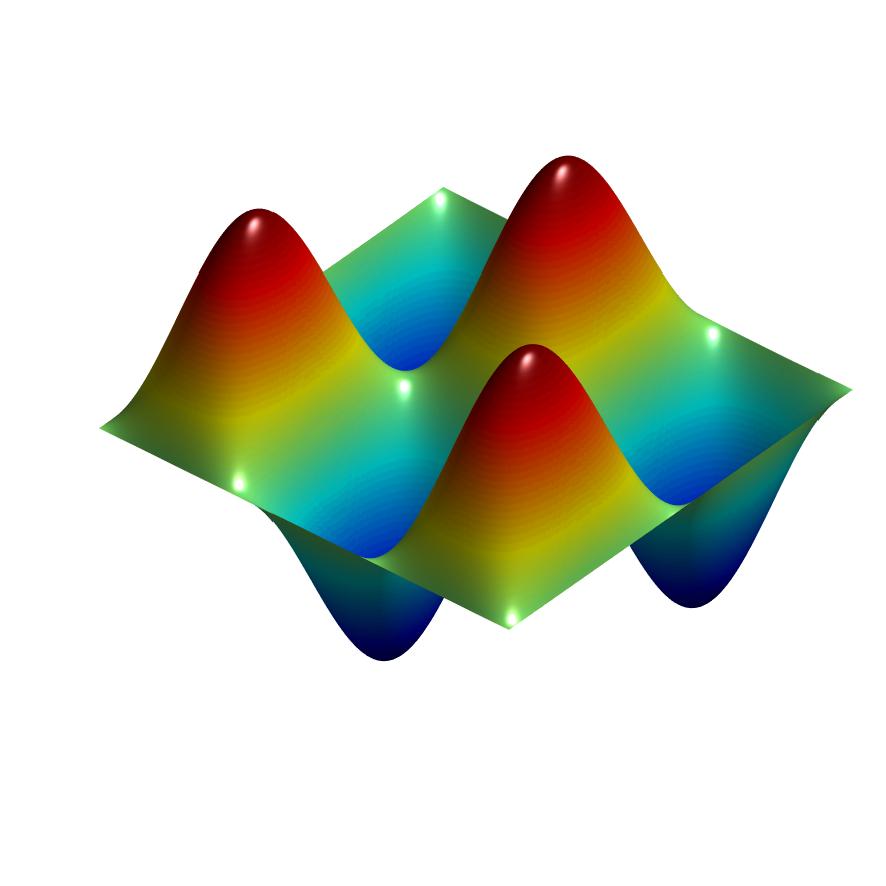}}

\put(2,0.75){\includegraphics[width=35mm,trim=1.5cm 1.5cm 1.5cm 1.5cm,clip]{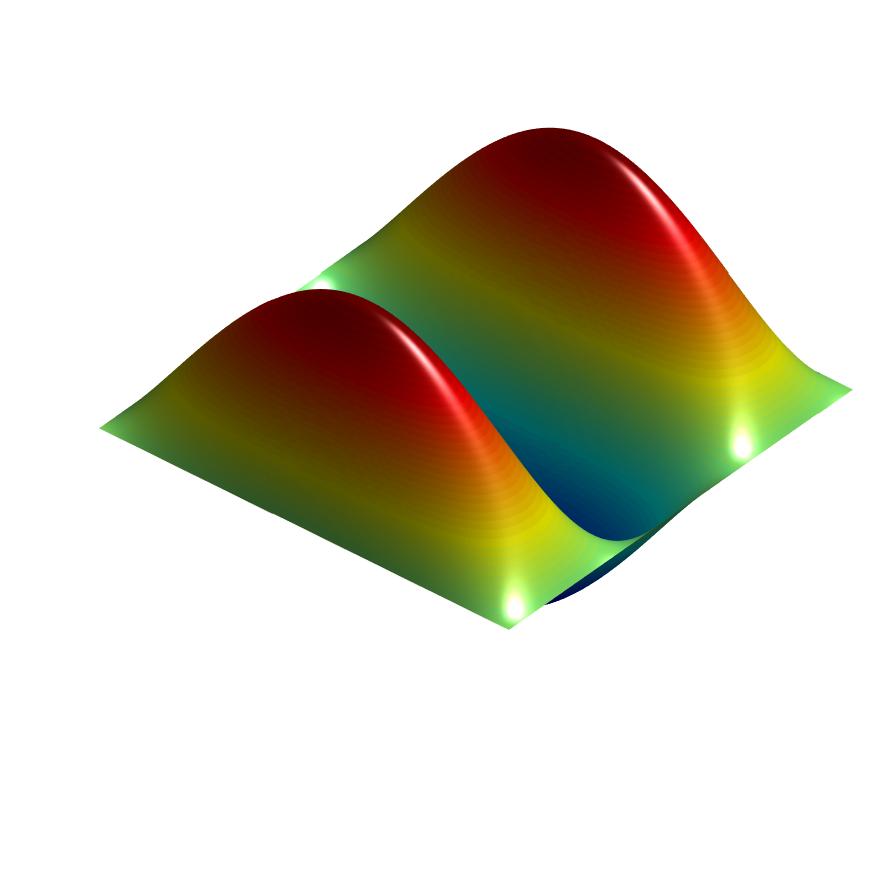}}
\put(5.5,0.75){\includegraphics[width=35mm,trim=1.5cm 1.5cm 1.5cm 1.5cm,clip]{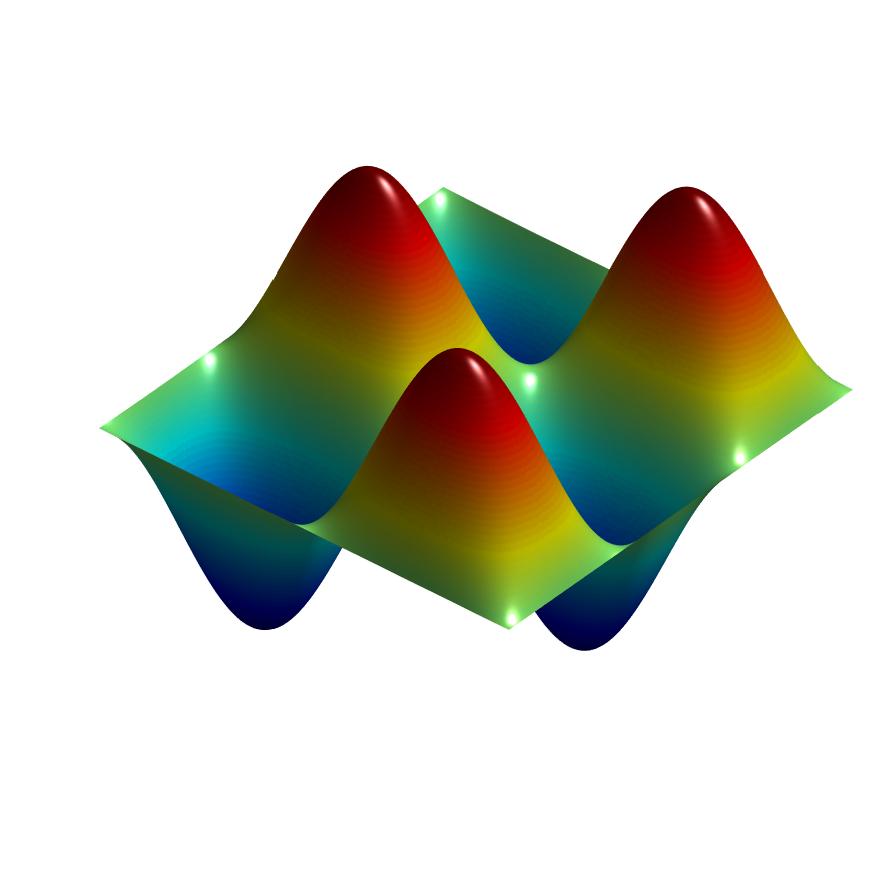}}
\put(9,0.75){\includegraphics[width=35mm,trim=1.5cm 1.5cm 1.5cm 1.5cm,clip]{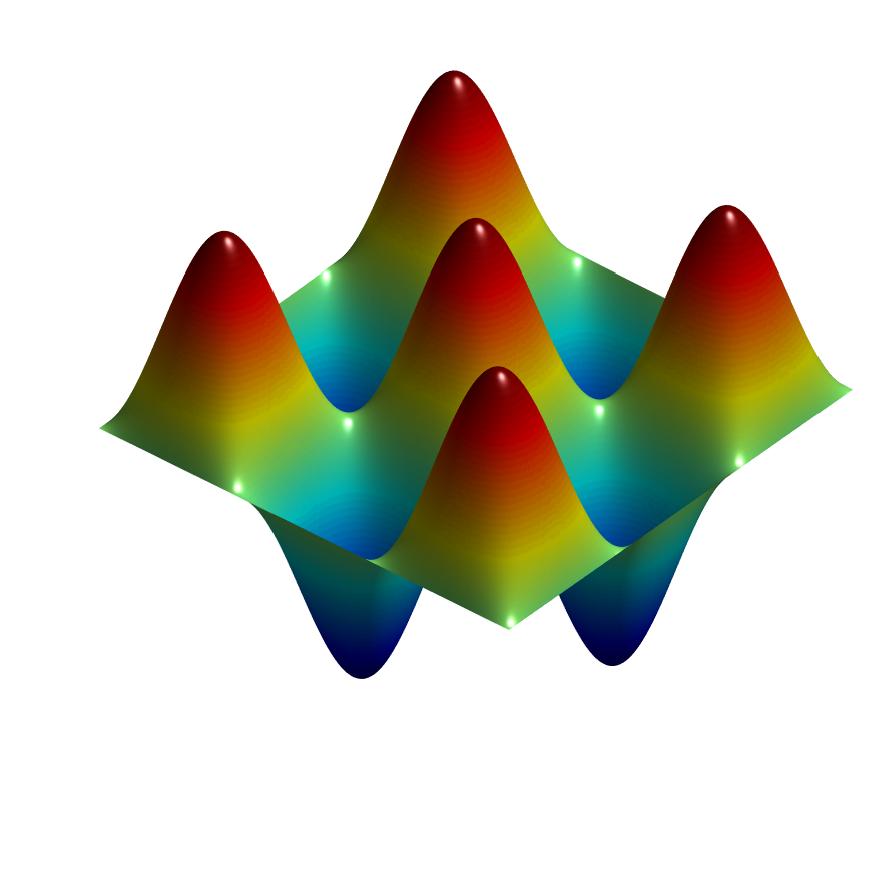}}
\put(2.5,1){{\small $\hat{f}_{(3,1)}=0.35136$}}
\put(6.,1){{\small $\hat{f}_{(3,2)}=0.45677$}}
\put(9.5,1){{\small $\hat{f}_{(3,3)}=0.59732$}}

\put(2.5,4.5){{\small $\hat{f}_{(2,1)}=0.17568$}}
\put(6.,4.5){{\small $\hat{f}_{(2,2)}=0.28109$}}
\put(9.5,4.5){{\small $\hat{f}_{(2,3)}=0.45677$}}

\put(2.5,7.5){{\small $\hat{f}_{(1,1)}=0.07027$}}
\put(6.,7.5){{\small $\hat{f}_{(1,2)}=0.17568$}}
\put(9.5,7.5){{\small $\hat{f}_{(1,3)}=0.35136$}}

\end{picture}
\vspace{-15mm}
\caption{Vibrating square plate under zero pre-load: First 9 mode shapes and natural frequencies $\hat{f}_{(m,n)}=\hat{\omega}_{(m,n)}/2\pi$ in [THz], $\hat{\omega}_{(m,n)}=\omega_{(m,n)} \big |_{\bE^{(0)}=\bolds{0}}$. $m$ and $n$ are the number of half waves in the mode shapes along $\hat{\bx}$ and $\hat{\by}$ directions. The boundary is simply supported.}
\label{f:rect_simply_mode_shapes}
\end{center}
\end{figure}
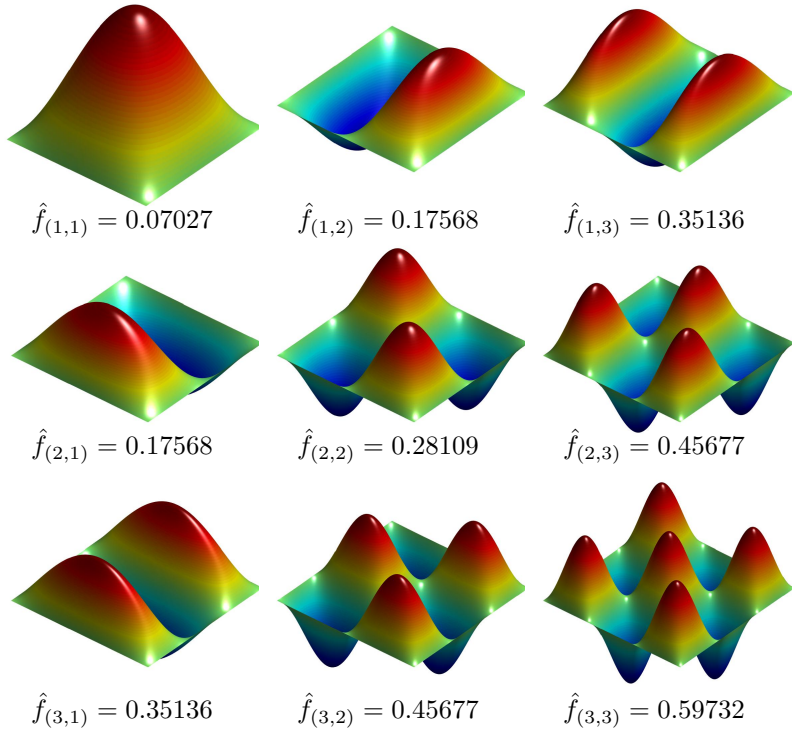

\begin{figure}
        \centering
    \includegraphics[height=60mm]{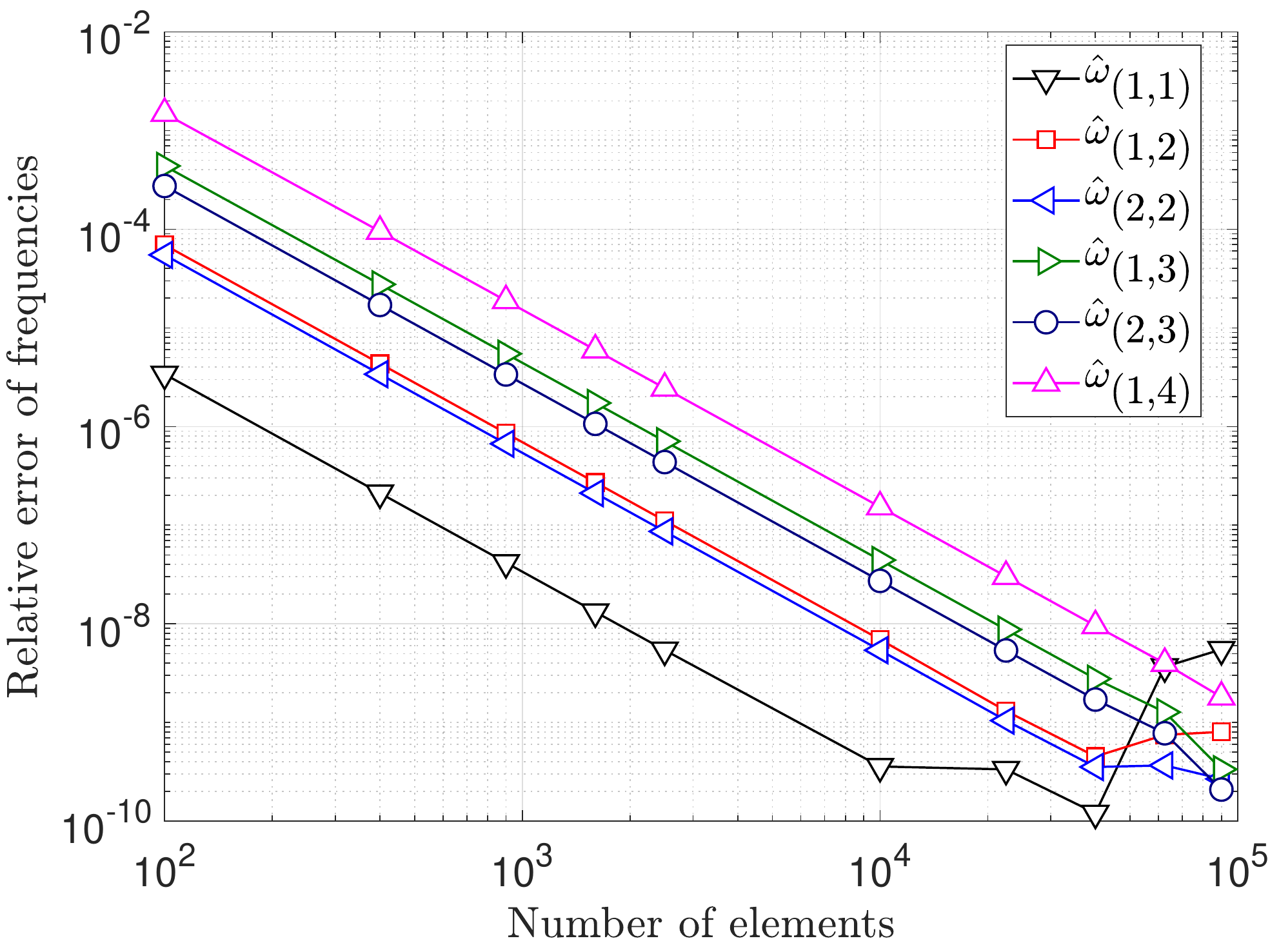}
    \caption{Vibrating square plate under zero pre-load: Relative error of the natural frequencies $\hat{\omega}_{(m,n)}=\omega_{(m,n)} \big |_{\bE^{(0)}=\bolds{0}}$ vs. mesh refinement. The analytical solution of (\ref{e:freq_lin_rect_simp}) is used as reference. $m$ and $n$ are defined in Fig.~\ref{f:rect_simply_mode_shapes}. The boundary is simply supported.}
    \label{f:Rect_simply_natural_freqeuncy_conv}
\end{figure}

\begin{figure}
    \begin{subfigure}{0.49\textwidth}
        \centering
    \includegraphics[height=58mm]{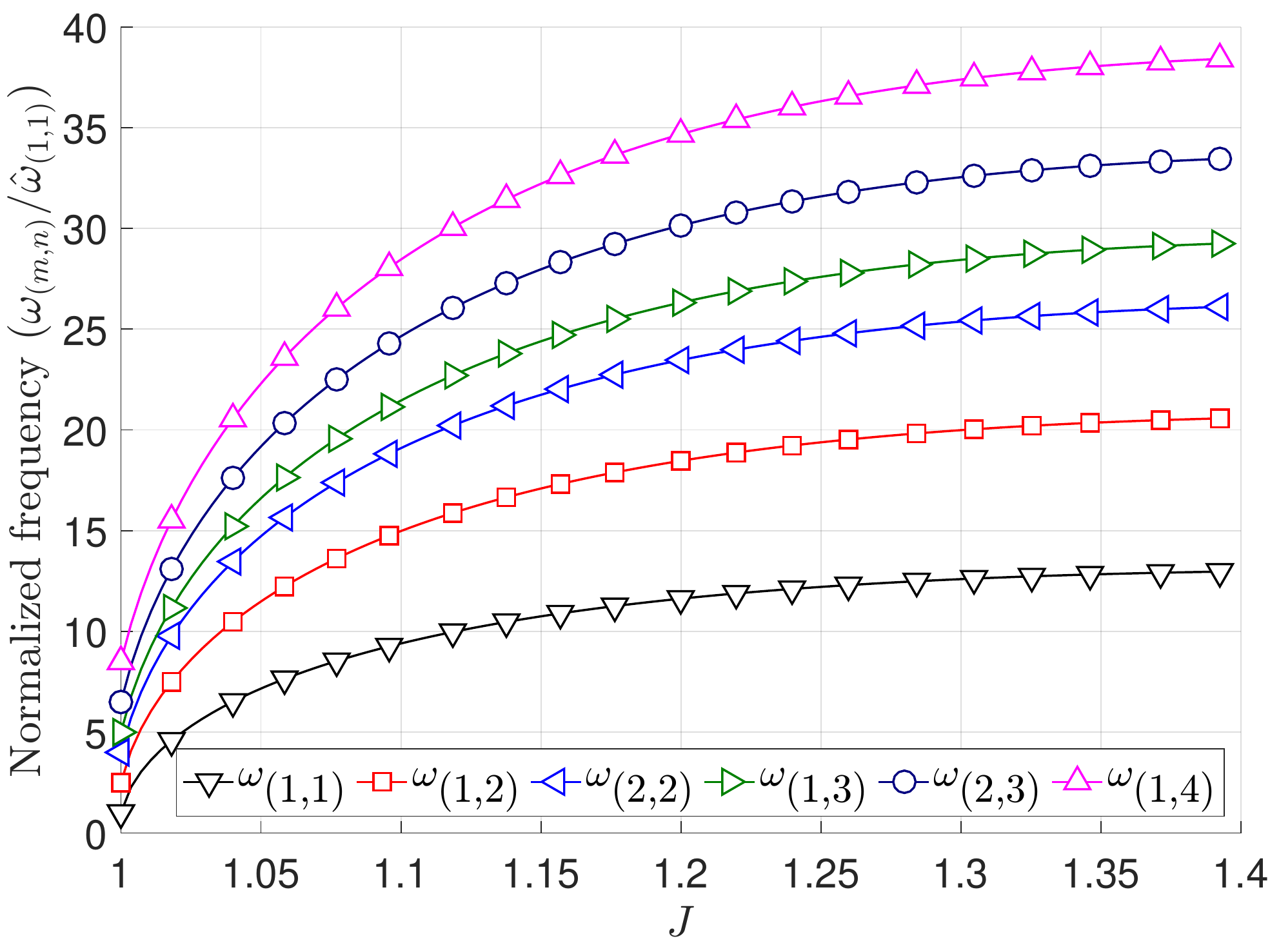}
        \subcaption{}
        \label{f:Rect_simply_freqeuncy_variation_biaxial}
    \end{subfigure}
    \begin{subfigure}{0.49\textwidth}
        \centering
    \includegraphics[height=58mm]{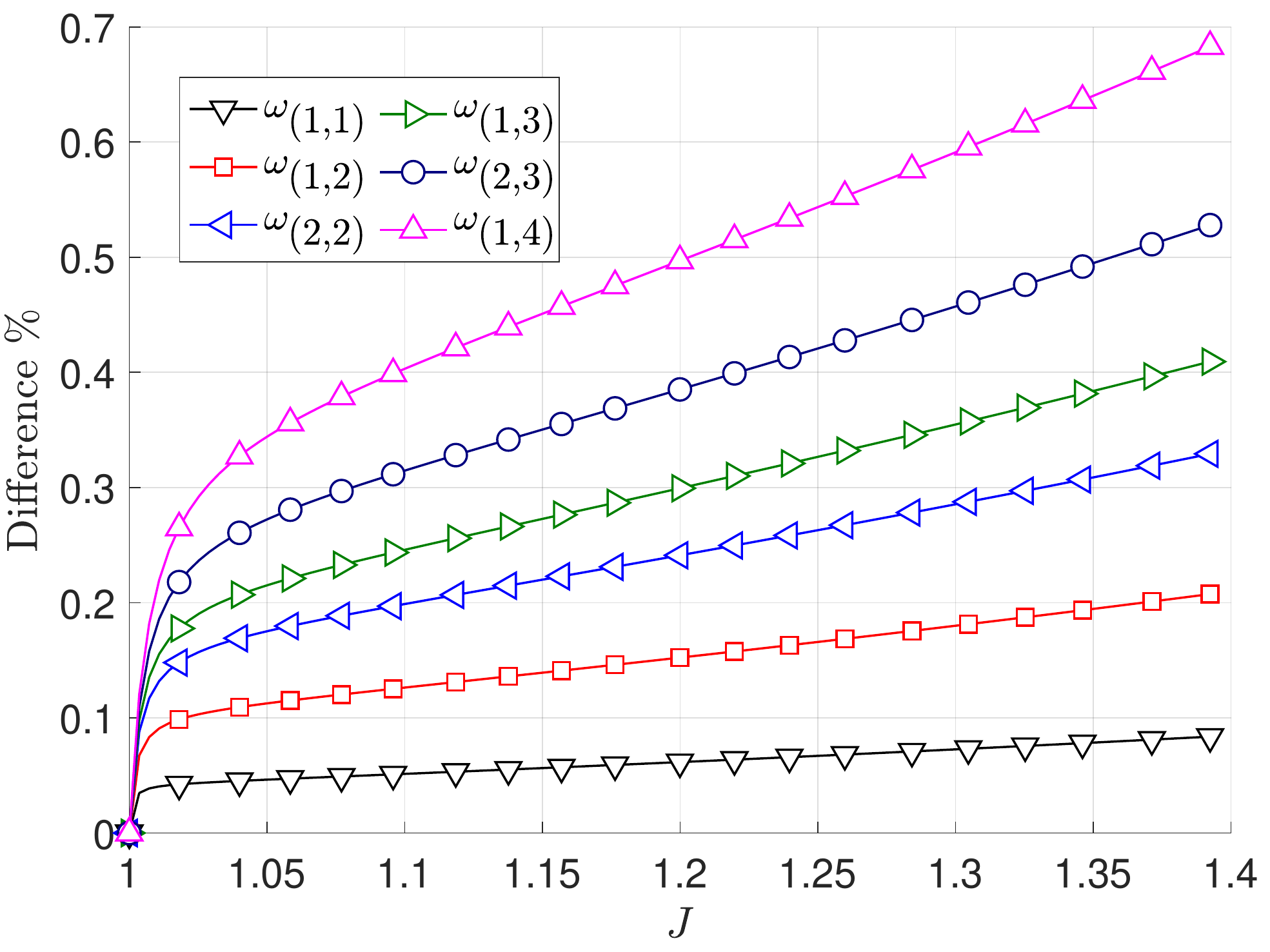}
        \subcaption{}
        \label{f:Rect_simply_freqeuncy_variation_biaxial_difference_with_ana}
    \end{subfigure}
    \caption{Vibrating square plate under pure dilatation: (\subref{f:Rect_simply_freqeuncy_variation_biaxial}) Variation of the frequencies and (\subref{f:Rect_simply_freqeuncy_variation_biaxial_difference_with_ana}) difference of the numerical results and the analytical solution of Eq.~\eqref{e:analytical_sol_rect_nonlinear_modal} in dependence of the surface stretch $J$. The plate has an edge length of 5 nm. $m$ and $n$ in $\omega_{(m,n)}$ are defined in Fig.~\ref{f:rect_simply_mode_shapes}. The results are normalized by $\hat{\omega}_{(1,1)}= \omega_{(1,1)}\big |_{J=1}$ (the first natural frequency). \textcolor{cgn}{The boundary is simply supported.}}
\end{figure}

\begin{figure}
    \begin{subfigure}{0.495\textwidth}
        \centering
    \includegraphics[height=60mm]{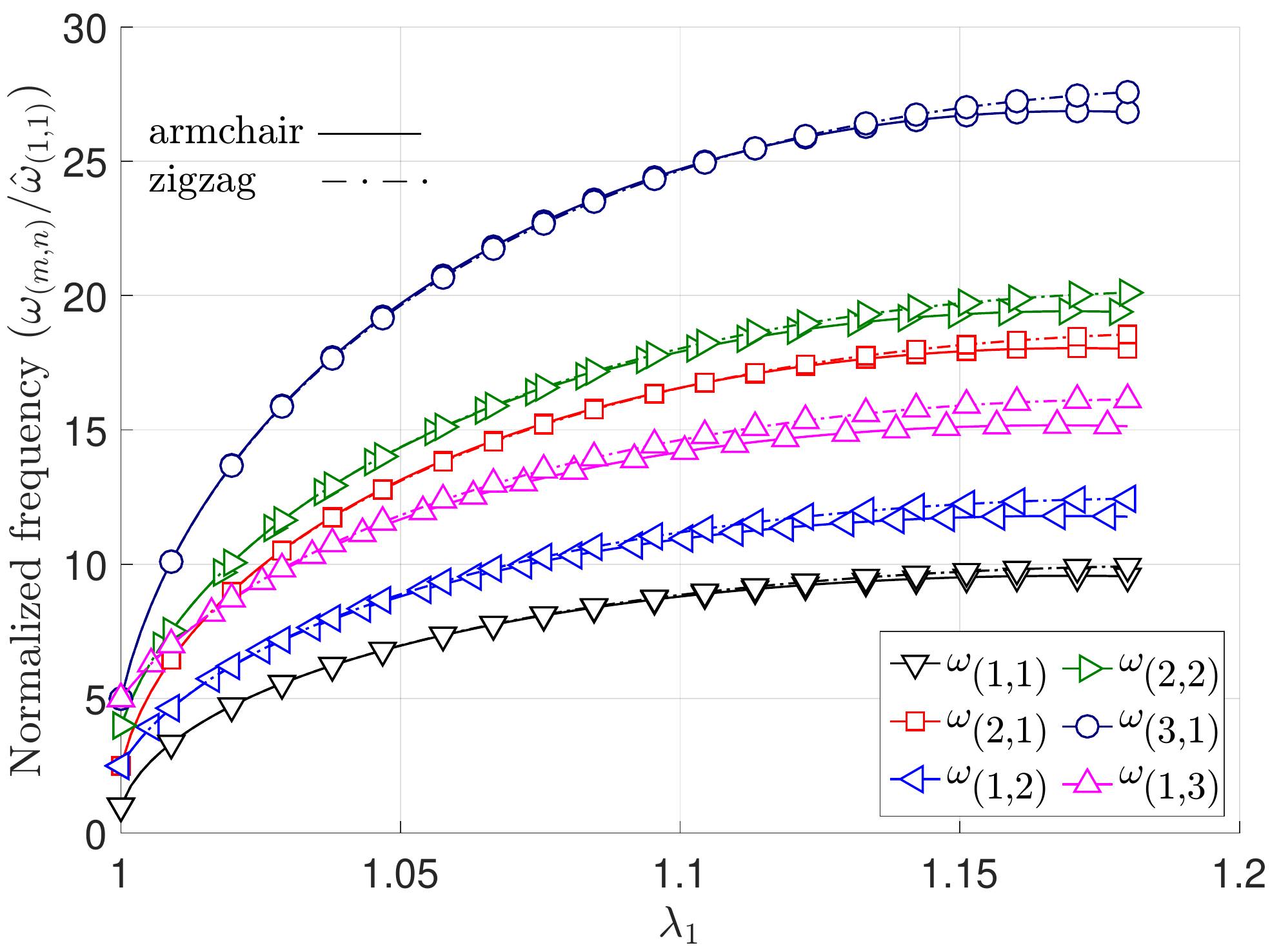}
        \subcaption{}
        \label{f:Rect_simply_freqeuncy_variation_uni_1to6}
    \end{subfigure}
    \begin{subfigure}{0.495\textwidth}
        \centering
    \includegraphics[height=60mm]{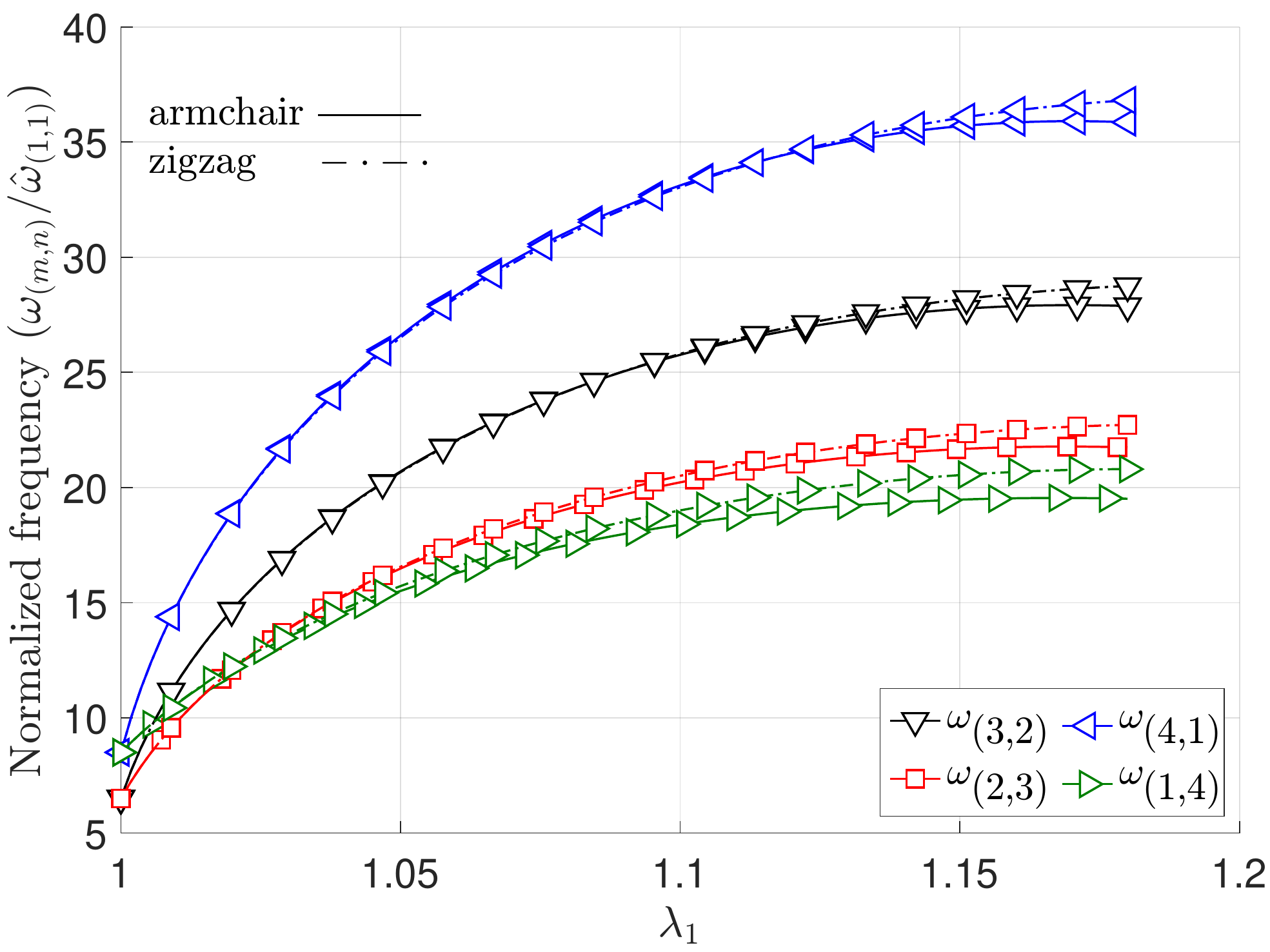}
        \subcaption{}
        \label{f:Rect_simply_freqeuncy_variation_uni_7to10}
    \end{subfigure}
    \caption{Vibrating square plate under uniaxial stretch:
            (\subref{f:Rect_simply_freqeuncy_variation_uni_1to6}) Frequencies 1 to 6;
            (\subref{f:Rect_simply_freqeuncy_variation_uni_7to10}) frequencies 7 to 10. $m$ and $n$ in $\omega_{(m,n)}$ are defined in Fig.~\ref{f:rect_simply_mode_shapes}. The edge length of the plate is 5 nm. The results are normalized by $\hat{\omega}_{(1,1)}= \omega_{(1,1)}\big |_{\lambda_1=1}$ (the first natural frequency). The plate is either stretched along the armchair or zigzag directions. \textcolor{cgn}{The boundary is simply supported.}}
    \label{f:Rect_simply_freqeuncy_variation_uni}
\end{figure}

First, a linear modal analysis is conducted for a simply supported square plate under zero pre-load, and the mode shapes and natural frequencies are obtained. \textcolor{cgn}{The mode shapes, natural frequencies and numbering of modes of the unloaded system are illustrated in Fig.~\ref{f:rect_simply_mode_shapes}}.
A convergence study for mesh refinement is conducted and reported in Fig.~\ref{f:Rect_simply_natural_freqeuncy_conv}.
As seen, the discretization error does not decrease beyond a certain refinement, since the condition number of the stiffness matrix becomes too large for an accurate solution of the problem. For the following simulation results a FE mesh with $80\times80$ quadratic NURBS elements is considered to ensure convergence.\\
Second, the modal analysis of a square plate under nonlinear pure dilatation and uniaxial stretch is investigated. An isotropic square plate has repeated frequencies under pure dilatation. Under pure dilatation, the material response is isotropic, the frequencies are increasing monotonically and the order of modes does not change, see Fig.~\ref{f:Rect_simply_freqeuncy_variation_biaxial}. \textcolor{cgn2}{Fig.~\ref{f:Rect_simply_freqeuncy_variation_biaxial_difference_with_ana} shows a comparison of the results with the analytical solution of Eq.~\eqref{e:analytical_sol_rect_nonlinear_modal}. The comparison shows agreement for zero pre-stretch ($J = 1$). As $J$ increases, differences show up due to the infinitesimal strain approximation used in the analytical solution.} The material has an anisotropic response under uniaxial loading. The frequencies increase faster if the specimen is stretched along the zigzag direction rather than in other directions. In uniaxial stretch, the repeated frequencies become distinct and the order of modes can change during loading\textcolor{cgn}{, e.g. $\omega_{(1,3)}$  is larger than $\omega_{(2,1)}$ in the unstretched structure, but beyond a certain stretch the frequencies reorder and $\omega_{(2,1)}$ become larger than $\omega_{(1,3)}$ (see Fig.~\ref{f:Rect_simply_freqeuncy_variation_uni}a).} In both cases, the frequencies increase up to a maximum, and the model becomes unstable if it is deformed further due to a loss of ellipticity of the elasticity tensor (e.g. vanishing shear modulus for pure dilatation). \textcolor{cgn}{The strain for vanishing shear modulus ($\mu(\varepsilon_a)=0$) can be analytically obtained as $\varepsilon_a=1/\beta\,\ln(\mu_0/\mu_1)$}. It should be mentioned that the classical formula for the stress-dependent frequencies \citep{ugural2009_Stresses_in_Beams_Plates} cannot capture the behavior correctly since it assume linear material behavior and thus predicts a linear frequency increase with the stretch, which is very different from the nonlinear behavior seen in Figs.~\ref{f:Rect_simply_freqeuncy_variation_biaxial} and \ref{f:Rect_simply_freqeuncy_variation_uni}.


\subsection{Vibrating circular plates}
Like for the square plate, the linear modal analysis is conducted for circular plates. \textcolor{cgn}{First, the behavior for zero pre-load is investigated. The resulting mode shapes, natural frequencies and the numbering of modes are illustrated in Figs.~\ref{f:Circ_simply_mode_shapes} and \ref{f:Circ_clamped_mode_shapes}} for simply supported and clamped plates, respectively.
\begin{figure}
\begin{center} \unitlength1cm
\begin{picture}(18,6)
\put(1.3,3.25){\includegraphics[width=25mm,trim=5cm 1cm 5cm 5cm,clip]{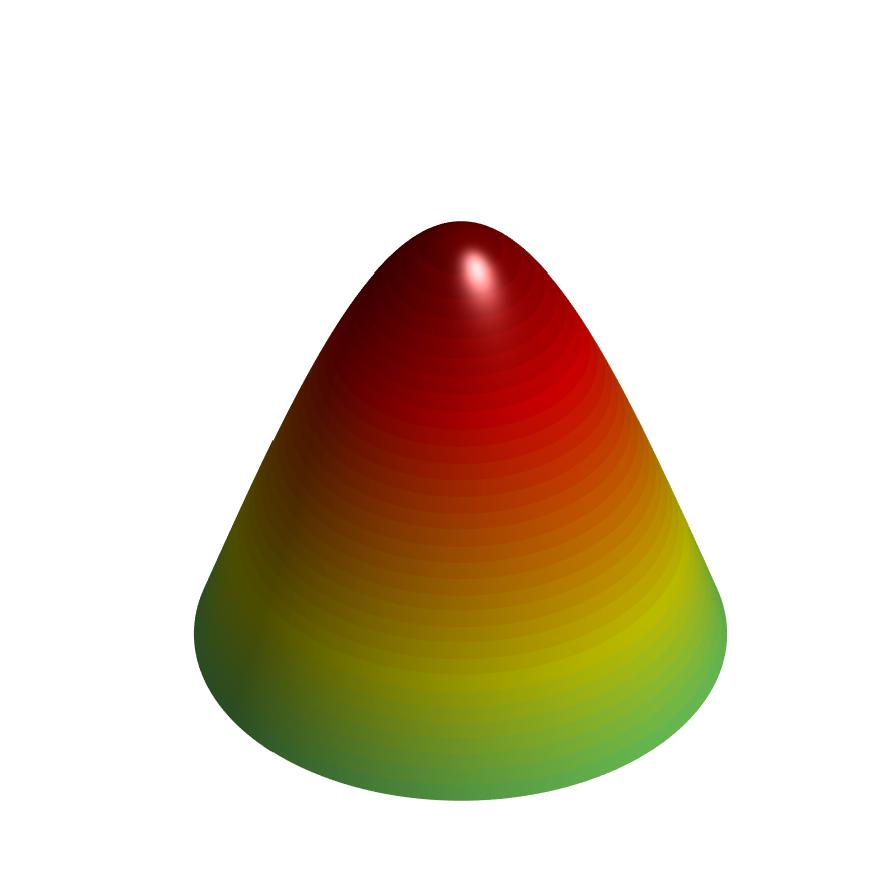}}
\put(4.3,3.25){\includegraphics[width=25mm,trim=5cm 2cm 5cm 4cm,clip]{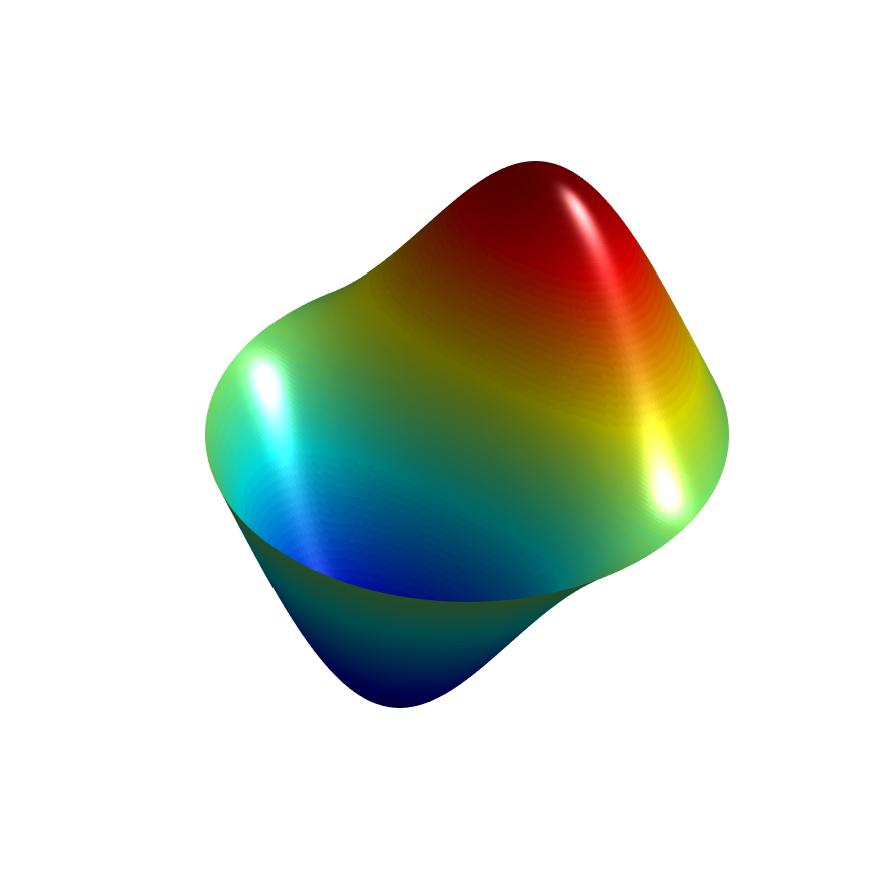}}
\put(7.9,3.25){\includegraphics[width=25mm,trim=5cm 2cm 5cm 4cm,clip]{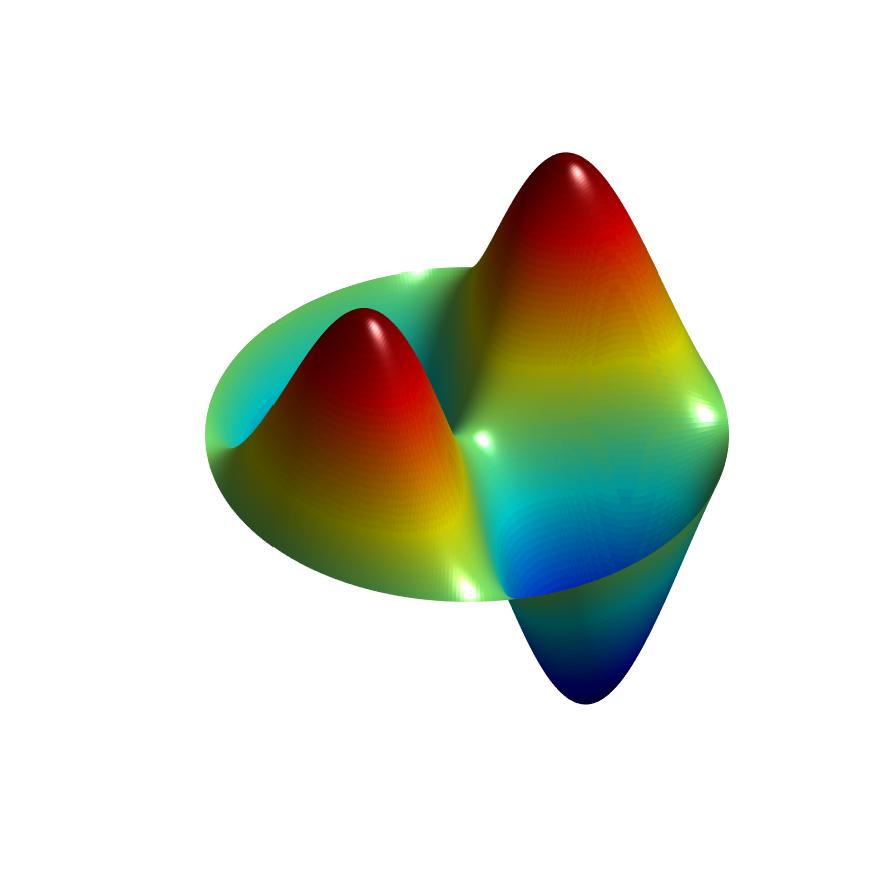}}
\put(11.05,3.25){\includegraphics[width=25mm,trim=5cm 2cm 5cm 4cm,clip]{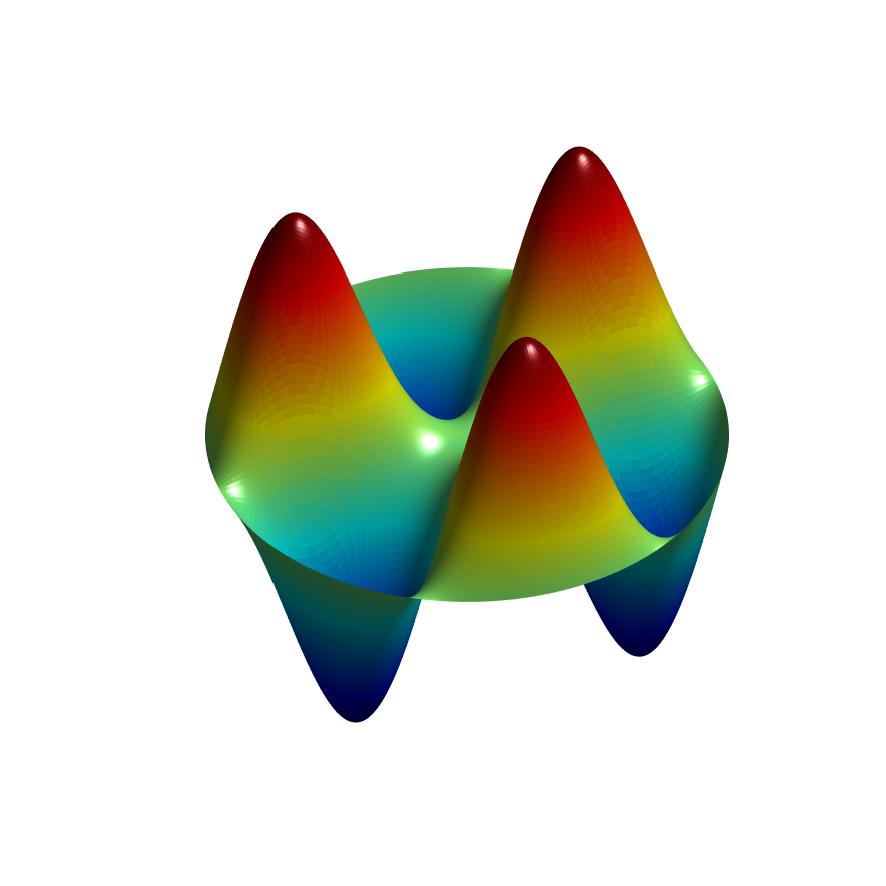}}
\put(1.3,0.5){\includegraphics[width=25mm,trim=5cm 4cm 5cm 6cm,clip]{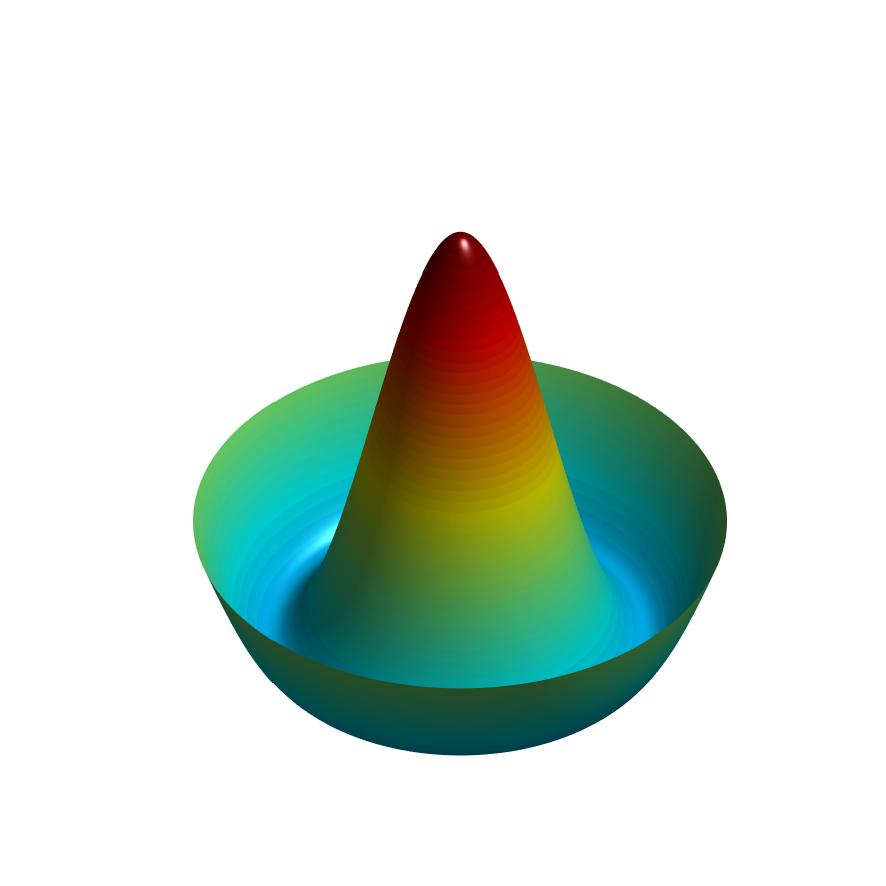}}
\put(4.3,0.25){\includegraphics[width=25mm,trim=5cm 4cm 5cm 6cm,clip]{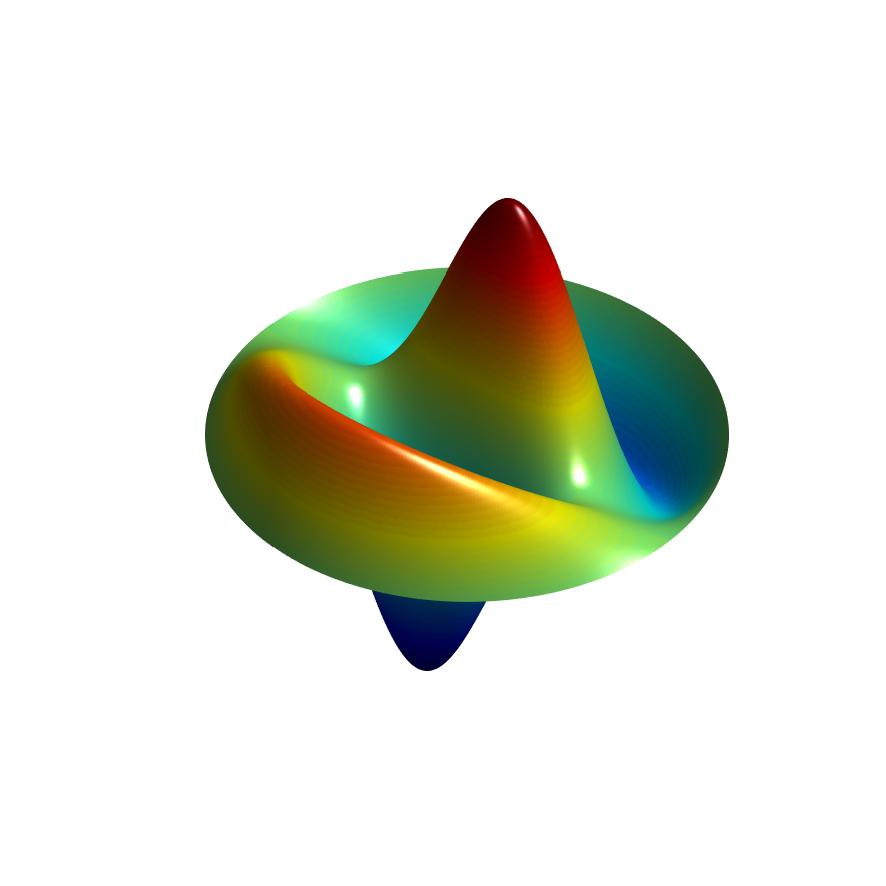}}
\put(7.9,0.25){\includegraphics[width=25mm,trim=5cm 4cm 5cm 6cm,clip]{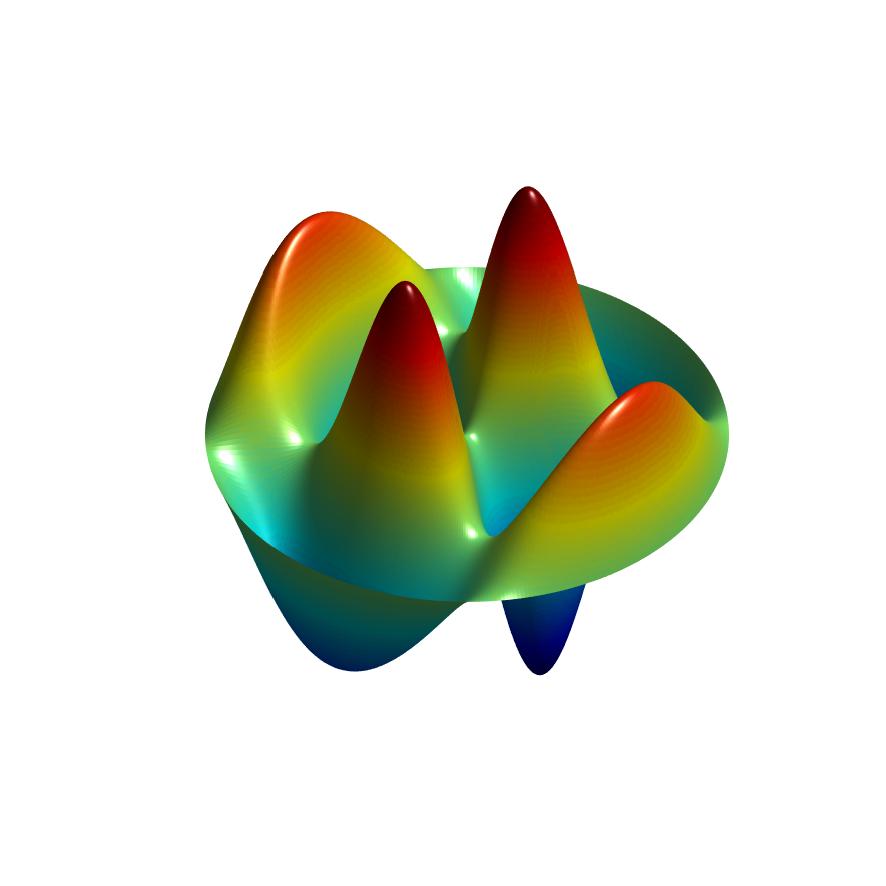}}
\put(11.05,0.25){\includegraphics[width=25mm,trim=5cm 4cm 5cm 6cm,clip]{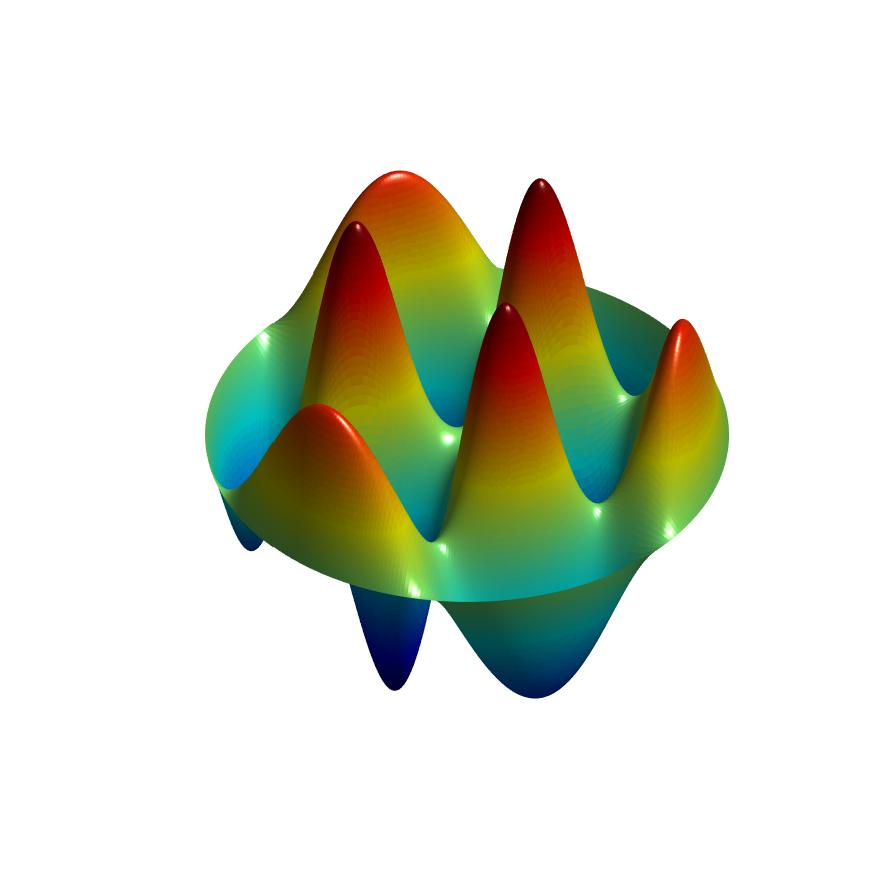}}

\put(1.3,0.15){{\small $\hat{f}_{(0,1)}=0.10453$}}
\put(4.5,0.15){{\small $\hat{f}_{(1,1)}=0.17136$}}
\put(8.05,0.15){{\small $\hat{f}_{(2,1)}=0.248427$}}
\put(11.4,0.15){{\small $\hat{f}_{(3,1)}=0.27008$}}

\put(1.3,3.1){{\small $\hat{f}_{(0,0)}=0.01581$}}
\put(4.5,3.1){{\small $\hat{f}_{(1,0)}=0.04806$}}
\put(8.05,3.1){{\small $\hat{f}_{(2,0)}=0.08987$}}
\put(11.4,3.1){{\small $\hat{f}_{(3,0)}=0.14099$}}

\end{picture}
\vspace{-4mm}
\caption{Vibrating circular plate at zero pre-load (simply supported boundary): First 8 mode shapes and natural frequencies $\hat{f}_{(m,n)}=\hat{\omega}_{(m,n)}/2\pi$ in [THz], $\hat{\omega}_{(m,n)}=\omega_{(m,n)} \big |_{\bE^{(0)}=\bolds{0}}$. $m$ and $n$ are the number of nodal diameters and circles, respectively. \textcolor{cgn}{A nodal line is a line of material points of the plate (excluding boundaries) that do not displace during vibration.}}
\label{f:Circ_simply_mode_shapes}
\end{center}
\end{figure}
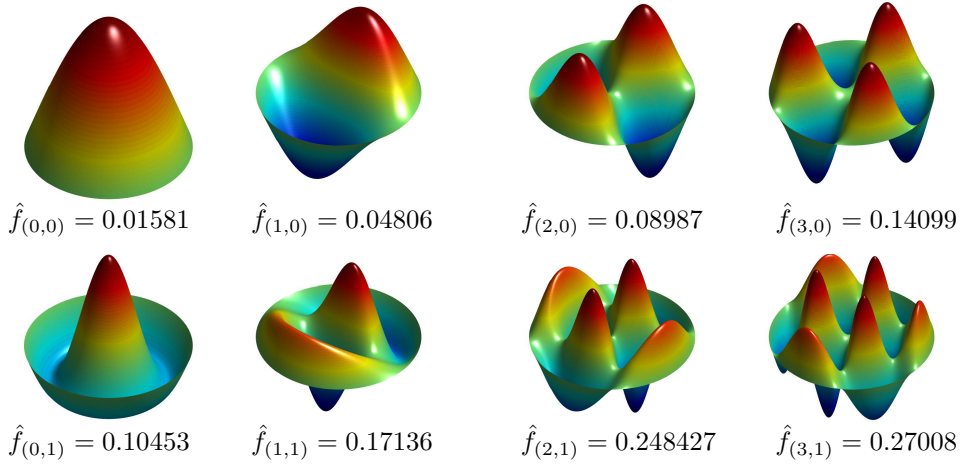
\begin{figure}
\begin{center} \unitlength1cm
\begin{picture}(18,6.0)
\put(1.3,3.5){\includegraphics[width=25mm,trim=5cm 3cm 5cm 5cm,clip]{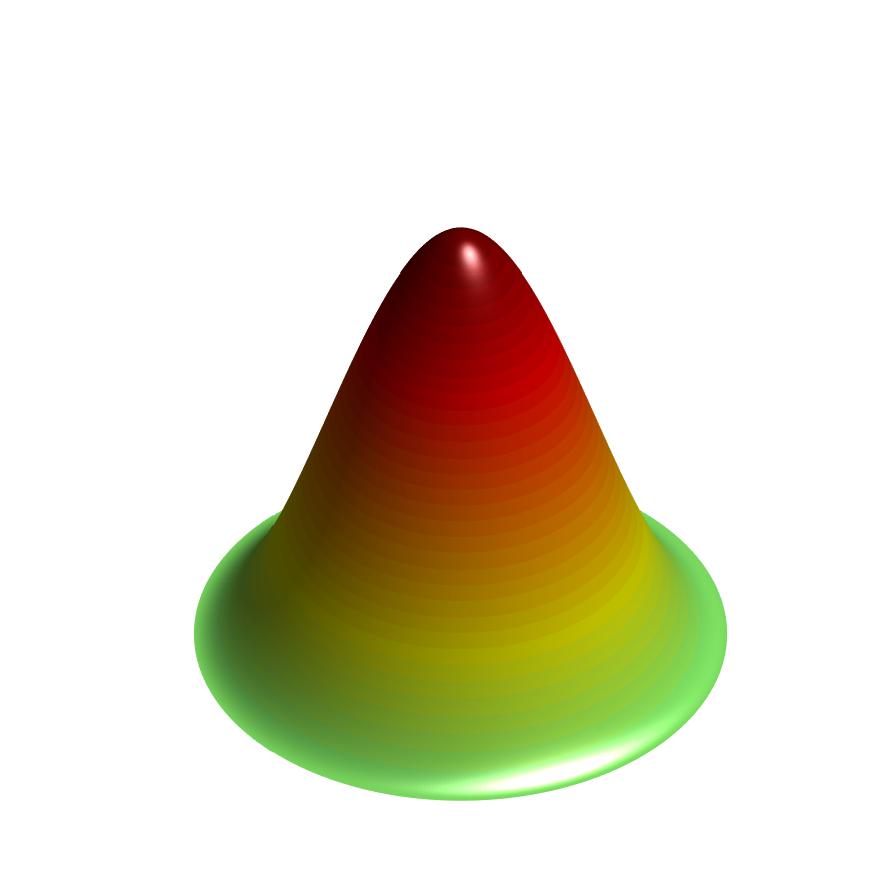}}
\put(4.3,3.25){\includegraphics[width=25mm,trim=5cm 3cm 5cm 5cm,clip]{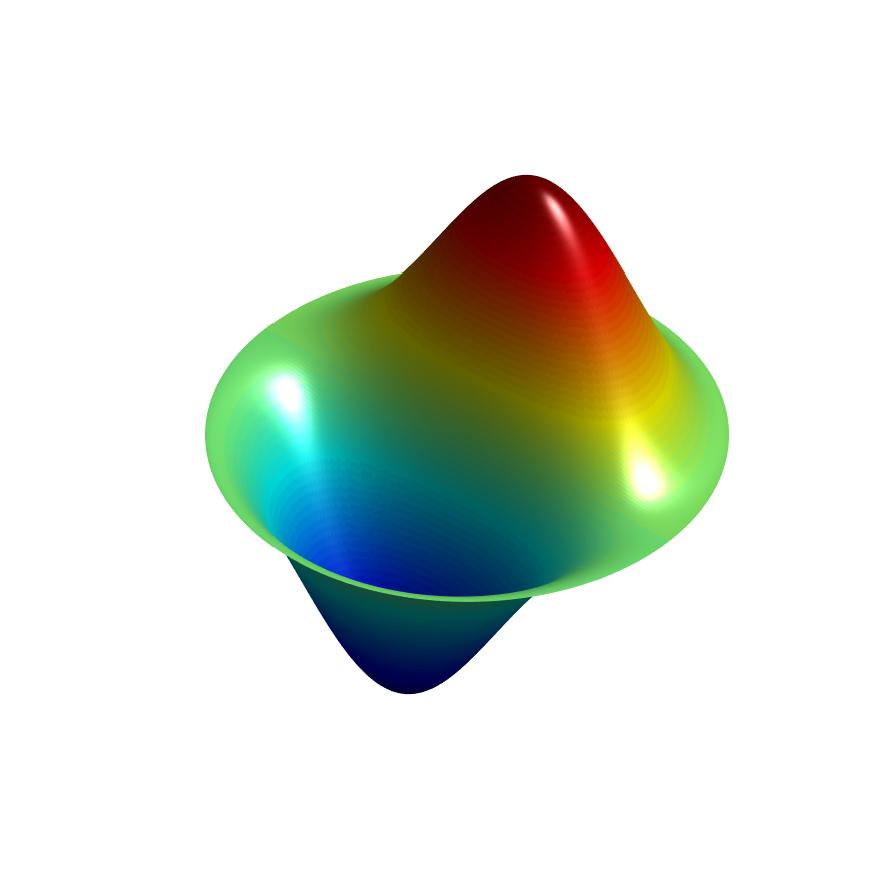}}
\put(7.9,3.25){\includegraphics[width=25mm,trim=5cm 3cm 5cm 5cm,clip]{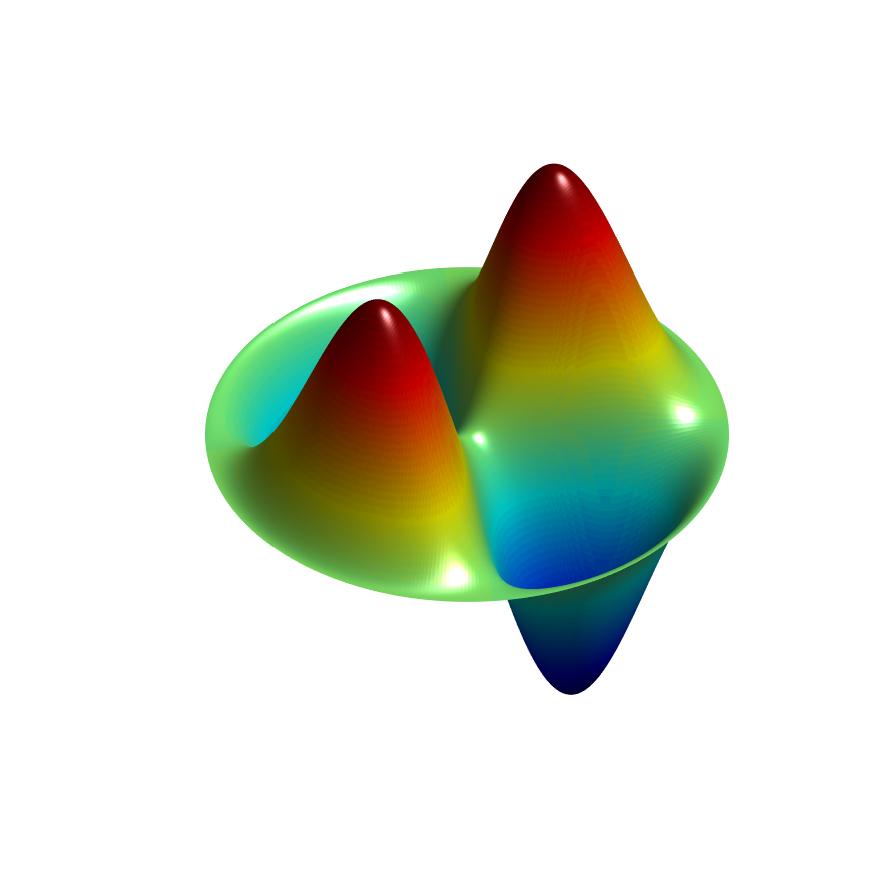}}
\put(11.6,3.25){\includegraphics[width=25mm,trim=5cm 3cm 5cm 5cm,clip]{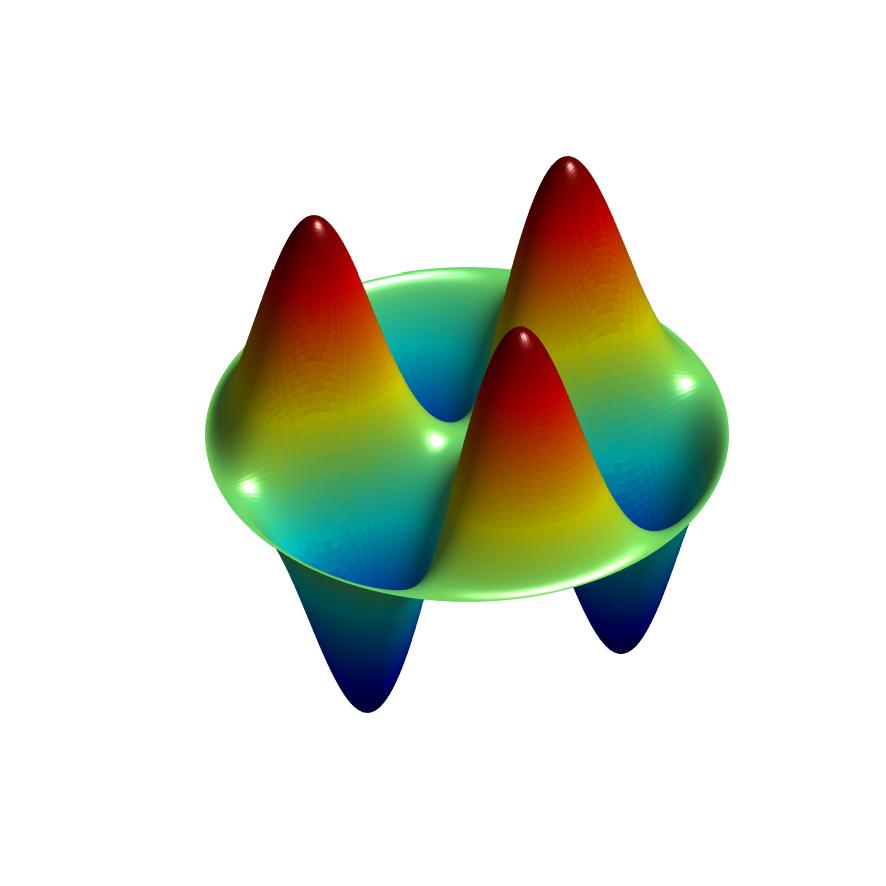}}
\put(1.3,0.5){\includegraphics[width=25mm,trim=5cm 3cm 5cm 7cm,clip]{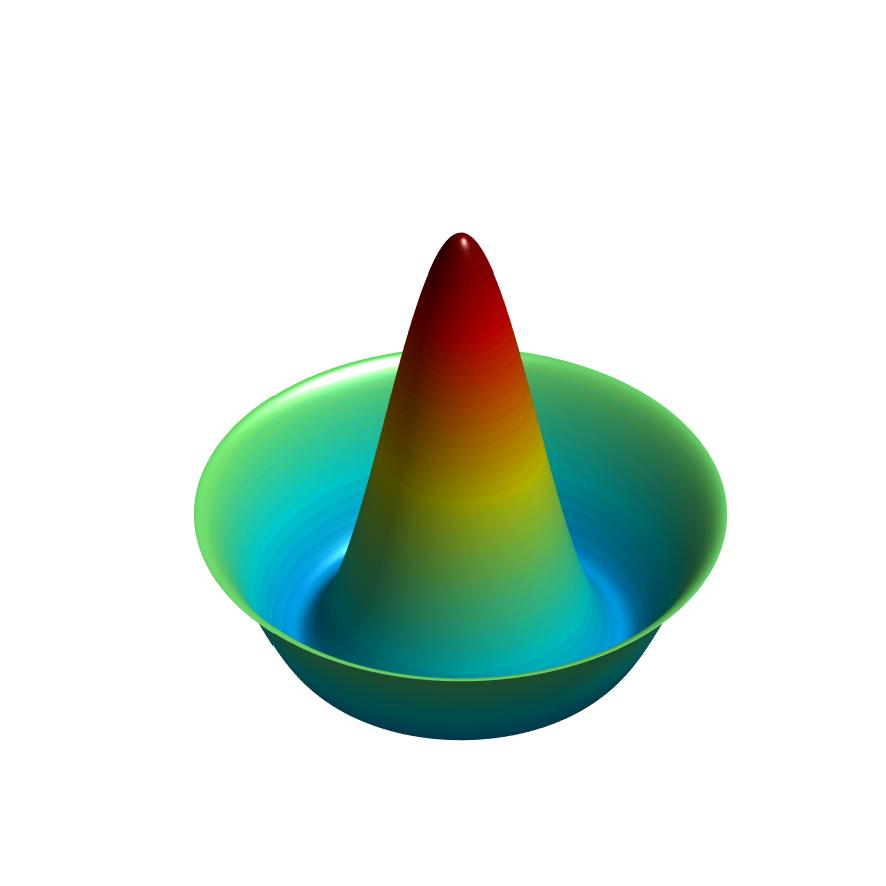}}
\put(4.3,0.3){\includegraphics[width=25mm,trim=5cm 4cm 5cm 6cm,clip]{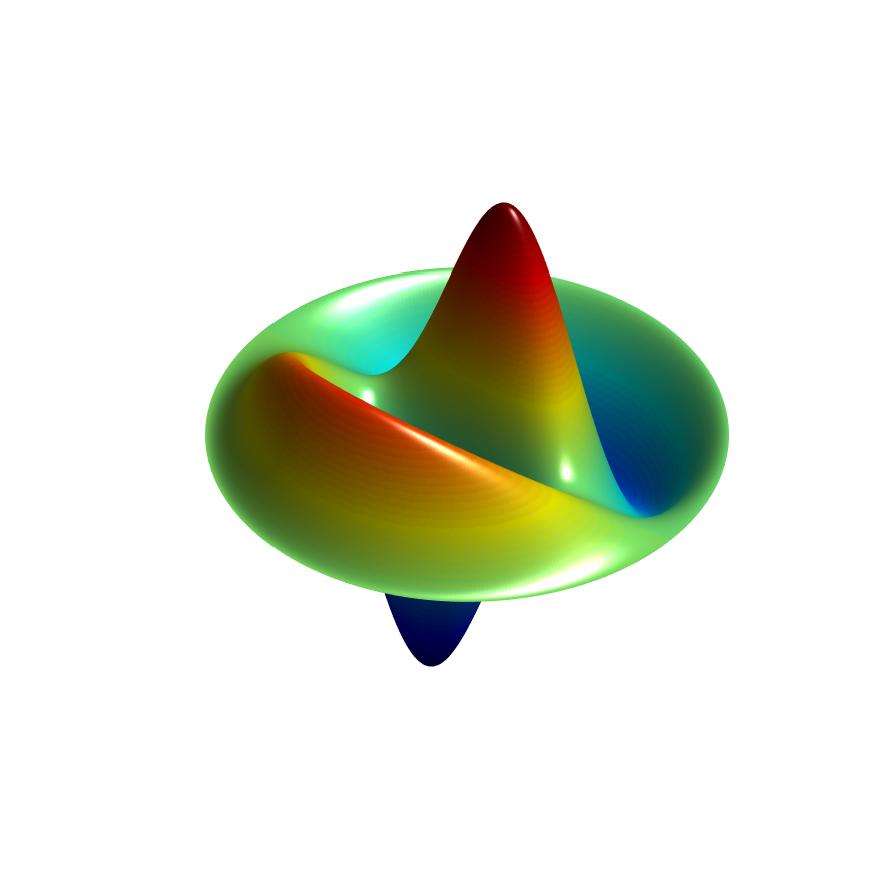}}
\put(7.9,0.3){\includegraphics[width=25mm,trim=5cm 4cm 5cm 6cm,clip]{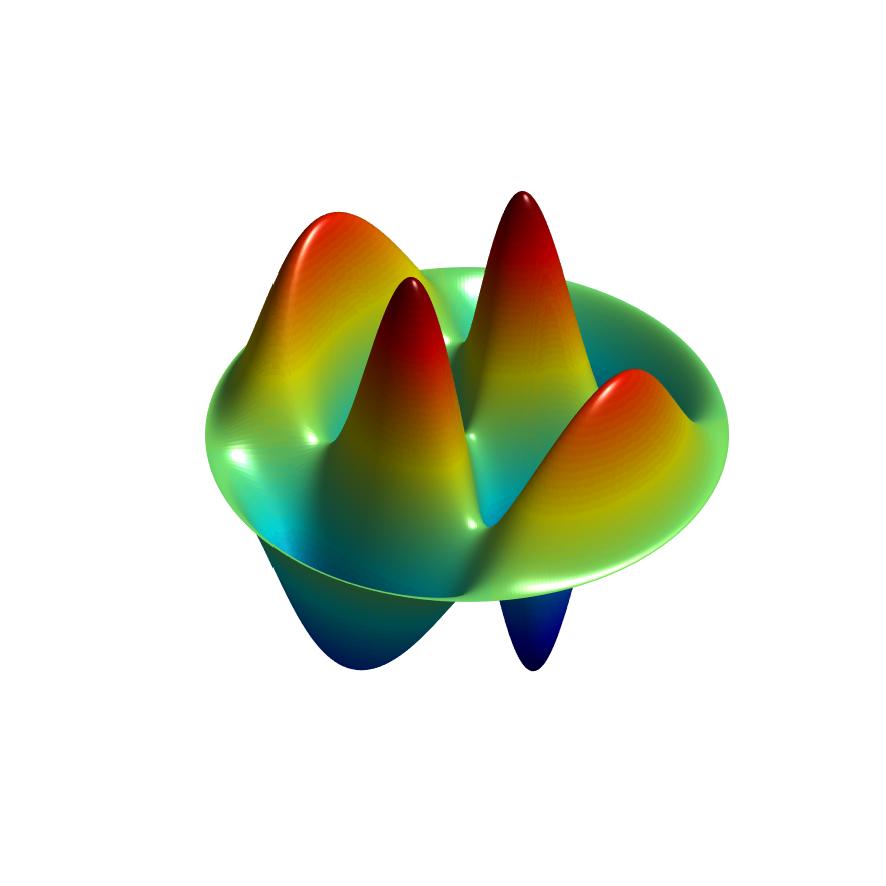}}
\put(11.6,0.3){\includegraphics[width=25mm,trim=5cm 4cm 5cm 6cm,clip]{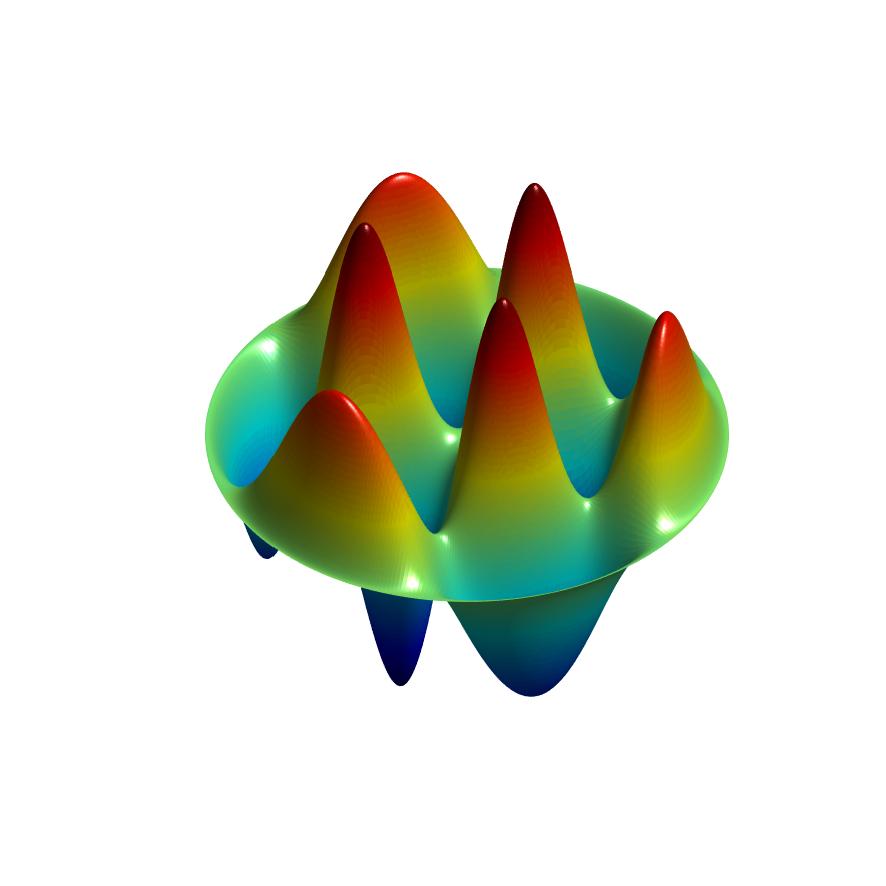}}

\put(1.3,0.3){{\small $\hat{f}_{(0,1)}=0.14158$}}
\put(4.5,0.3){{\small $\hat{f}_{(1,1)}=0.21655$}}
\put(8,0.3){{\small $\hat{f}_{(2,1)}=0.30112$}}
\put(11.6,0.3){{\small $\hat{f}_{(3,1)}=0.323038$}}

\put(1.3,3.1){{\small $\hat{f}_{(0,0)}=0.03636$}}
\put(4.5,3.1){{\small $\hat{f}_{(1,0)}=0.07568$}}
\put(8.,3.1){{\small $\hat{f}_{(2,0)}=0.12416$}}
\put(11.6,3.1){{\small $\hat{f}_{(3,0)}=0.18167$}}

\end{picture}
\vspace{-4mm}
\caption[]{Vibrating circular plate at zero pre-load (clamped boundary): First 8 mode shapes and natural frequencies $\hat{f}_{(m,n)}=\hat{\omega}_{(m,n)}/2\pi$ in [THz], $\hat{\omega}_{(m,n)}=\omega_{(m,n)} \big |_{\bE^{(0)}=\bolds{0}}$. $m$ and $n$ are the number of nodal diameters and circles, respectively. \textcolor{cgn}{A nodal line is a line of material points of the plate (excluding boundaries) that do not displace during vibration.}}
\label{f:Circ_clamped_mode_shapes}
\end{center}
\end{figure}
A convergence study for mesh refinement is conducted and reported in Fig.~\ref{f:Circ_natural_freqeuncy_conv}.
\begin{figure}
    \begin{subfigure}{.495\textwidth}
        \centering
    \includegraphics[height=60mm]{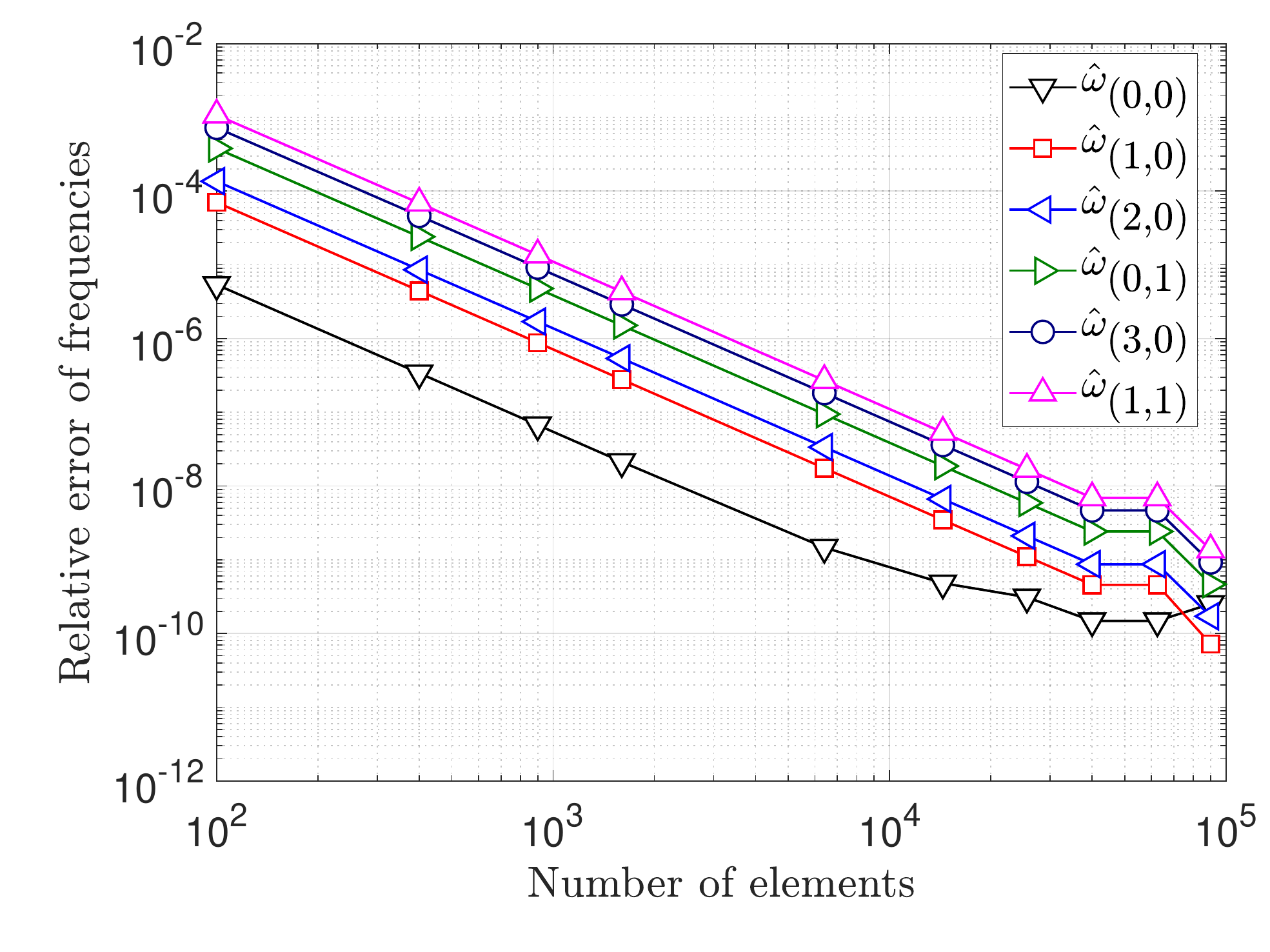}
        \subcaption{}
        \label{f:Circ_simply_natural_freqeuncy_conv}
    \end{subfigure}
        \begin{subfigure}{.495\textwidth}
        \centering
    \includegraphics[height=60mm]{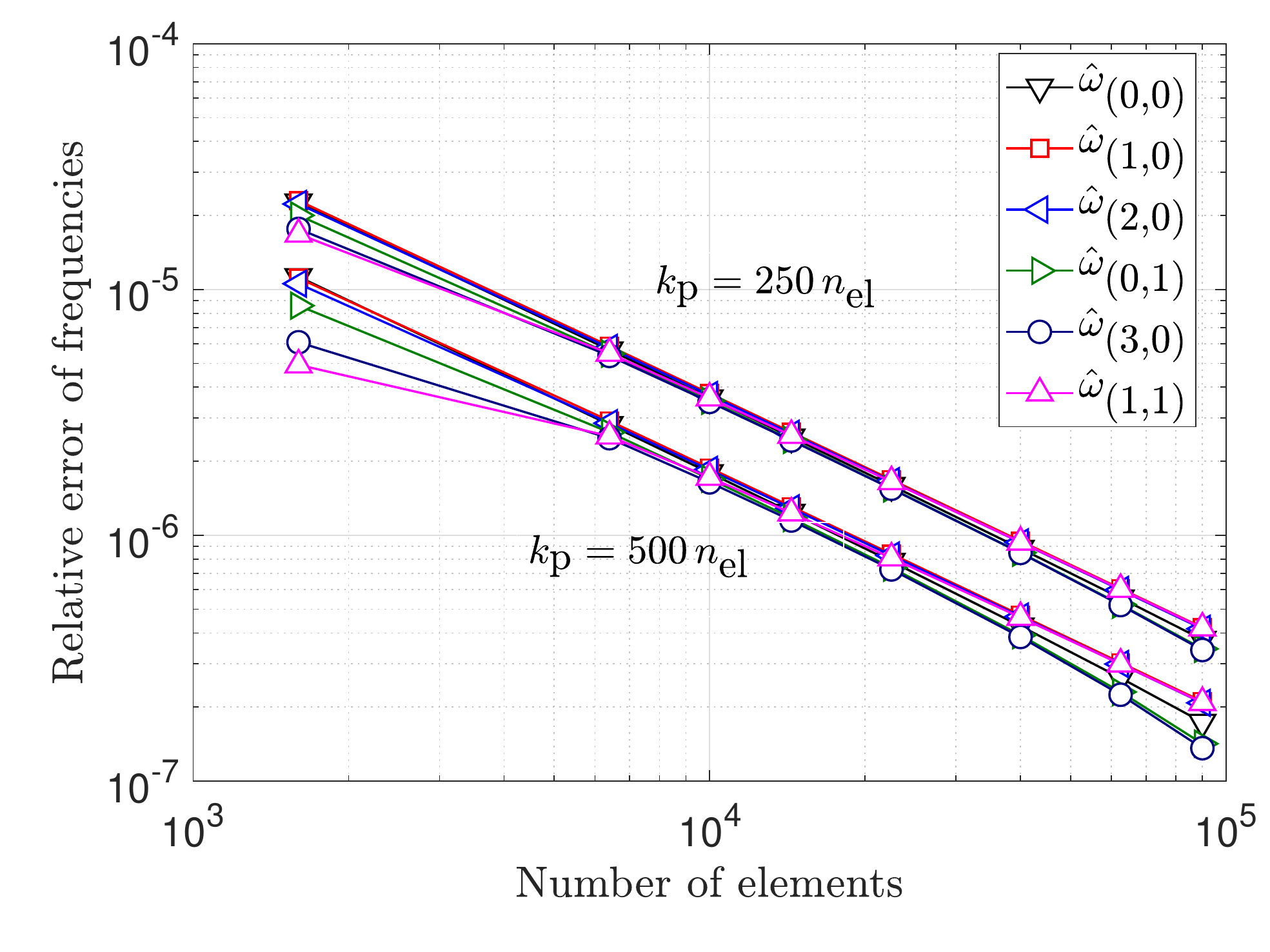}
        \subcaption{}
        \label{f:Circ_clamp_natural_freqeuncy_conv}
    \end{subfigure}
    \caption{Vibrating circular plate under zero pre-load: Relative error of the natural frequencies for (\subref{f:Circ_simply_natural_freqeuncy_conv}) simply supported boundary; (\subref{f:Circ_clamp_natural_freqeuncy_conv}) clamped boundary. The analytical solutions of (\ref{e:freq_lin_circ_clamp}) and (\ref{e:freq_lin_circ_simp}) are used as reference. $m$ and $n$ in $\hat{\omega}_{(m,n)}=\omega_{(m,n)} \big |_{\bE^{(0)}=\bolds{0}}$ are defined in Figs.~\ref{f:Circ_simply_mode_shapes} and \ref{f:Circ_clamped_mode_shapes}. The rotational dofs are fixed along the boundary by means of a penalty term using different penalty parameters $k_\mathrm{p}$, see \citet{Duong2016_01} for details.}
    \label{f:Circ_natural_freqeuncy_conv}
\end{figure}Due to ill-conditioning, the discretization error does not decrease beyond a certain refinement level. This is similar to the behavior of the square plate. For the following simulation results a FE mesh with $80\times80$ quadratic NURBS elements is considered to ensure convergence.\\
Second, the vibrational behavior of a circular graphene plate under pure dilatation loading is investigated. In Fig.~\ref{f:Circ_freqeuncy_variation_biaxial}, the variation of the frequencies under pure dilatational loading is presented. As shown, the frequencies increase monotonically with $J$. The structure will become unstable if the loading is increased to far. Like for the square plate, the frequency dependency on deformation is nonlinear and cannot be captured correctly by analytical formulas based on linear elasticity. \\
Finally, a circular graphene plate in contact with an adhesive substrate is investigated. Locally, the substrate is modeled as a half space at each contact point and the atomic interaction is modeled via the Lennard Jones (L-J) potential. The half space potential can be written as
\eqb{lll}
\Psi_{\text{(VdW)h}} \is \ds -\Gamma\left[\frac{3}{2}\left(\frac{h_0}{r}\right)^3-\frac{1}{2}\left(\frac{h_0}{r}\right)^9\right]~,
\eqe
where $h_0$, $\Gamma$ and $r$ are the equilibrium distance, the interfacial adhesion energy per unit area and the normal distance of a surface point to the substrate. The derivation of the half space potential, contact force and stiffness can be found in \citet{sauer_a_contact_mechanics_model_quasi}, \citet{Sauer_formulaton_and_analysis} and \citet{Aitken_Effects_of_mismatch_strain}. In Fig.~\ref{f:Circ_freqeuncy_variation_adhesive}, the \textcolor{cgn}{boundary conditions}, the variation of the frequencies with increasing adhesion energy, and the first mode shape are shown. The locations with sudden frequency changes (indicated by dashed lines in Fig.~\ref{f:Circ_adhesive_freqeuncy_variation}) are local instabilities where the frequency suddenly drops to zero and quickly recovers again, as the enlargement in Fig.~\ref{f:Circ_adhesive_freqeuncy_variation_zoom} shows. A similar issue has been reported for loading of graphene on an adhesive substrate by \citet{Kumar2016_01}. For these adhesion energies, the graphene surface can come very close to the substrate leading to negative eigenvalues of the contact stiffness matrix so that the eigenvalues of the total problem become very small. A large change in the frequencies and deformation can be seen at those frequencies. The softening effect of the stiffness results in the local mode shapes that are shown in Fig.~\ref{f:circ_SS_mode_softening}.
\begin{figure}
    \begin{subfigure}{0.495\textwidth}
        \centering
    \includegraphics[height=60mm]{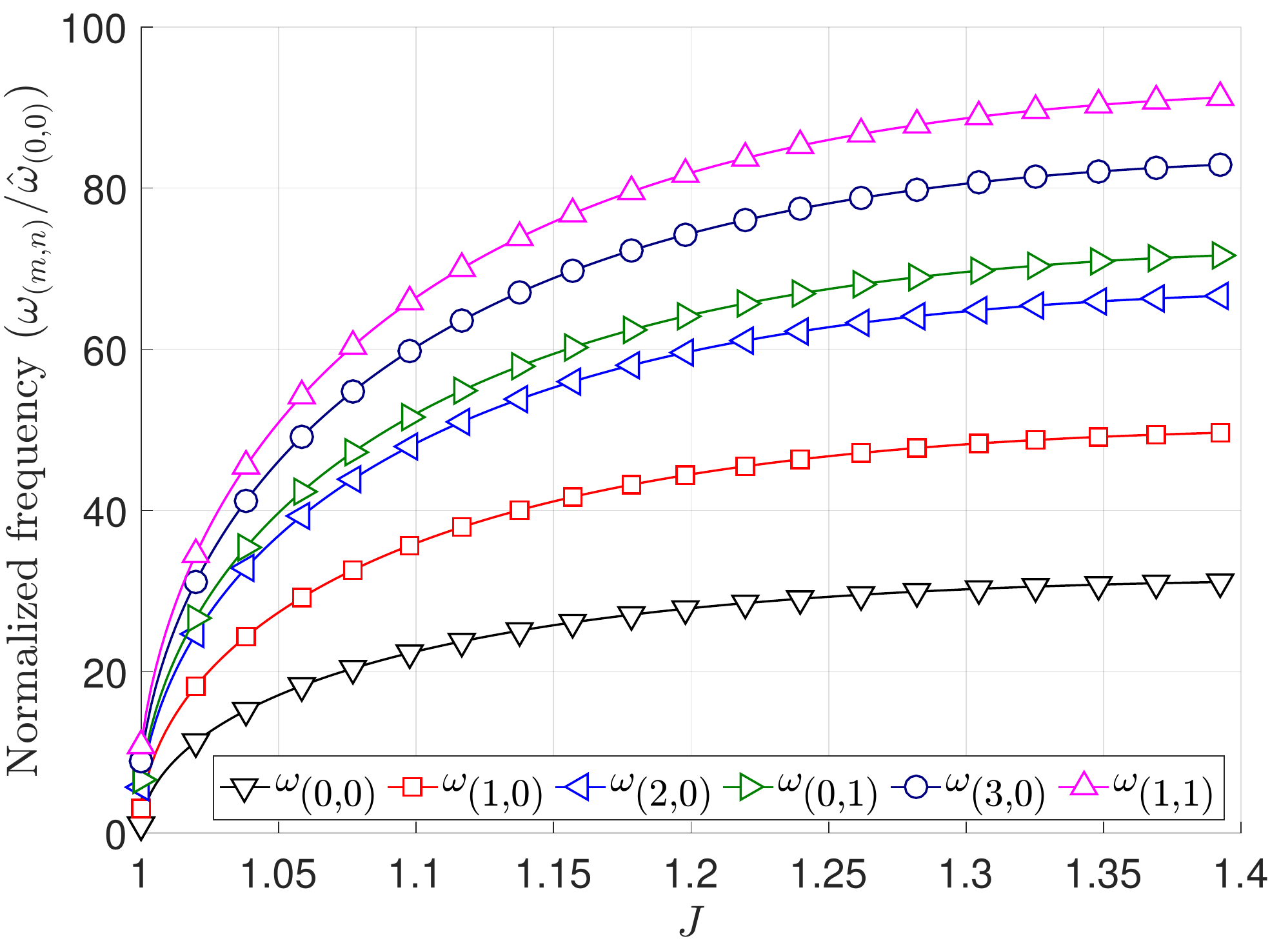}
        \subcaption{}
        \label{f:Circ_simply_freqeuncy_variation_biaxial}
    \end{subfigure}
    \begin{subfigure}{.495\textwidth}
        \centering
    \includegraphics[height=60mm]{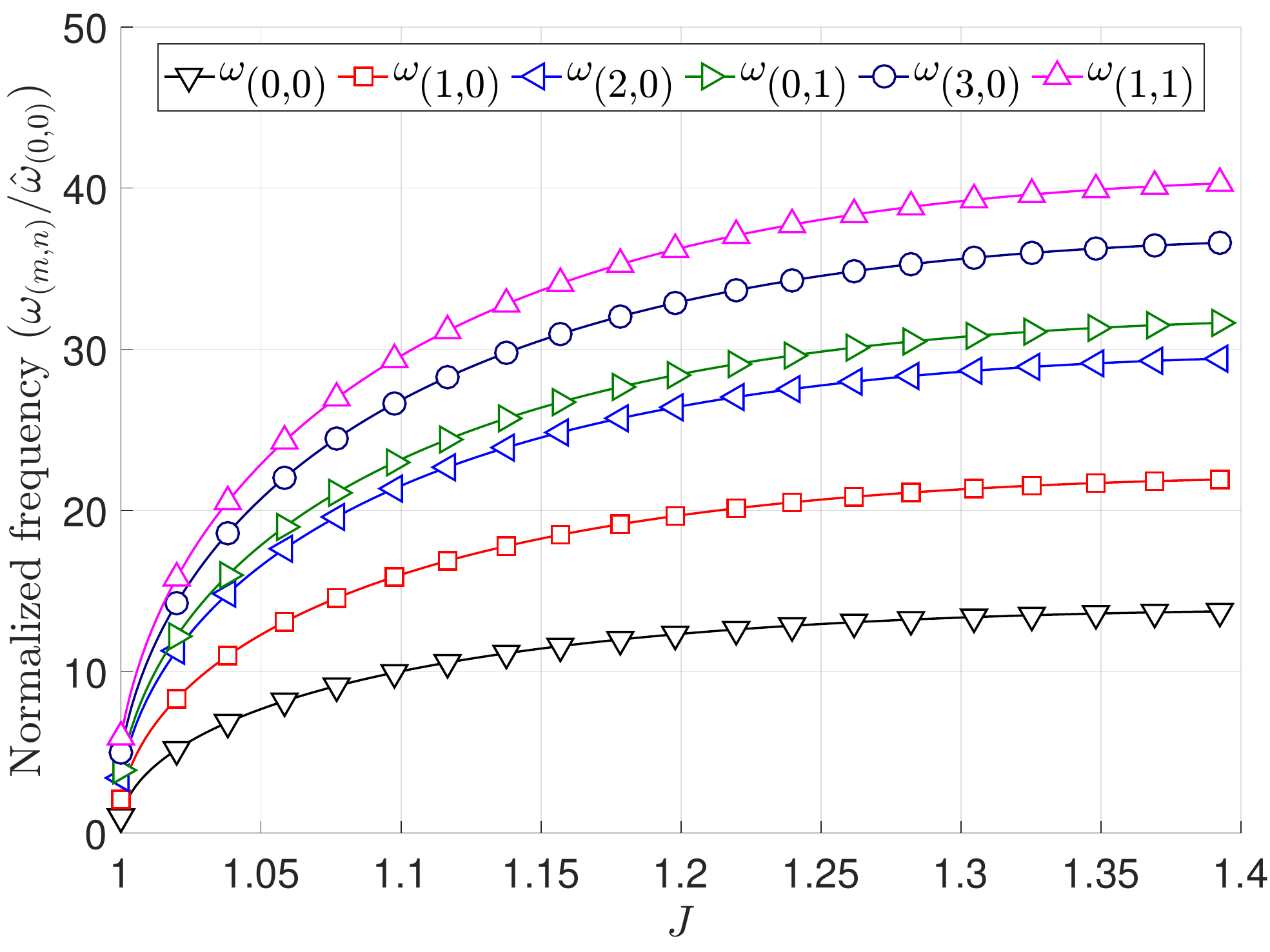}
        \subcaption{}
        \label{f:Circ_clamp_freqeuncy_variation_biaxial}
    \end{subfigure}
    \caption{Vibrating circular plate under pure dilatation: Variation of the bending frequencies with the surface stretch $J$ for (\subref{f:Circ_simply_freqeuncy_variation_biaxial}) simply supported boundary; (\subref{f:Circ_clamp_freqeuncy_variation_biaxial}) clamped boundary. $m$ and $n$ in $\omega_{(m,n)}$ are defined in Figs.~\ref{f:Circ_simply_mode_shapes} and \ref{f:Circ_clamped_mode_shapes}. The radius of the disk is 5 nm and the results are normalized by $\hat{\omega}_{(0,0)}= \omega_{(0,0)}\big |_{J=1}$ (the first natural frequency).}
    \label{f:Circ_freqeuncy_variation_biaxial}
\end{figure}

\begin{figure}
    \begin{subfigure}{0.495\textwidth}
    \centering
    \includegraphics[height=60mm]{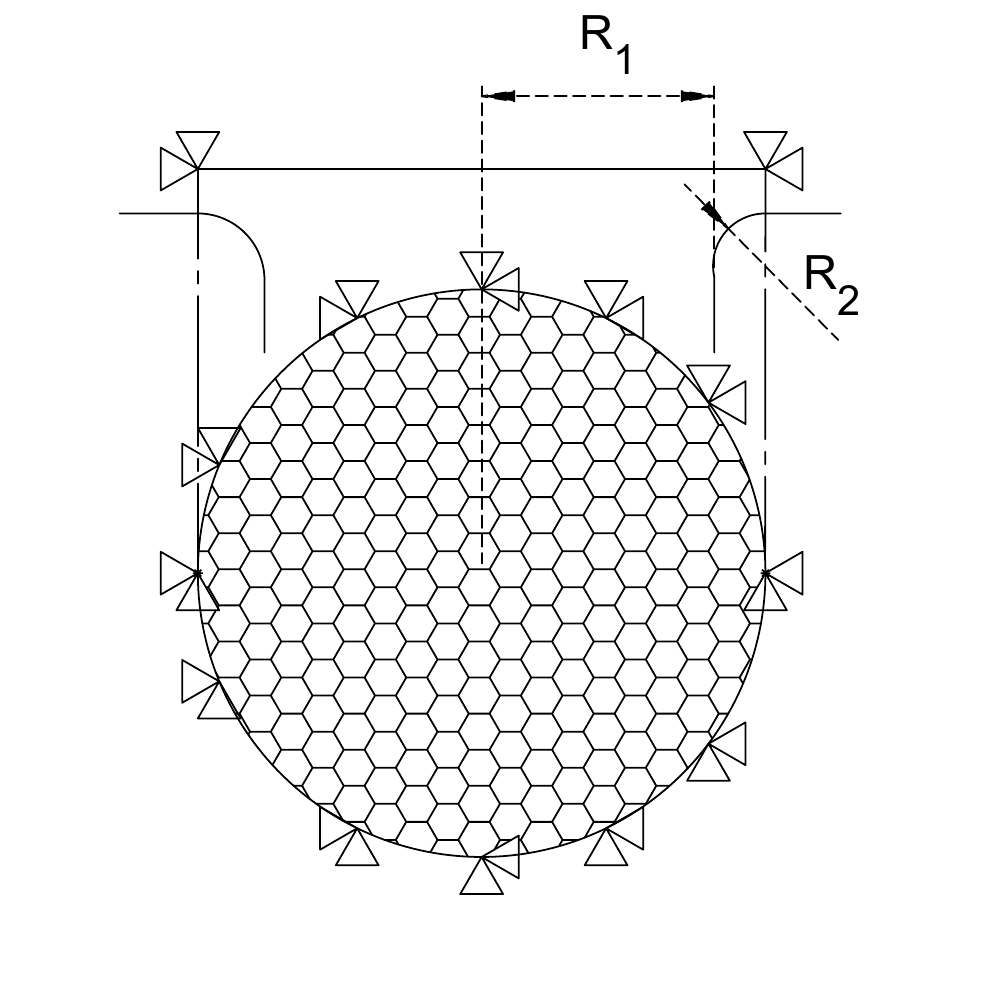}
        \subcaption{}
        \label{f:Full_disk_adhesive_substrate_modal}
    \end{subfigure}
        \begin{subfigure}{0.495\textwidth}
    \centering
    \includegraphics[height=60mm]{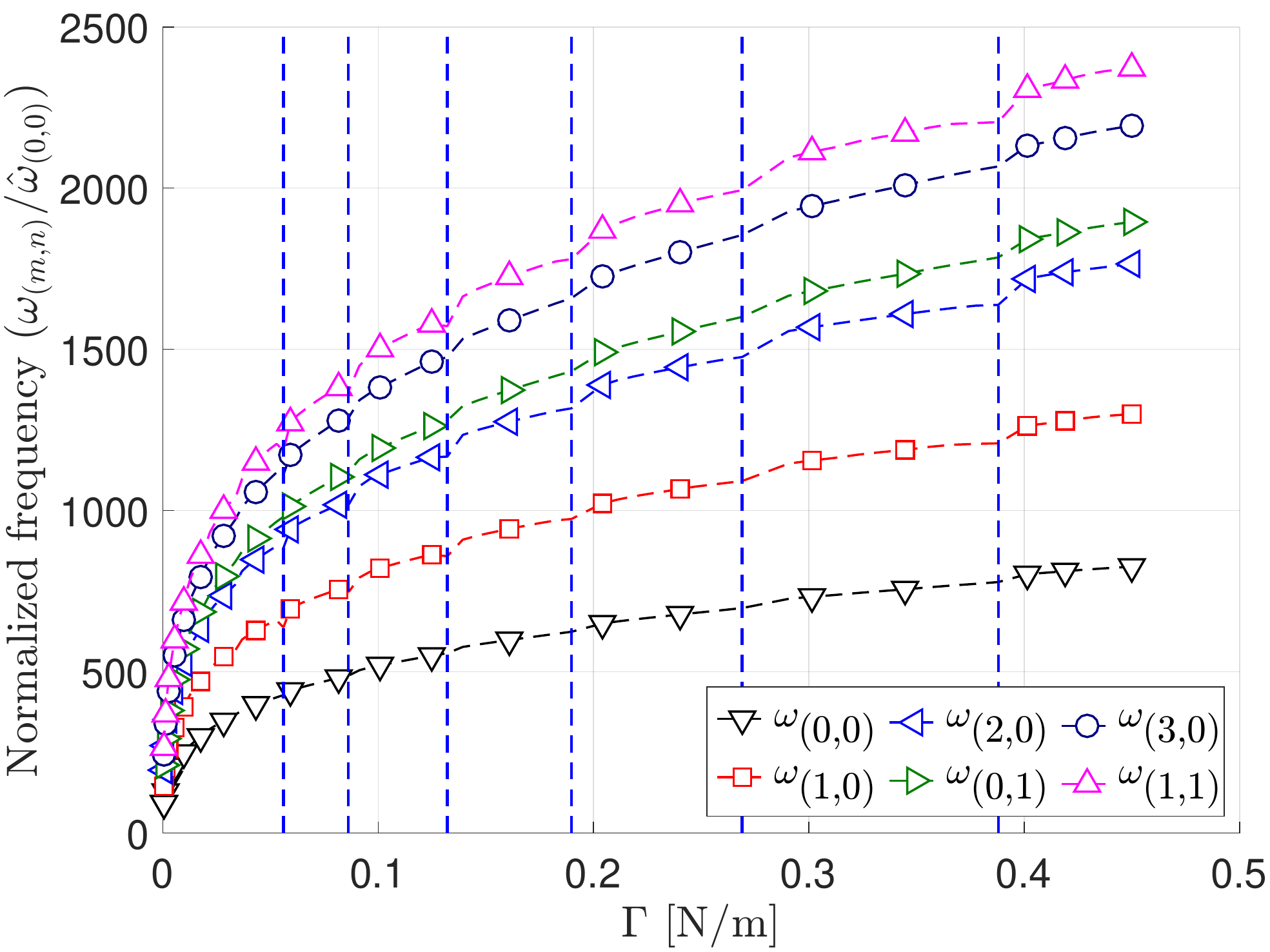}
        \subcaption{}
        \label{f:Circ_adhesive_freqeuncy_variation}
    \end{subfigure}\\
    \begin{subfigure}{0.495\textwidth}
        \centering
    \includegraphics[width=80mm,trim=15cm 15cm 14cm 15cm,clip]{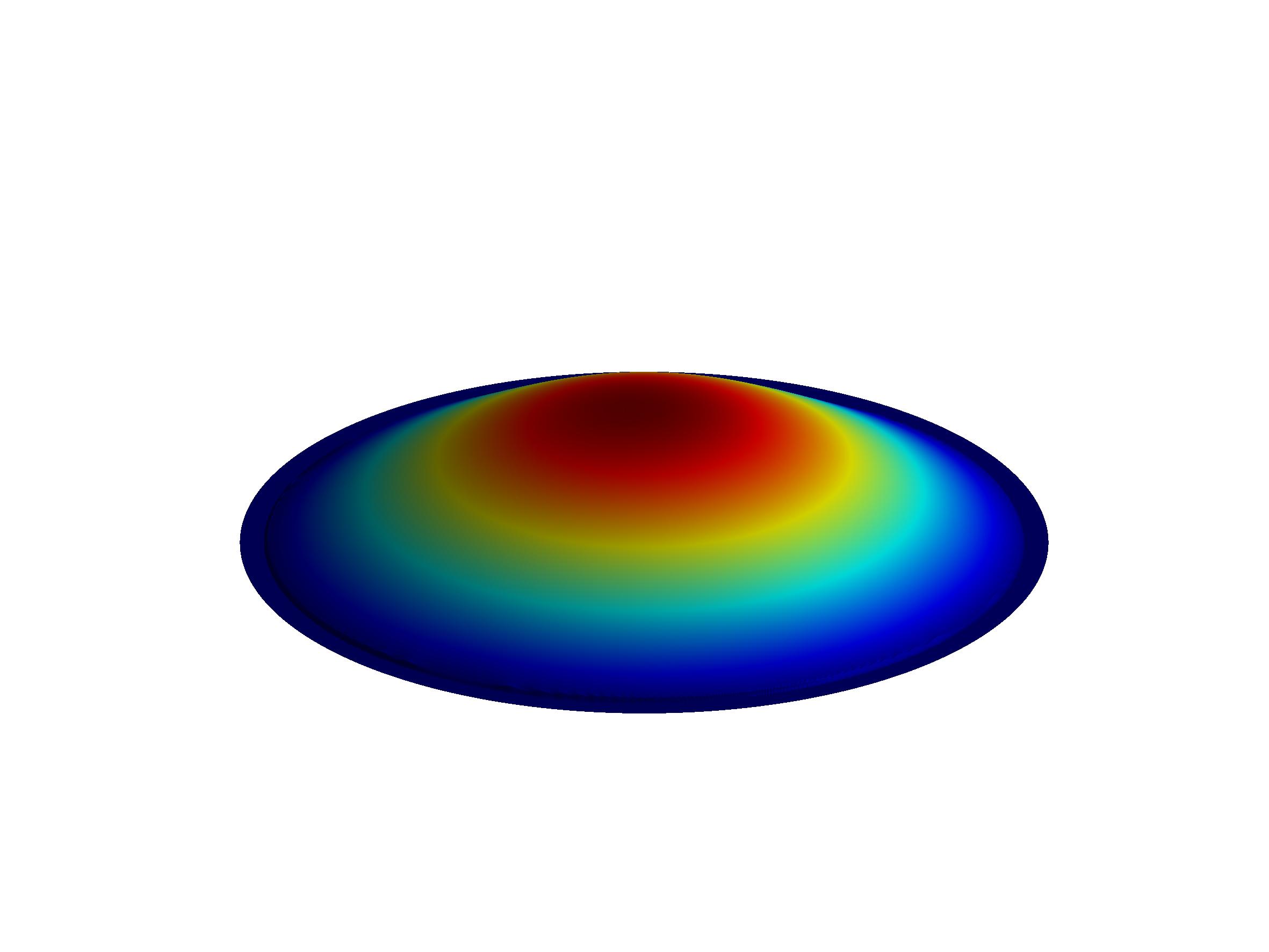}
        \subcaption{}
        \label{f:Circ_adhesive_w0_0}
    \end{subfigure}
    \begin{subfigure}{0.495\textwidth}
        \centering
    \includegraphics[width=80mm,trim=3cm 3cm 20cm 15cm,clip]{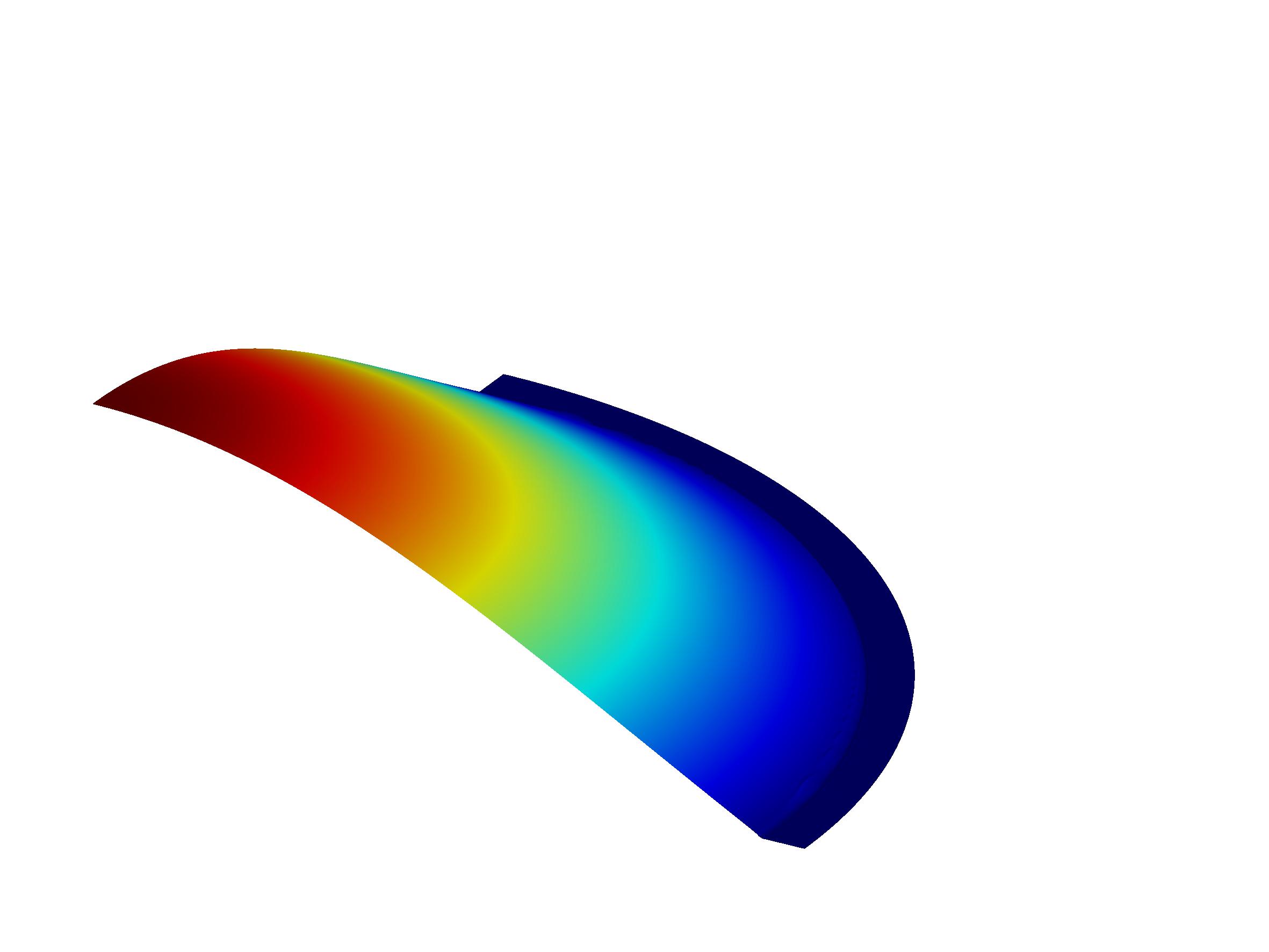}
        \subcaption{}
        \label{f:Circ_adhesive_w0_0_section}
    \end{subfigure}
    \caption{Vibrating circular plate on an adhesive substrate: (\subref{f:Full_disk_adhesive_substrate_modal}) Illustration of the boundary conditions; (\subref{f:Circ_adhesive_freqeuncy_variation}) variation of the frequencies with the adhesion strength $\Gamma$; (\subref{f:Circ_adhesive_w0_0}) first mode shape $\omega_{(1,1)}$; (\subref{f:Circ_adhesive_w0_0_section}) cut of (\subref{f:Circ_adhesive_w0_0}). $m$ and $n$ in $\omega_{(m,n)}$ are defined in Fig.~\ref{f:Circ_simply_mode_shapes}. The radii of the cavity $R_1$ and the fillet $R_2$ are 450 nm and 50 nm. The results are normalized by $\hat{\omega}_{(0,0)}=\omega_{(0,0)}\big |_{\Gamma=0}$ (the first natural frequency). Instability points are indicated by dashed vertical lines.}
    \label{f:Circ_freqeuncy_variation_adhesive}
\end{figure}

\begin{figure}
    \begin{subfigure}{0.495\textwidth}
        \centering
    \includegraphics[height=50mm,trim=12cm 25cm 3cm 25cm,clip]{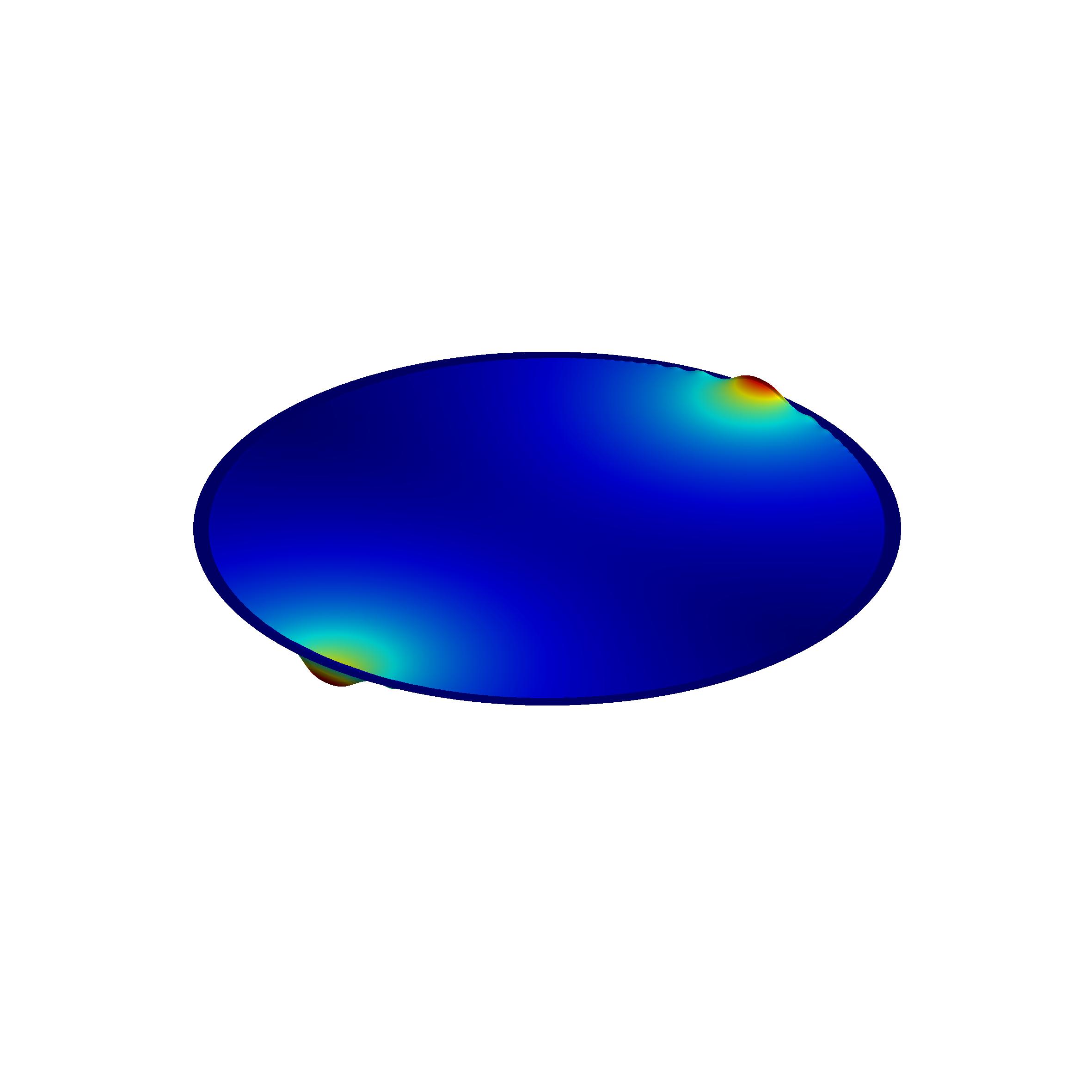}
    \caption{}
    \label{f:circ_SS_mode_instabiliy}
    \end{subfigure}
        \begin{subfigure}{0.495\textwidth}
        \centering
    \includegraphics[height=50mm]{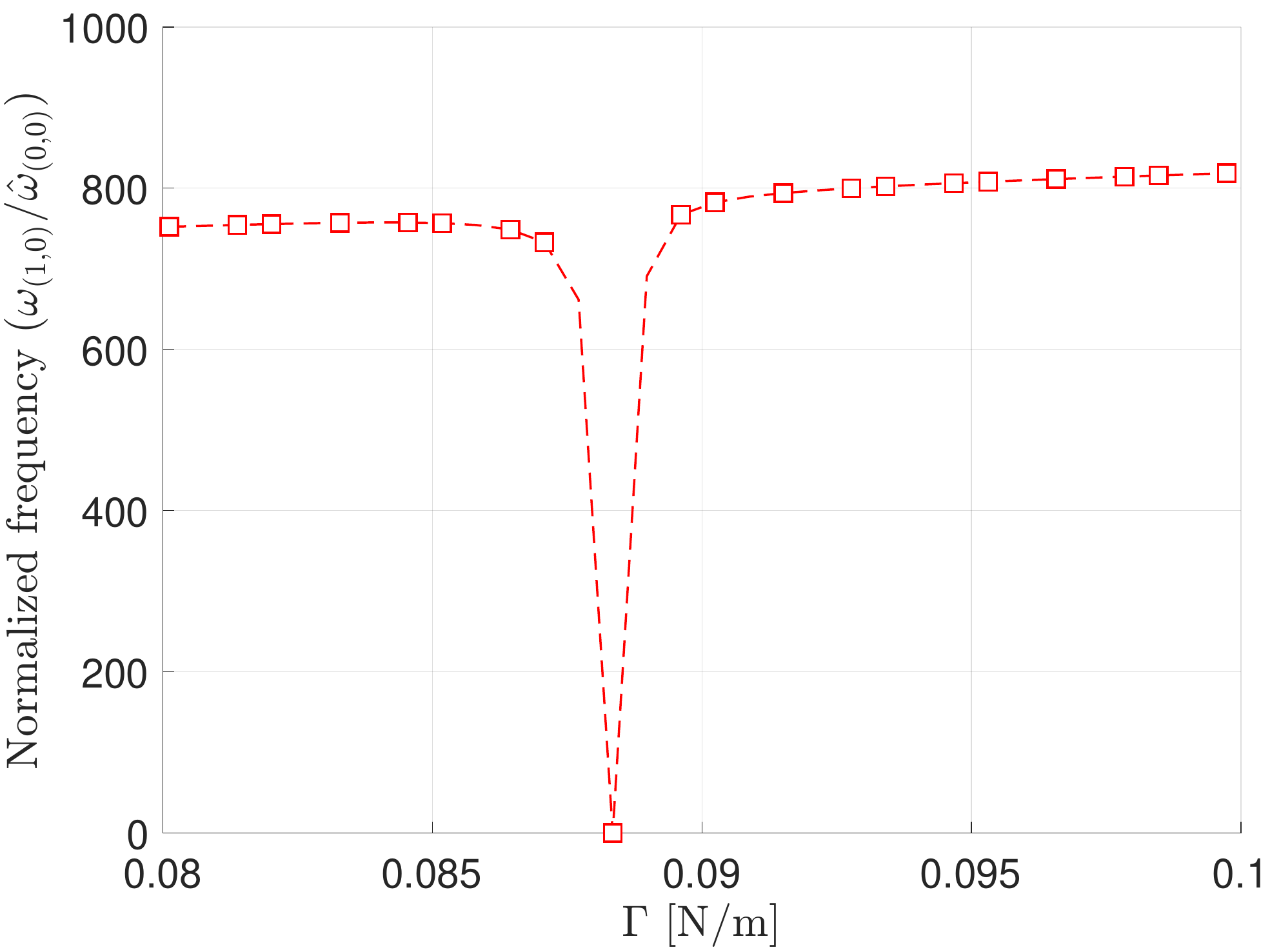}
    \caption{}
    \label{f:Circ_adhesive_freqeuncy_variation_zoom}
    \end{subfigure}
    \caption{Vibrating circular plate on an adhesive substrate: (\subref{f:circ_SS_mode_instabiliy}) Local mode near the instability point $\Gamma=0.08834~\mathrm{N/m}$; (\subref{f:Circ_adhesive_freqeuncy_variation_zoom}) enlargement of (\ref{f:Circ_adhesive_freqeuncy_variation}) for $\omega_{(1,0)}$.}
    \label{f:circ_SS_mode_softening}
\end{figure}

\subsection{Vibrating carbon nanotubes}
\begin{figure}[h]
\begin{center} \unitlength1cm
\begin{picture}(18,4)
\put(0,0.){\includegraphics[width=38mm,trim=22cm 0cm 22cm 0cm,clip]{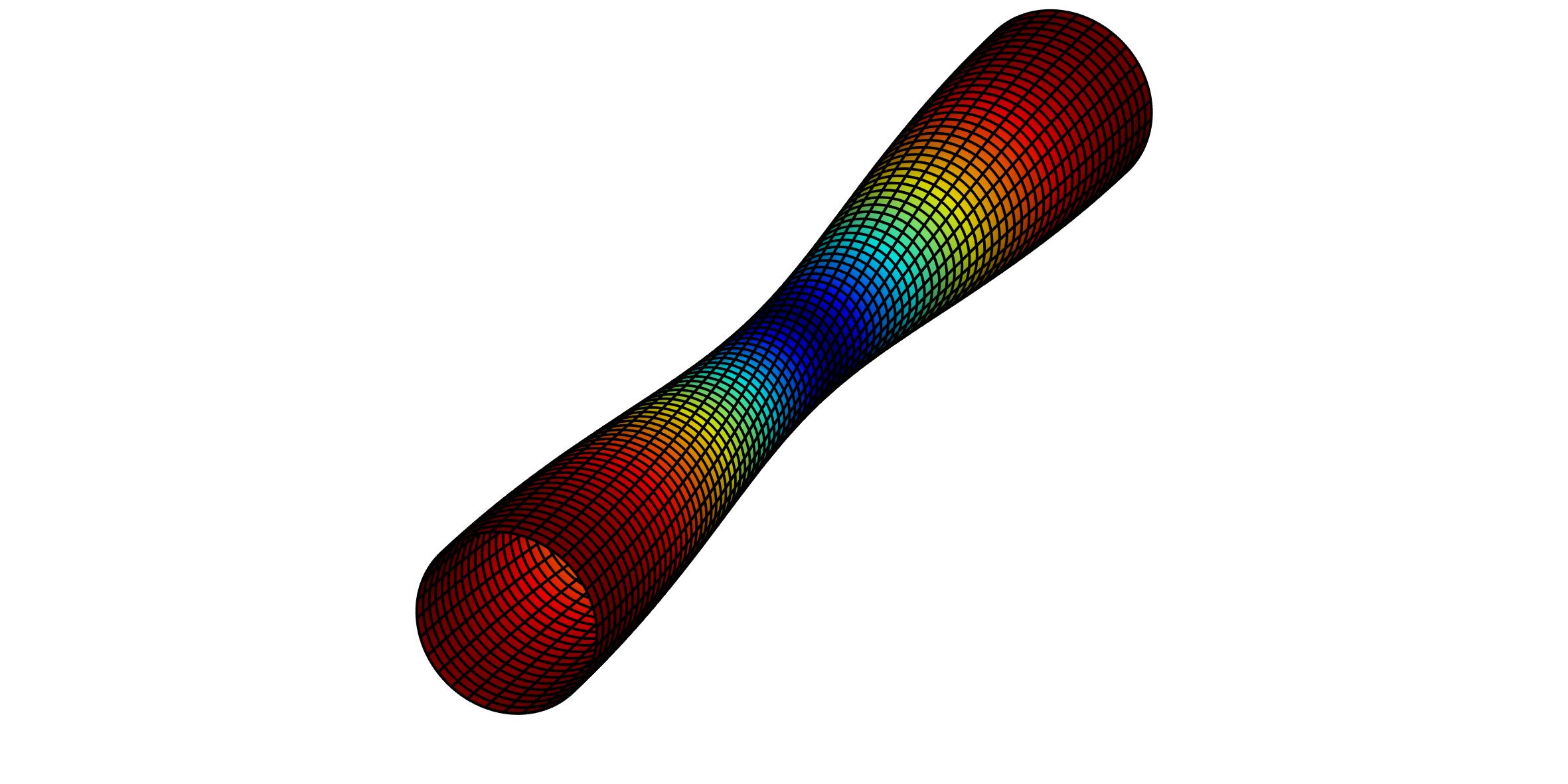}}
\put(4,0.){\includegraphics[width=38mm,trim=22cm 0cm 22cm 0cm,clip]{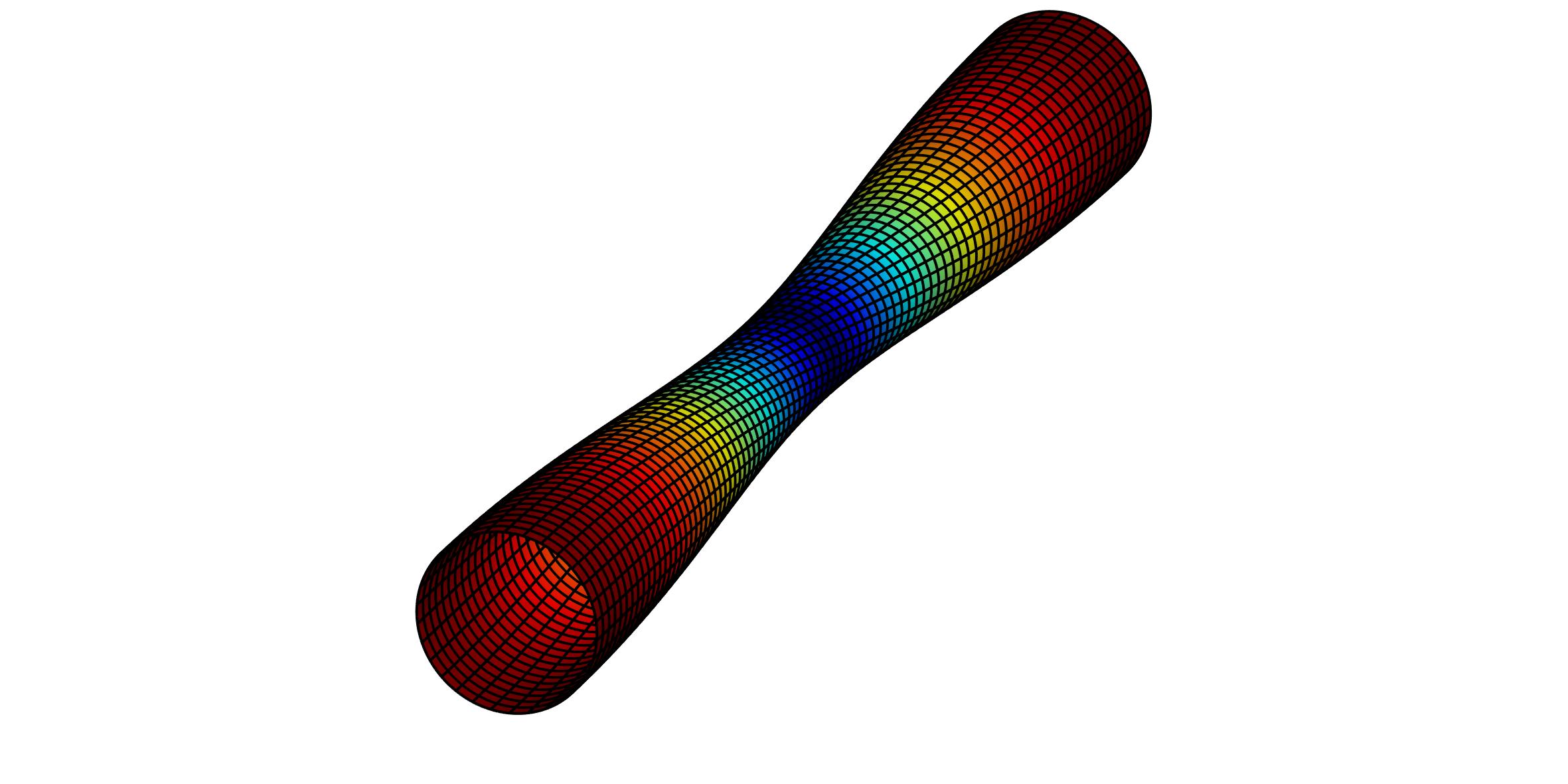}}
\put(8,0.){\includegraphics[width=38mm,trim=22cm 0cm 22cm 0cm,clip]{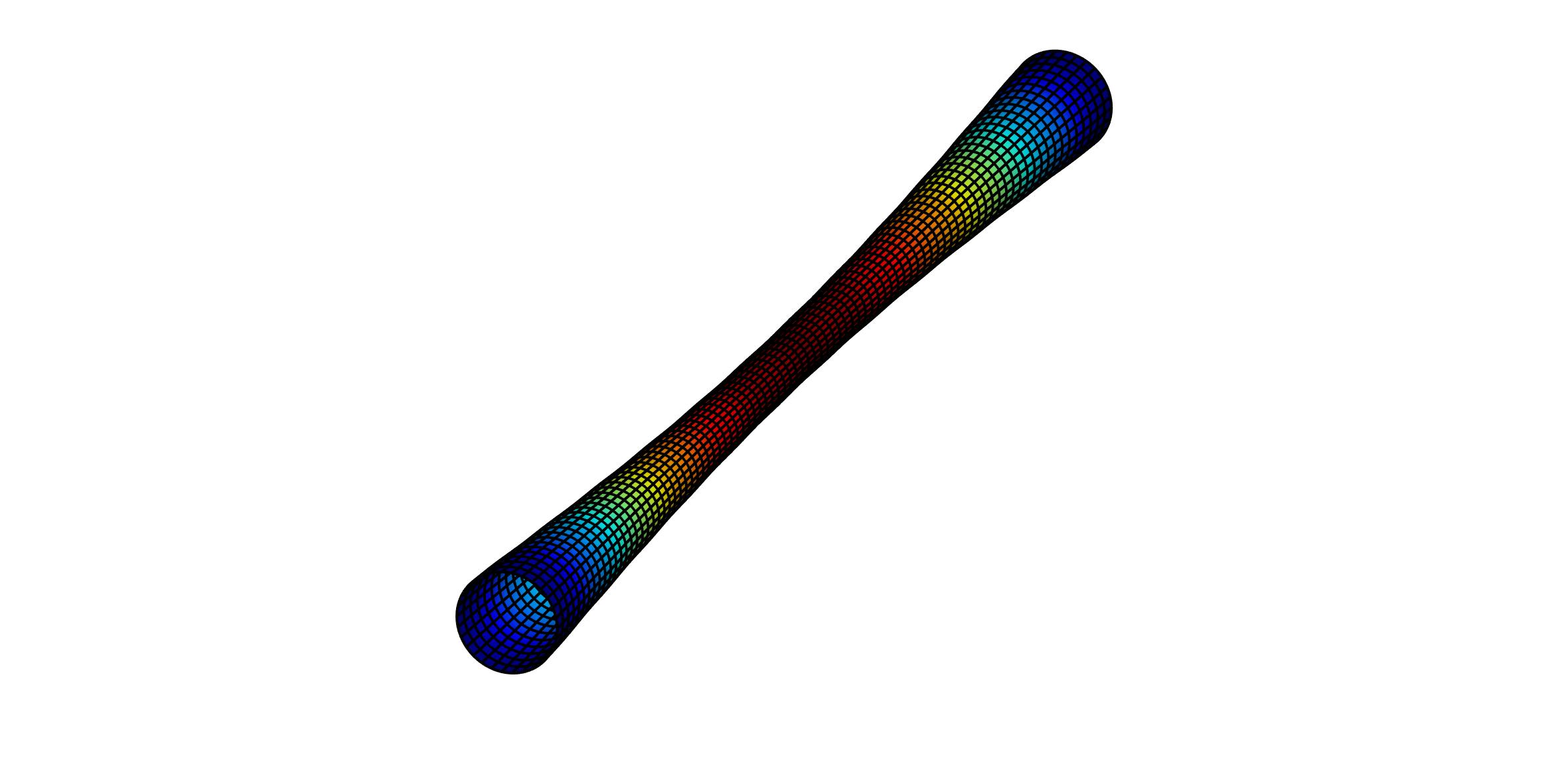}}
\put(12,0){\includegraphics[width=38mm,trim=22cm 0cm 22cm 0cm,clip]{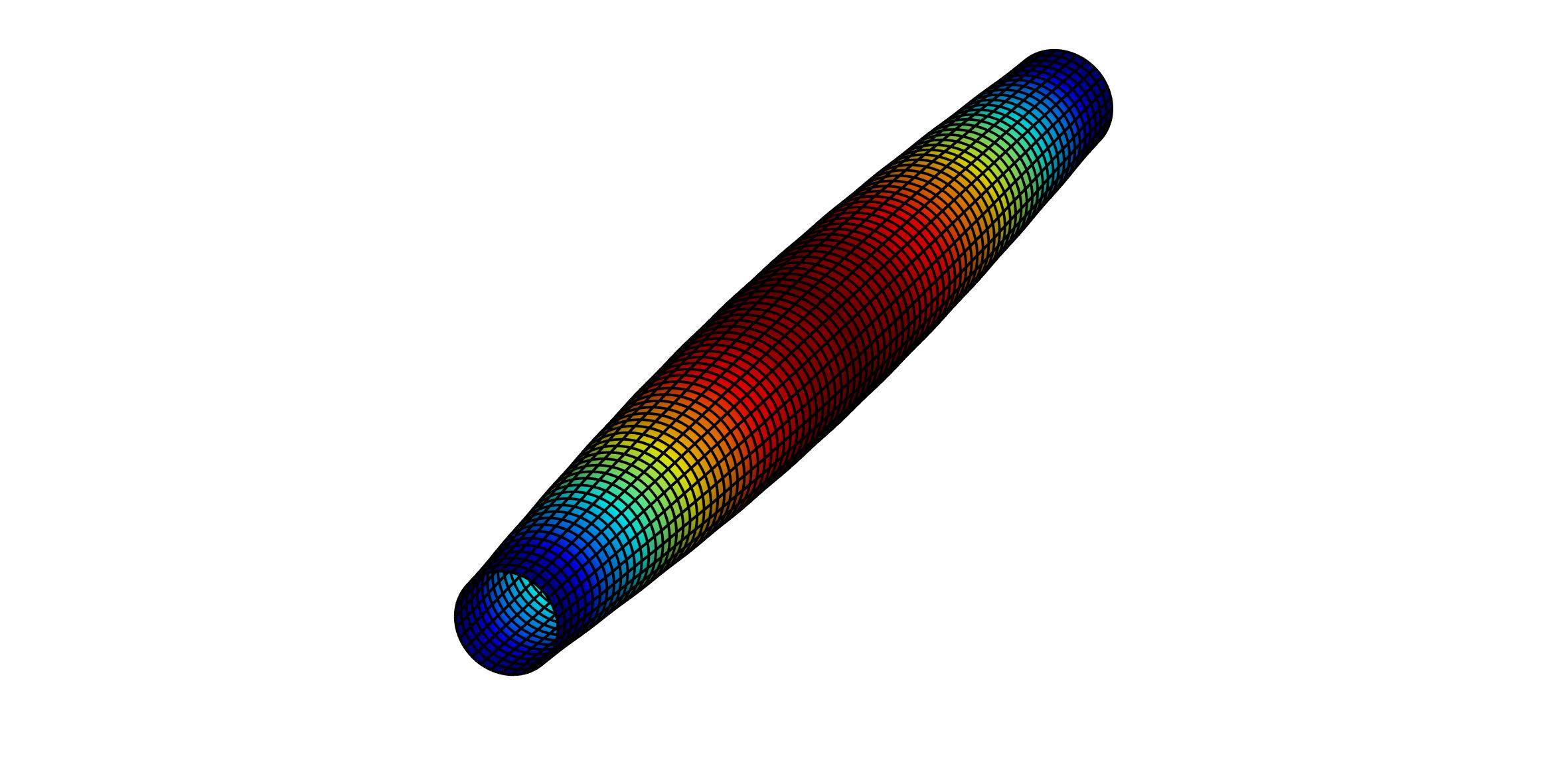}}

\put(3.8,0){a}
\put(11.8,0){b}

\end{picture}
\caption{Vibrating CNT with free boundary: Mode shapes:
            (a) Torsion mode 1 (TM1) occurring at 0.41545 THz,
            (b) radial breathing mode 1 (RB1) occurring at 5.13513 THz. \textcolor{cgn2}{The figures are colored by the norm of the eigenvectors.}}
\label{f:CNT_SS_radial_torsion_geo}
\end{center}
\end{figure}

\begin{figure}[h]
        \centering
    \includegraphics[height=50mm,trim=5cm 5cm 3cm 5cm,clip]{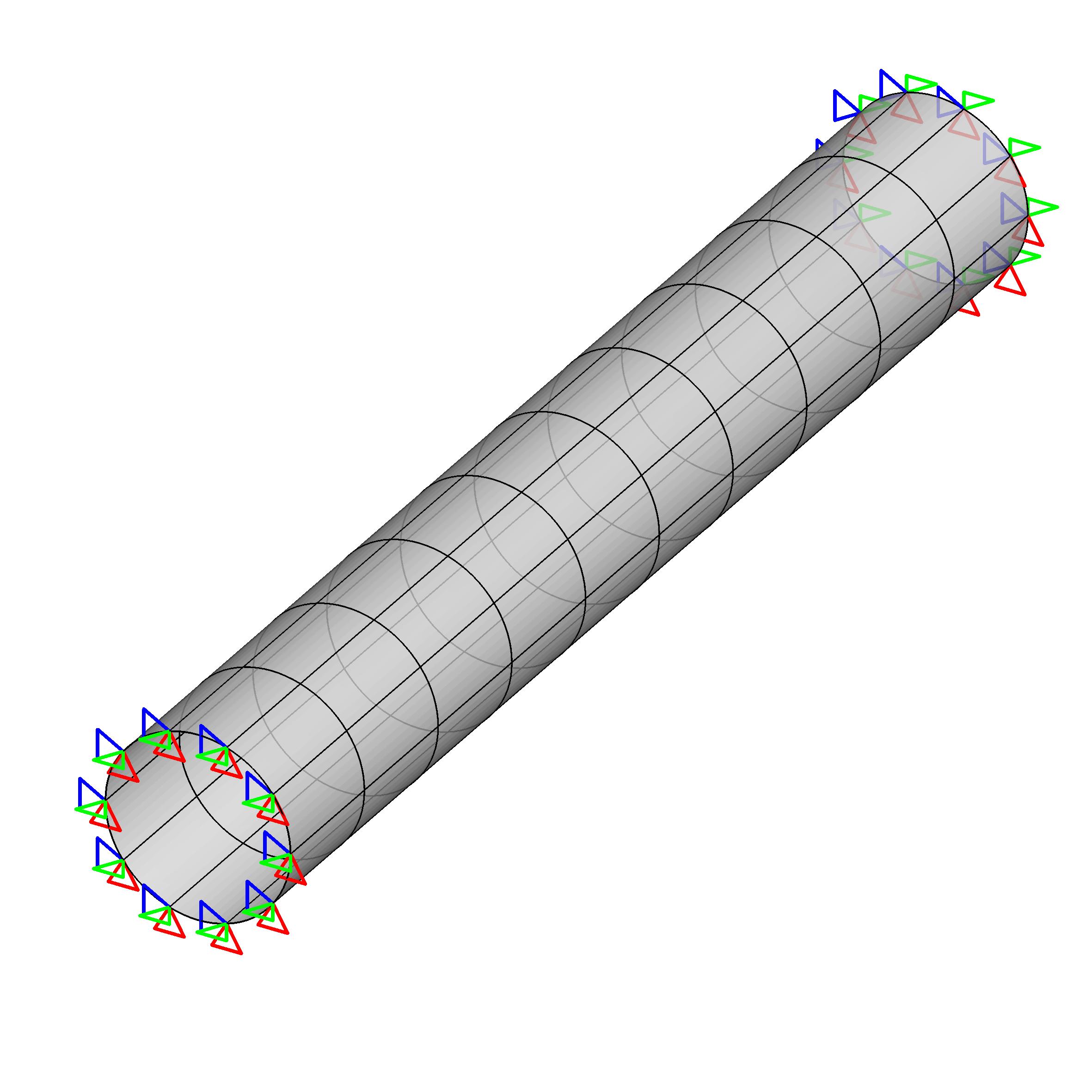}
  \caption{Vibrating CNT: Boundary conditions and finite element mesh.}
    \label{f:CNT_SS_BC}
\end{figure}

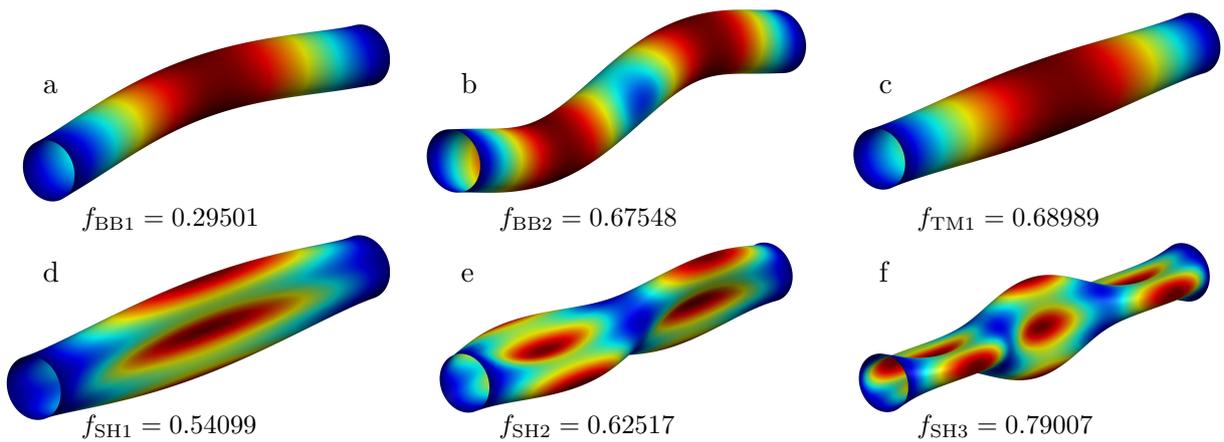
\begin{figure}[h]
\begin{center} \unitlength1cm
\begin{picture}(18,6)
\put(0,3.){\includegraphics[height=28mm,trim=5cm 20cm 0cm 20cm,clip]{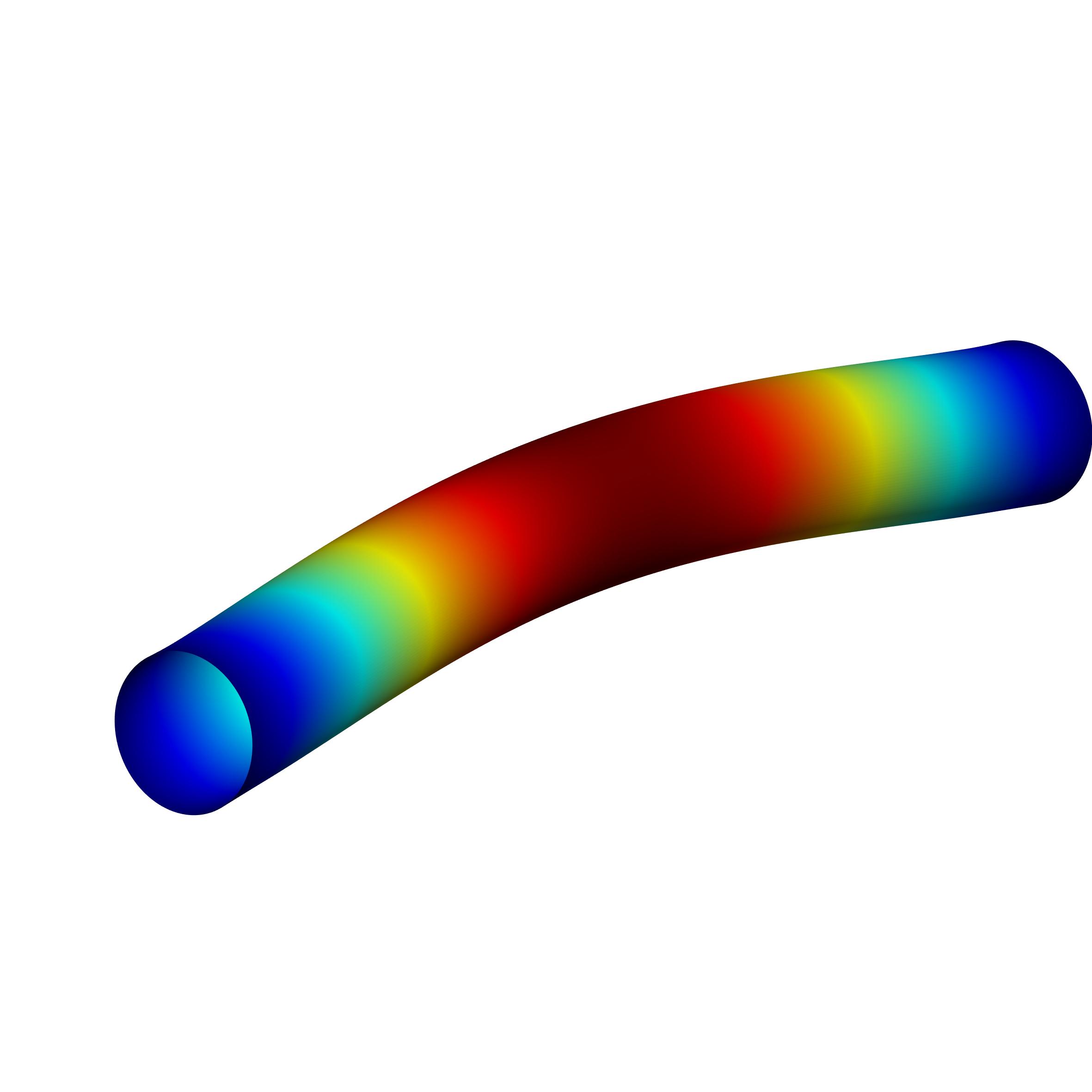}}
\put(5.5,3.){\includegraphics[height=28mm,trim=5cm 20cm 0cm 20cm,clip]{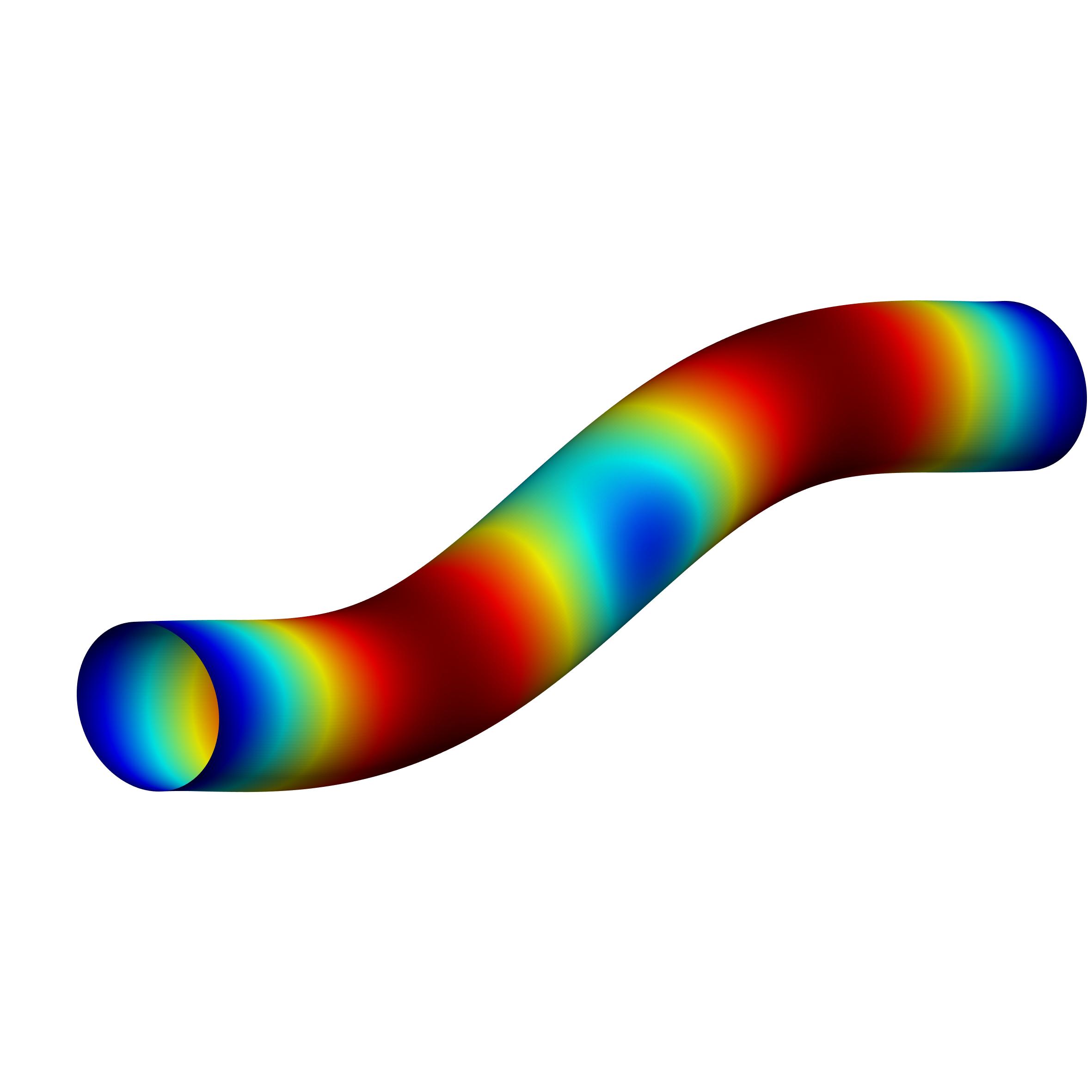}}
\put(11,3.){\includegraphics[height=28mm,trim=5cm 20cm 0cm 20cm,clip]{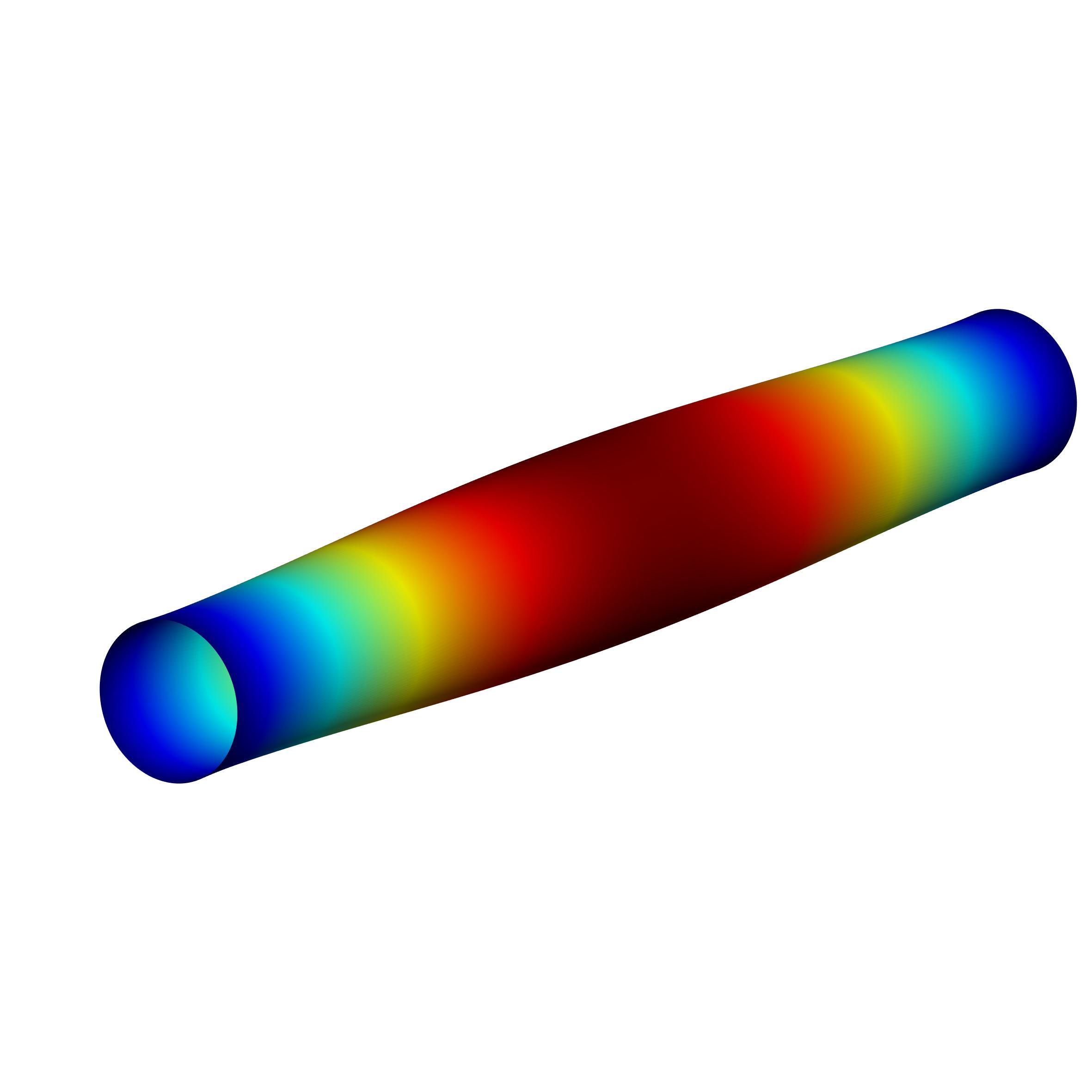}}
\put(0,0){\includegraphics[height=28mm,trim=5cm 20cm 0cm 20cm,clip]{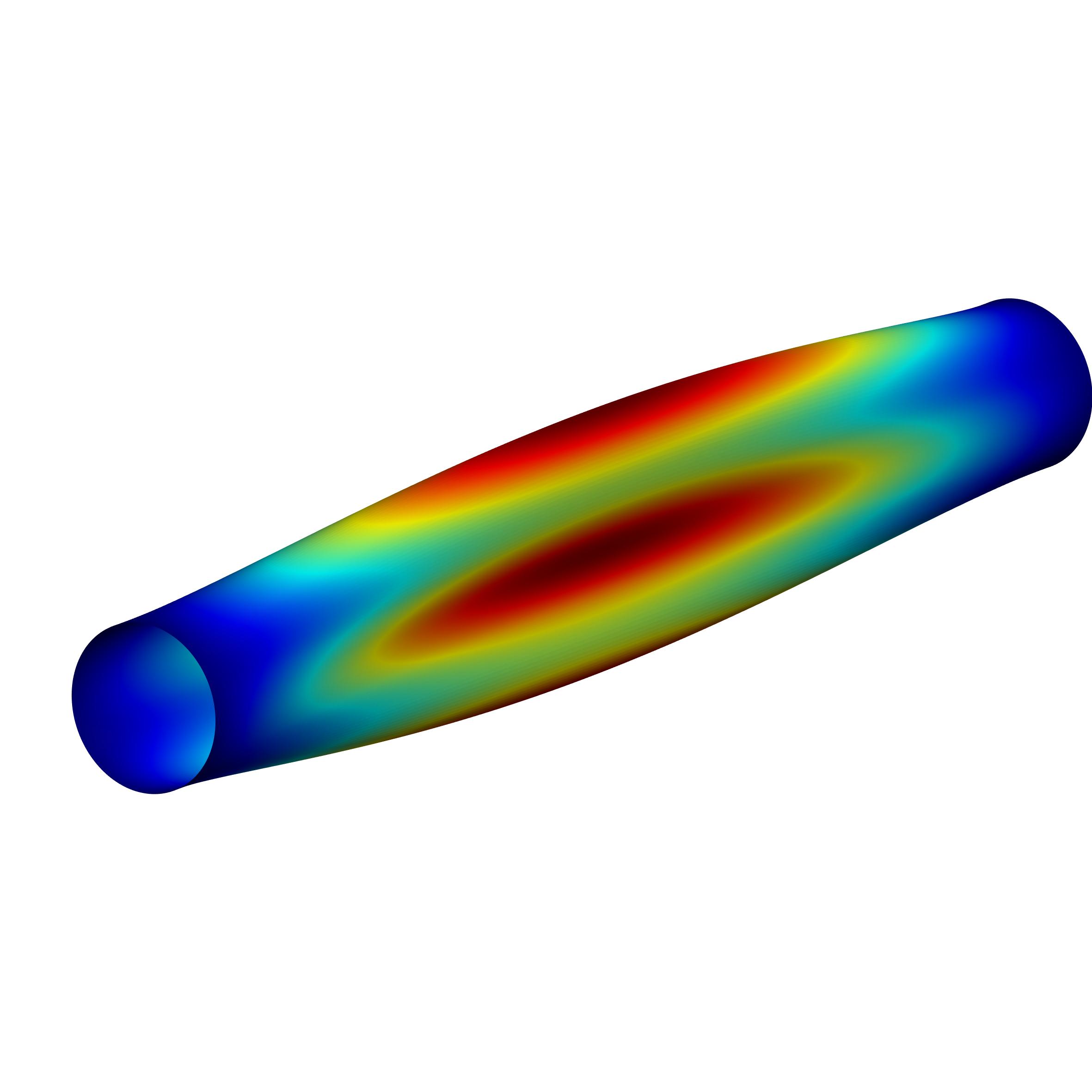}}
\put(5.5,0){\includegraphics[height=28mm,trim=5cm 20cm 0cm 20cm,clip]{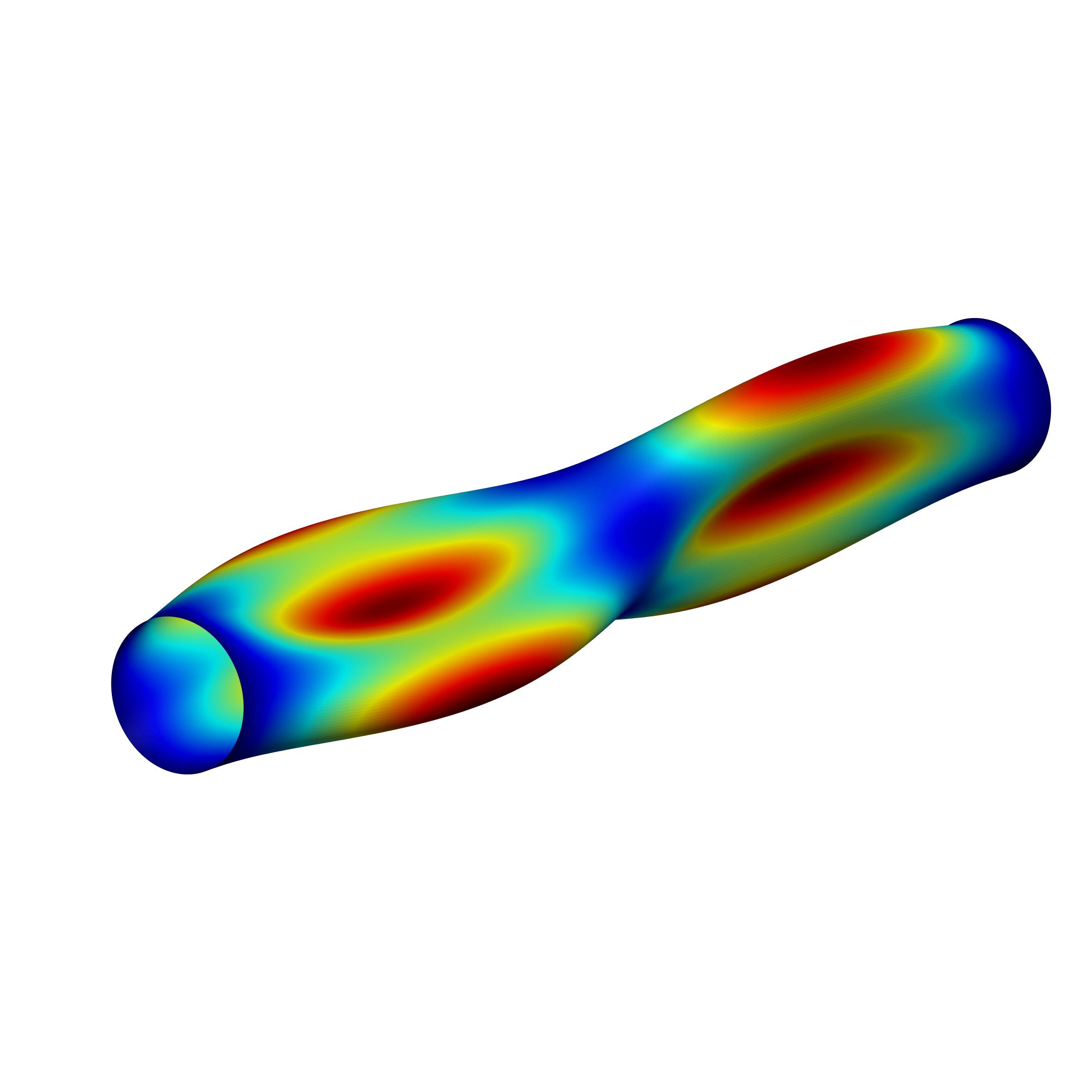}}
\put(11,0){\includegraphics[height=28mm,trim=5cm 20cm 0cm 20cm,clip]{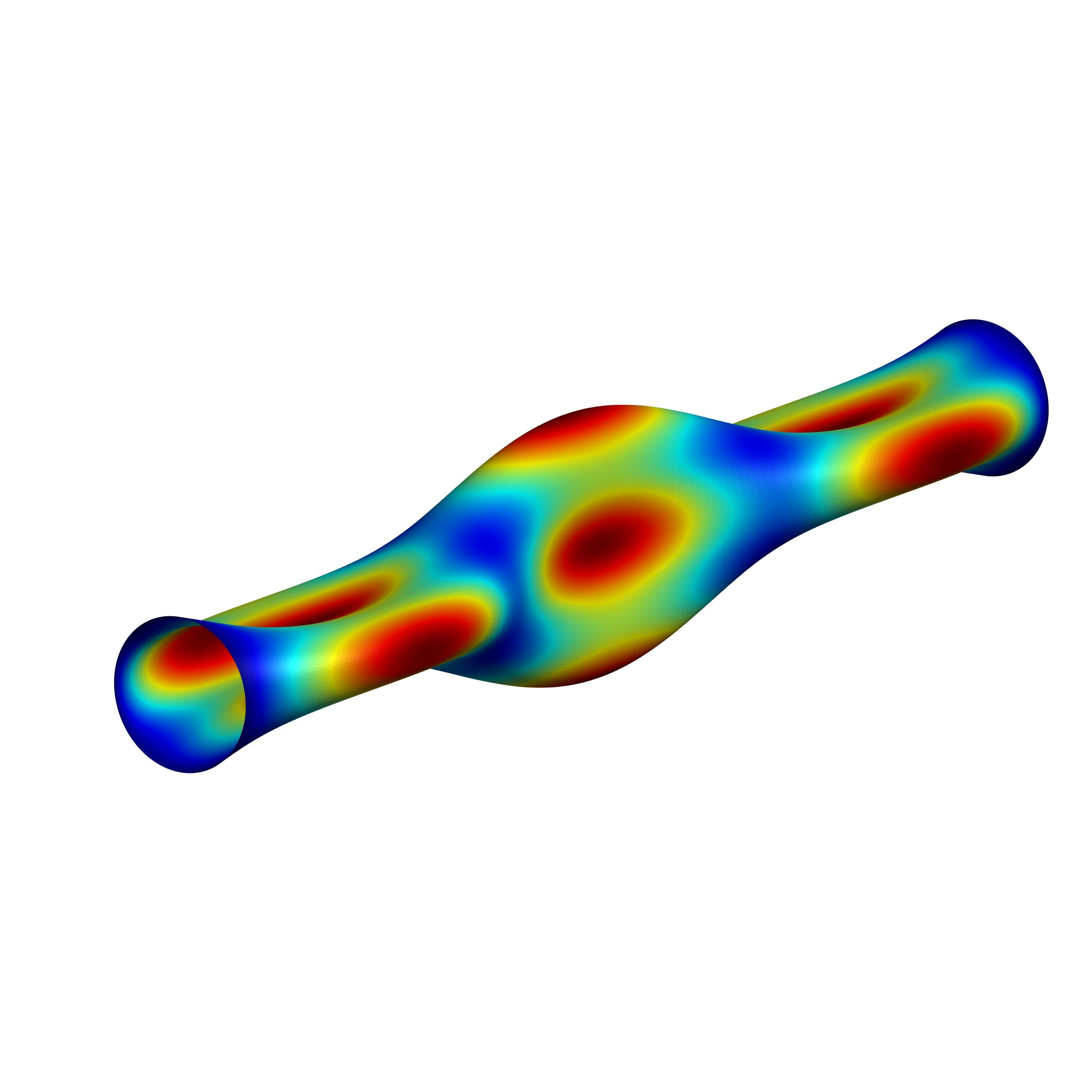}}

\put(0.5,4.5){a}
\put(6,4.5){b}
\put(11.5,4.5){c}
\put(.5,2){d}
\put(6,2){e}
\put(11.5,2){f}

\put(1,0.0){{\small $f_{\text{SH1}}=0.54099$}}
\put(6.5,0.0){{\small $f_{\text{SH2}}=0.62517$}}
\put(12,0.0){{\small $f_{\text{SH3}}=0.79007$}}

\put(1,2.75){{\small $f_{\text{BB1}}=0.29501$}}
\put(6.5,2.75){{\small $f_{\text{BB2}}=0.67548$}}
\put(12,2.75){{\small $f_{\text{TM1}}=0.68989$}}

\end{picture}
\caption{Vibrating CNT at zero pre-tension: Mode shapes and natural frequencies $\hat{f}_{(m,n)}=\hat{\omega}_{(m,n)}/2\pi$ in [THz]:
            (a) Bending beam mode 1 (BB1),
            (b) bending beam mode 2 (BB2),
            (c) \textcolor{cgn}{torsion} mode 1 (TM1),
            (d) shell mode 1 (SH1),
            (e) shell mode 2 (SH2) and
            (f) shell mode 3 (SH3). CNT(10,10) with a length of 10 nm and simply supported boundary is used, where $m$ and $n$ in CNT$(n,m)$ are the chirality parameters. \textcolor{cgn2}{The figures are colored by the norm of the eigenvectors.}}
\label{f:CNT_SS_mode_shapes}
\end{center}
\end{figure}

\begin{figure}[h]
    \begin{subfigure}{0.495\textwidth}
        \centering
    \includegraphics[height=55mm]{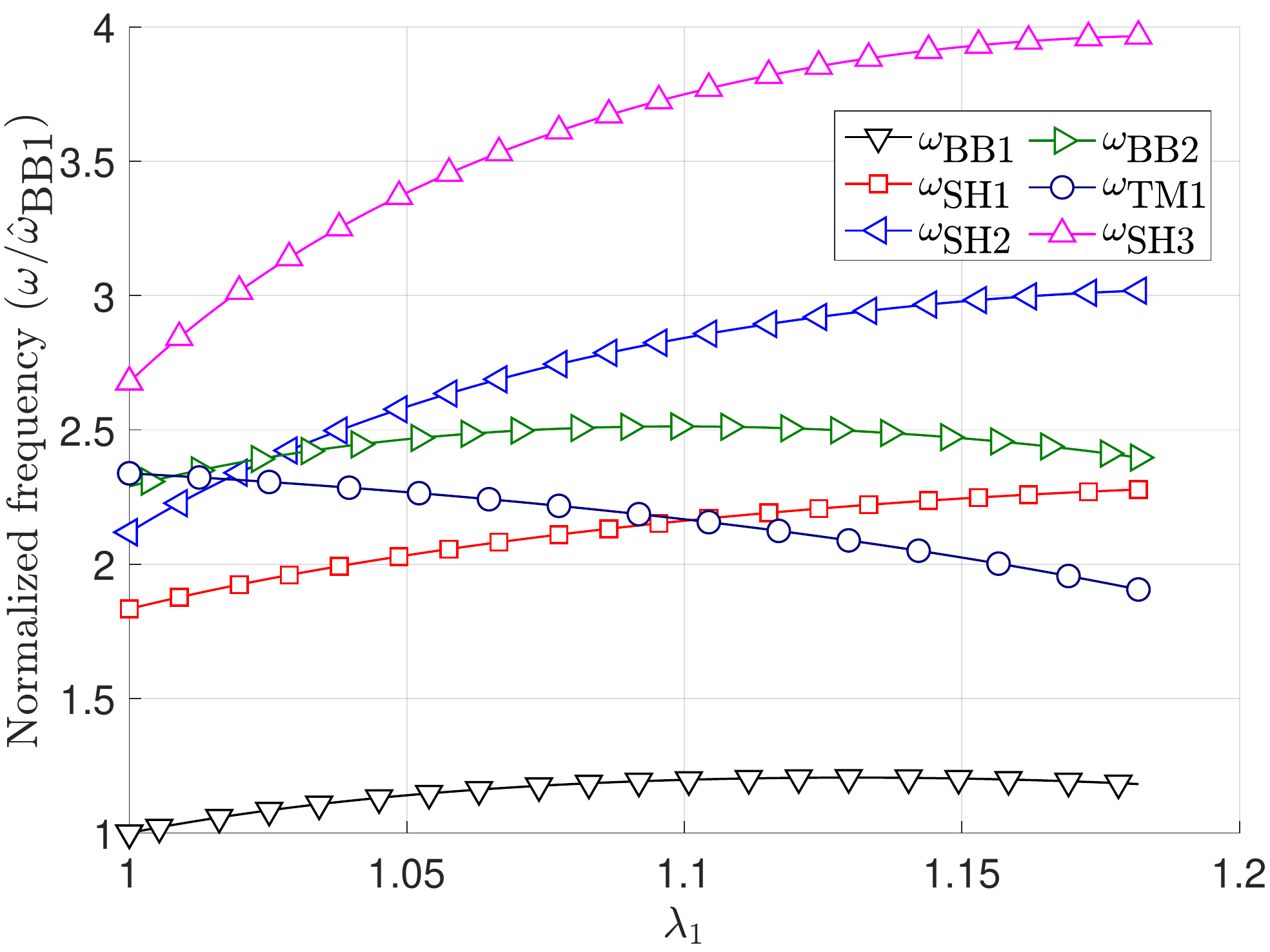}
        \subcaption{}
        \label{f:CNT_simply_simply_cn10_cm10_freqeuncy_variation_uni}
    \end{subfigure}
        \begin{subfigure}{0.495\textwidth}
        \centering
    \includegraphics[height=55mm]{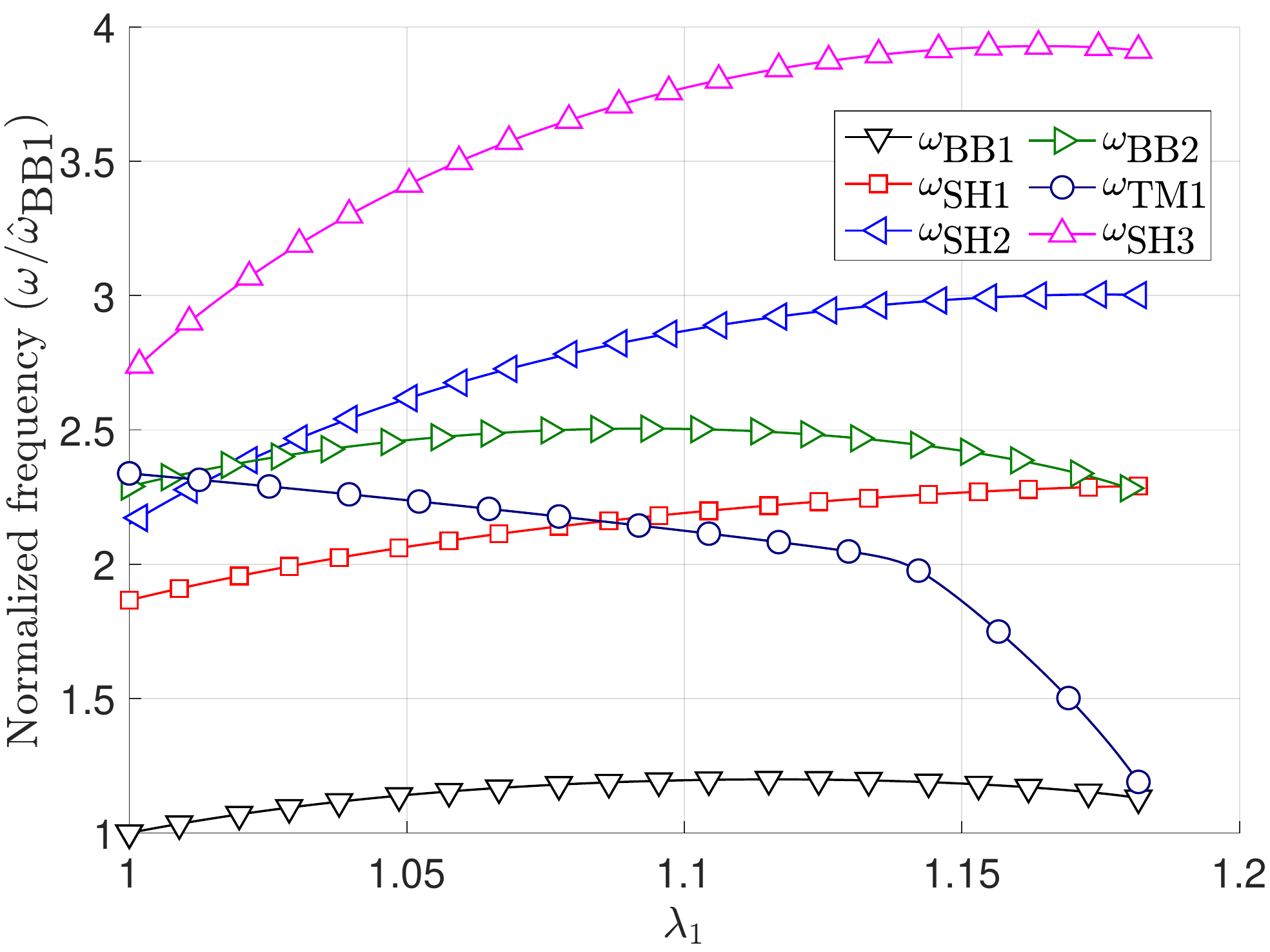}
        \subcaption{}
        \label{f:CNT_simply_simply_cn17_cm0_freqeuncy_variation_uni}
    \end{subfigure}
         \begin{subfigure}{1\textwidth}
        \centering
    \includegraphics[height=55mm]{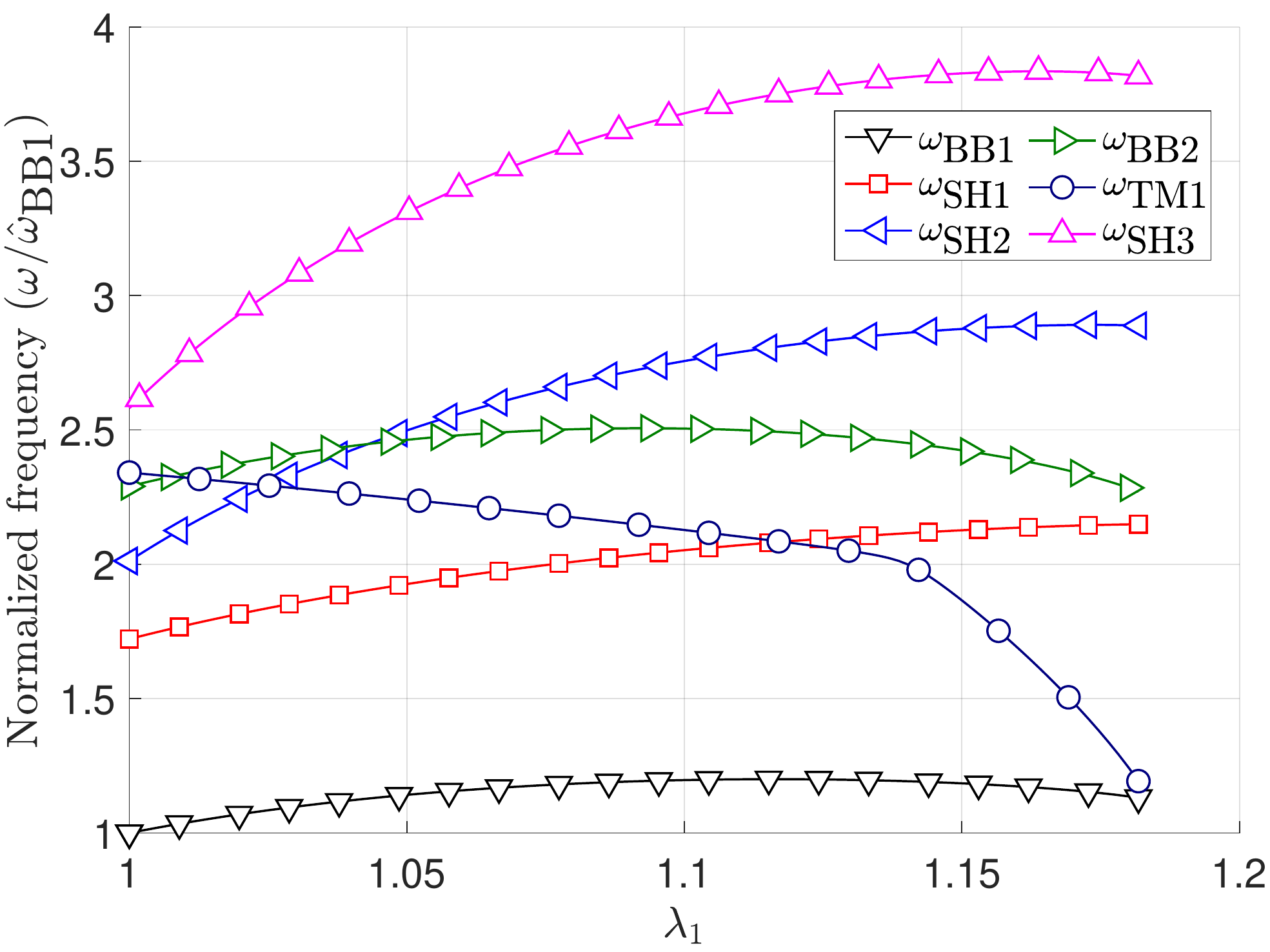}
        \subcaption{}
         \label{f:CNT_simply_simply_cn14_cm7_freqeuncy_variation_uni}
    \end{subfigure}
    \caption{Vibrating CNTs under axial tension: Variation of the frequencies versus stretch $\lambda_1$.
    (\subref{f:CNT_simply_simply_cn10_cm10_freqeuncy_variation_uni}) CNT(10,10),
    (\subref{f:CNT_simply_simply_cn17_cm0_freqeuncy_variation_uni}) CNT(17,0),
    (\subref{f:CNT_simply_simply_cn14_cm7_freqeuncy_variation_uni}) CNT(14,7). An aspect ratio of 7.3746 is used.
    BB, SH and TM stand for bending beam, shell and \textcolor{cgn}{torsion} modes. \textcolor{cgn}{The boundary is simply supported.}}\label{f:CNT_simply_simply_freqeuncy_variation_uni}
\end{figure}

\textcolor{cgn}{In this section, modal analysis of CNTs is conducted to calculate the frequencies of CNTs. The model is verified in a linear regime and then the simulation is extended to the nonlinear regime. For the following simulation results a FE mesh with $160\times160$ quadratic NURBS elements is considered to ensure convergence.\\
First, a linear modal analysis with free boundaries is conducted and the FE formulation is verified. The radial breathing modes can be easily computed in different numerical and experimental methods. In this mode, the radius of the CNT increases and there are no tangential displacements. The radial breathing mode should not be confused with the first torsion mode (see Fig.~\ref{f:CNT_SS_radial_torsion_geo}). The natural radial breathing modes of different CNTs for the proposed model are compared with other finite element, molecular dynamics, molecular mechanics, ab-initio and experimental results from the literature (Tab.~\ref{t:CNT_FF_radial_breathing_freq}). \textcolor{cgn2}{The quantum and experimental results of the radial breathing mode have been obtained by either using a periodic boundary condition in the axial direction or by considering very long CNTs, respectively. In Tab.~\ref{t:CNT_FF_radial_breathing_freq}, it has been confirmed that the radial breathing mode is converging when increasing the aspect ratio, so the proposed continuum results can be compared with atomistic and experimental results.} The new results are in a good agreement with those. The frequencies from the proposed model are higher than other numerical results, but lower than (and thus closer to) experimental results. The first radial mode has a frequency that is several times larger than the first torsion mode in the current example.}\\
Second, the influence of uniaxial stretch on the frequencies of CNTs is investigated. The CNTs are simply supported (Fig.~\ref{f:CNT_SS_BC}) and stretched in the axial direction. The mode shapes, natural frequencies and their naming are shown in Fig.~\ref{f:CNT_SS_mode_shapes} for zero axial pre-tension. Figure \ref{f:CNT_simply_simply_freqeuncy_variation_uni} shows that the frequency of the \textcolor{cgn}{torsion} and shell modes are monotonically decreasing and increasing with stretch, respectively. In contrast, the bending beam frequencies first increase and then decrease under further stretching. The frequencies of the higher shell modes increase more than the lower shell modes, and the frequencies of the bending beam modes are changing with a similar factor for all CNTs. \textcolor{cgn2}{CNT(10,10) is stretched in the zigzag direction, which is the strongest direction, while CNT(17,0) and CNT(14,7) are stretched in other directions that are weaker than the armchair direction. So the torsion modes of CNT(17,0) and CNT(14,7) show a rapid decrease beyond $\lambda_1=1.142$, but the decrease is more slowly for CNT(10,10). A similar behavior should be seen for CNT(10,10), if it could be stretched further. The CNTs will become unstable if there are stretched further.}

\begin{table}[h]
\centering
\begin{tabular}{l c c c c c c }
  \hline
  CNT($n$,$m$) & AR & PM & \cite{Batra2008_01} (MD, FE) & \cite{Gupta2010_01} (MM) & \cite{Lawler2005_01} (FP) & \cite{Kuzmany1998_01} (Exp)\\
  \hline
  CNT(10,10) & 5.669 & 5.14269 & NA & 5.02583 & NA & NA\\
  CNT(10,10) & 10 & 5.13660 & NA & 5.01628 & NA & NA\\
  CNT(10,10) & 15 & 5.13513	& 5.01783, 4.96518 & NA & NA & NA\\
  CNT(10,10) & $\infty$ & 5.13427	& 5.01783, 4.96518 & NA & 5.06649 & 5.30632\\
  CNT(20,20) & 5.669 & 2.57346 & NA & 2.51630 & NA & NA\\
  CNT(25,10) & 5.669 & 2.85468 & NA & 2.79265 & NA & NA\\
  CNT(30,0) & 5.669 & 2.47617 & NA & 2.42334 & NA & NA\\
  CNT(43,0) & 6.052 & 2.07282 & NA & 2.02818 & NA & NA\\
  \hline
\end{tabular}
  \caption{Vibrating CNT with free boundary: The natural radial breathing frequencies [THz] according to various studies. The speed of light is taken as $2.9979\times10^8$ m/s. AR = aspect ratio; PM = proposed model; MD = molecular dynamics; FE = three-dimensional finite elements; MM= molecular mechanics; FP = first principles; Exp = experimental; NA = not available.}
  \label{t:CNT_FF_radial_breathing_freq}
\end{table}

\begin{figure}
    \begin{subfigure}{0.495\textwidth}
        \centering
    \includegraphics[width=80mm]{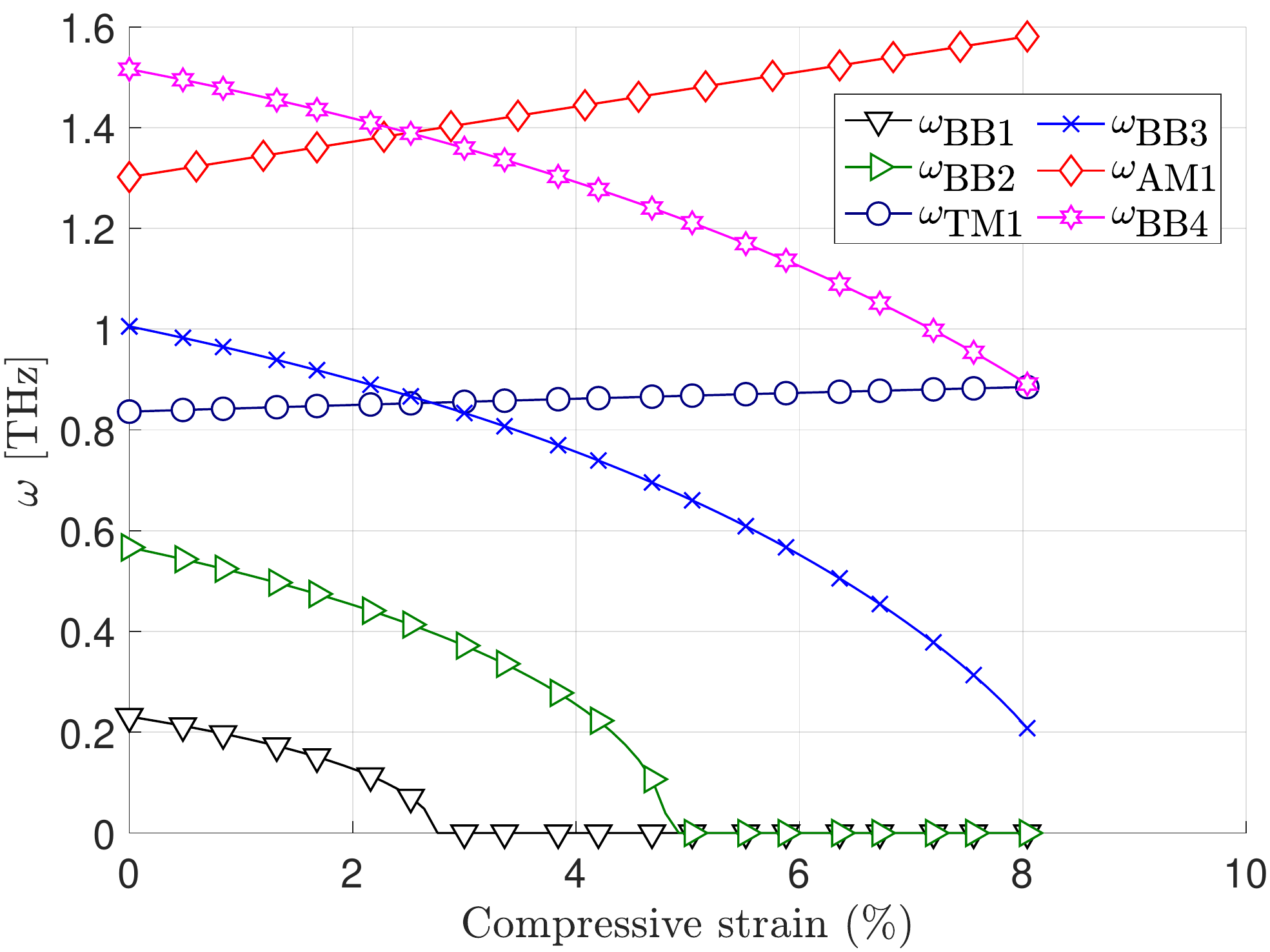}
    \caption{}
    \label{f:CNT_SS_cn7_cm0_AR15_m80_n80_compressed_freq}
    \end{subfigure}
        \begin{subfigure}{0.495\textwidth}
        \centering
    \includegraphics[width=80mm]{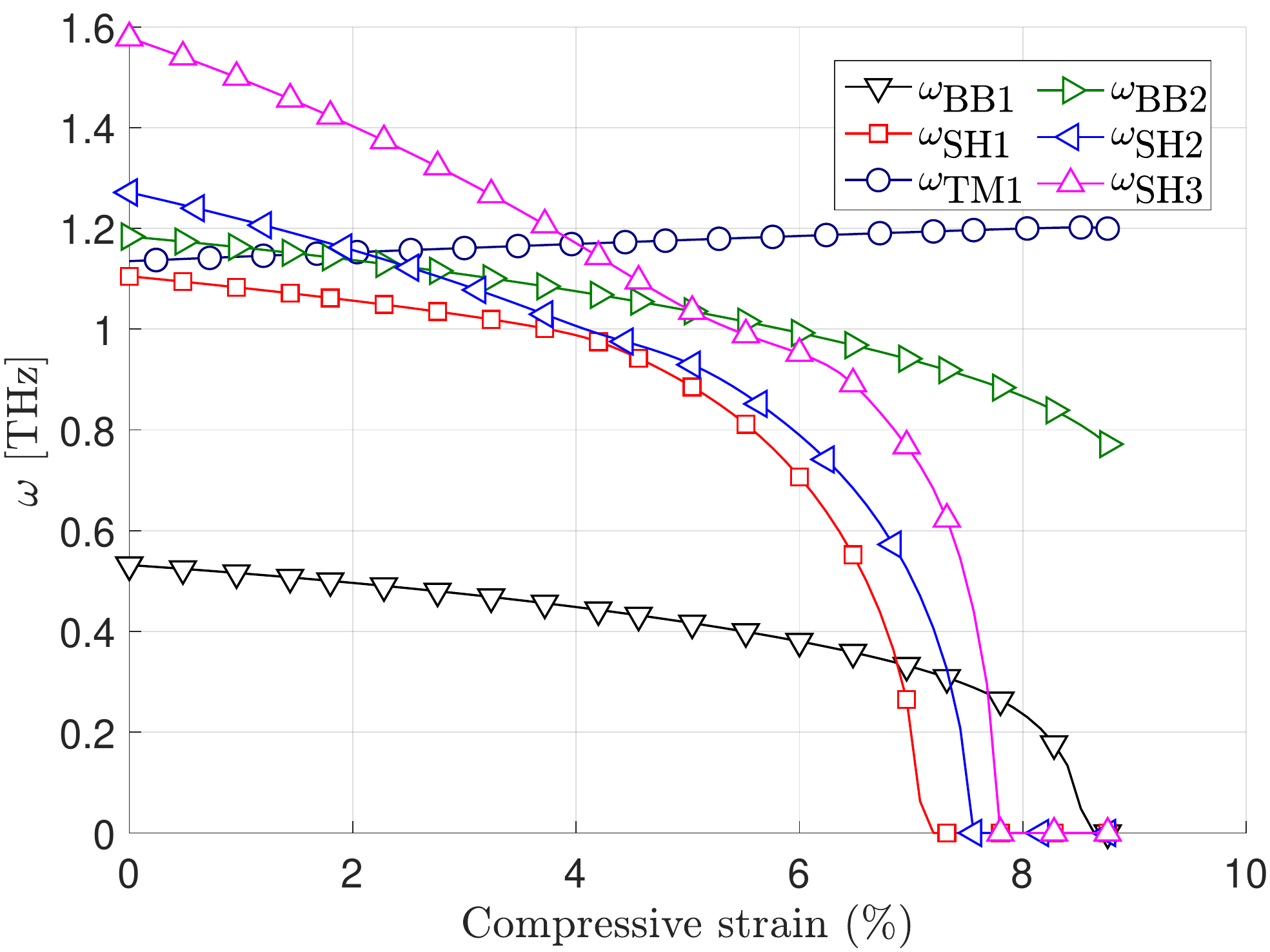}
    \caption{}
    \label{f:CNT_SS_cn7_cm7_AR6_4_m80_n80_compressed_freq}
    \end{subfigure}
    \caption{Vibrating CNTs under axial compression:
     Variation of the frequencies versus the axial strain for (\subref{f:CNT_SS_cn7_cm0_AR15_m80_n80_compressed_freq}) CNT(7,0) with AR=15,
                    (\subref{f:CNT_SS_cn7_cm7_AR6_4_m80_n80_compressed_freq}) CNT(7,7) with AR=6.4. AR= aspect ratio. AM1= axial mode 1, BB4= bending beam mode 4 (see Fig.~\ref{f:CNT_SS_mode_shapes}) for other mode shapes. The boundary is simply supported. Zero frequencies indicate buckling.}
    \label{f:CNT_SS_compressed_freq}
\end{figure}

\textcolor{cgn}{Finally, the buckling of CNTs under compressive axial strains is investigated. CNTs buckle in bending beam or shell modes depending on their aspect ratio. The frequencies of the first and second bending beam mode become zero at $2.7\%$ and $5.4\%$ strain for CNT(7,0) with aspect ratio 15, indicating buckling (Fig.~\ref{f:CNT_SS_compressed_freq}). \citet{Gupta2016_01} obtain axial buckling strains of $2.2\%$ and $4.3\%$ for the same aspect ratio, which are lower than the current results. The frequencies of modes SH1, SH2, SH3, BB1 become zero at $7.14\%$, $7.56\%$, $7.8\%$ and $8.58\%$ strain for CNT(7.7) with aspect ratio 6.4. \citet{Yakobson1996_01} and \citet{Gupta2016_01} obtain the buckling strains $5\%$ and $9.12\%$ for the shell and bending beam modes for the same CNT.}

\FloatBarrier
\section{\textcolor{cgn}{Conclusions}}\label{s:conclusion}
The nonlinear vibrational properties of graphene-based structures are determined with a new rotation-free shell formulation based on isogeometric finite elements. A hyperelastic material model is used to describe graphene-based structures under large deformations, accounting for nonlinear compressing, stretching and bending. The frequencies are affected strongly by those nonlinearities. Additionally, the nonlinearities of substrate interaction have a strong effect on the frequencies. A calculation based on linear elasticity can underestimate the frequencies, whereas the proposed nonlinear material model avoids this problem. It is therefore important to include the effects of pre-stretch or substrate adhesion in the modal analysis. The results of the current research can be used in the detailed design of graphene-based devices such as MEMS and NEMS. These calculations are essential to control resonance and stability. \textcolor{cgn}{The modal analysis for square and circular plates under pure dilatation, uniaxial stretch, and substrate adhesion is conducted. The natural frequencies are verified with the analytical solutions for zero pre-load. The natural radial breathing mode of a CNT is compared with results from the literature and there is good agreement. The variation of CNT frequencies with uniaxial stretch and compression is obtained and the influence of chirality is investigated. The strains where the frequencies vanish correspond to the strain of various buckling modes. Further, the influence of chirality on the variation of the frequencies is studied. In the present study the nonlinearities due to large initial deformations, contact and constitution are considered. Another source of nonlinearity is tension modulation \citep{Avanzini2012_01,bank2009_01}. Tension modulation describes tension that is constant in space, but varies in the time. The proposed model can be applied to many other graphene-based structures. An interesting example are the vibration of carbon nanocones \citep{FirouzAbadi2011_01,Gandomani2016_01}. They can be studied in future work based the present finite element model.}
\FloatBarrier
\section*{Acknowledgement}{Financial support from the German Research Foundation (DFG) through grant GSC 111 is
gratefully acknowledged.}

\appendix

\section{Analytical solution of natural frequencies}\label{s:analytical_sol_modal}
The natural frequencies for simply supported square and clamped/simply supported circular plates are analytically obtained for the Canham bending energy model. Here, it is shown that the shell theories based on the classical Koiter bending energy model \citep{leissa1969Vibration_of_Plates,hagedorn2007Vibrations_and_Waves, blevins2015formulas_for_Dynamics} and Canham bending energy have the same differential equation but they have two different characteristic equations for a simply supported circular plate. This is due to the moment boundary conditions and Poisson's ratio. The characteristic equations, which are obtained from the Koiter and Canham models, are identical for circular clamped plates. In addition, they are identical for square simply supported plates. In this section, first the mass density of graphene is calculated. Then, the characteristic equations for the different geometries and boundary conditions are derived. These solutions are used in the verification of the numerical method.
\subsection{Graphene density}
\begin{figure}
        \centering
   \includegraphics[height=55mm]{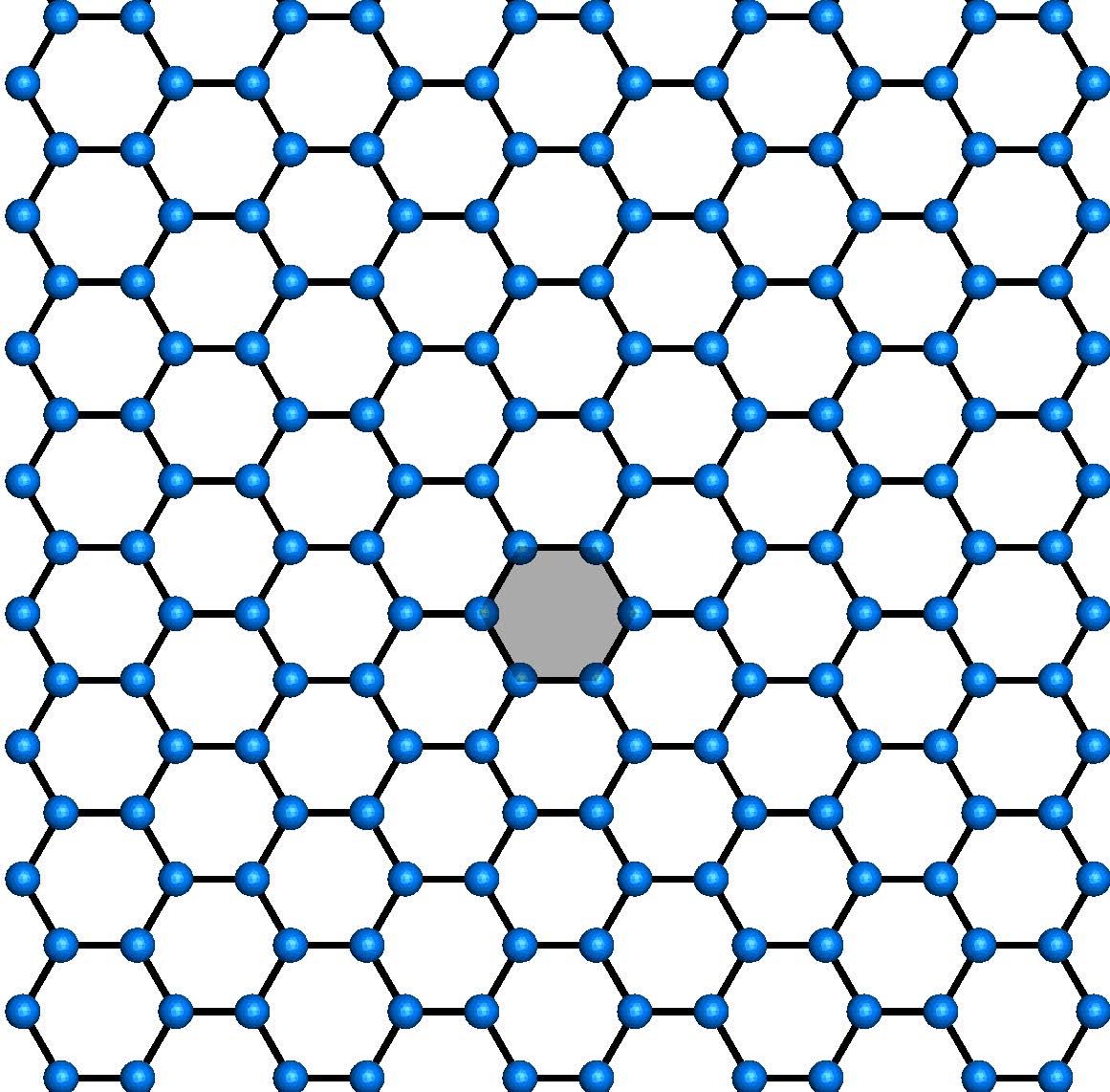}
   \caption{Representative area element (RAE) for the computation of the mass density.}
   \label{f:graphene_RVE}
\end{figure}

A hexagonal Representative Area Element~(RAE) is selected to compute the mass density~(Fig.~\ref{f:graphene_RVE}). The area of the RAE can be computed as
\eqb{lll}
\ds A_{\mathrm{RAE}} \is \ds \frac{3\sqrt{3}}{2}\,a^2~,
\eqe
where $a$ is the length of the carbon-carbon bond. In addition, each carbon atom is shared among three RAEs and so there are $6\times\frac{1}{3}=2$ full atoms per the RAE. Finally, the mass density in the reference configuration is computed as
\eqb{lll}
\ds \rho_{0}  \is \ds \frac{2m}{A_{\mathrm{RAE}}} = 0.76106 \times 10^{-6} \frac{\mathrm{kg}}{\mathrm{m}^2}~,
\eqe
where $m$ is the mass of a carbon atom and $\rho_{0}$ can be connected to the surface density in the current configuration via the area stretch $J$ as
\eqb{lll}
\rho_{0} \is J\,\rho~.
\eqe
\subsection{Equilibrium Equation}
In this section, the plate equilibrium equation is derived for the Canham bending model \citep{CANHAM1970_01,Belay2016_01} in Cartesian coordinates. Then, it is transformed to operator form to be used in a cylindrical coordinate.\\
The Canham bending energy density per current area can be written as
\eqb{lll}
w_{\mathrm{b}} = c\,\left(2H^{2}-\kappa\right)~,
\eqe
where $H$ and $\kappa$ are the mean and Gaussian curvatures and $c$ is the bending modulus. If the strains are assumed to be infinitesimal and the shell is flat in the reference configuration, the bending deflection can be described by $z(x,y)$ and the bending strain energy density becomes
\eqb{lll}
W_{\mathrm{b}} \is \ds c\,\left[\frac{1}{2}\left(\kappa_{xx}^2+\kappa_{yy}^2\right)+\kappa_{xy}^2\right]~,
\eqe
where the Cartesian components of the curvature are
\eqb{lll}
\ds \kappa_{xx} \dis z_{,xx}~,\\[3mm]
\ds \kappa_{yy} \dis z_{,yy}~,\\[3mm]
\ds \kappa_{xy} \dis z_{,xy}~,
\label{e:curva_cart_def}
\eqe
where ${}_{,x}$ and ${}_{,y}$ denote partial differentiation w.r.t.~$x$ and $y$.
The Cartesian components of the bending moment are
\eqb{lll}
M_{xx} = \ds -\pa{W_{\mathrm{b}}}{\kappa_{xx}} = -c\,\kappa_{xx}~,\\[3mm]
M_{yy} = \ds -\pa{W_{\mathrm{b}}}{\kappa_{yy}} = -c\,\kappa_{yy}~,\\[3mm]
M_{xy} = \ds -\pa{W_{\mathrm{b}}}{\kappa_{xy}} = -c\,\kappa_{xy}~.
\label{e:moment_cart}
\eqe
The final equilibrium equation takes the well-known form \citep{ugural2009_Stresses_in_Beams_Plates}
\eqb{lll}
\ds \nabla^4 (w) + \frac{\rho}{c}\,w_{,tt} \is \ds \frac{p}{c}~.\\
\label{e:equib_cart_w}
\eqe

\subsection{Rectangular plate: simply supported}
The boundary conditions for a simply supported rectangular plate are
\eqb{l}
\ds w=0;~M_{xx}=-c\paq{w}{x}=0~\text{at}~x=\pm a~,\\[3mm]
\ds w=0;~M_{yy}=-c\paq{w}{y}=0~\text{at}~y=\pm b~.\\
\eqe
where $a$ and $b$ are the half length and half width of the plate. The solution of Eq.~(\ref{e:equib_cart_w}) can be decomposed into the harmonic time independent and time dependent parts as
\eqb{l}
\ds w(x, y, t) = W(x, y)\,e^{-i\hat{\omega} t}~.
\eqe
The final solution is \citep{leissa1969Vibration_of_Plates,ugural2009_Stresses_in_Beams_Plates}
\eqb{l}
\ds w(x, y, t) = \sum\limits_{m,n=1}^{\infty} {A_{(m,n)}\,\sin\left(\frac{m\pi x}{a}\right)\,\sin\left(\frac{n\pi y}{b}\right)\,e^{-i\hat{\omega}_{(m,n)}t}}~,
\eqe
where $A_{(m,n)}$ can be calculated by Fourier transformation and $\hat{\omega}_{(m,n)}$ are the natural vibration frequencies given by
\eqb{l}
\ds \hat{\omega}_{(m,n)} = \pi^2\,\left(\frac{m^2}{a^2}+\frac{n^2}{b^2}\right)\,\sqrt{\frac{c}{\rho}};~~\text{with}~~m,n=1,2,3,...~~.
\label{e:freq_lin_rect_simp}
\eqe
\textcolor{cgr2}{Including the effect of in-plane stresses gives
\eqb{lll}
\ds \omega_{(m,n)}^2 \is \ds \hat{\omega}^{2}_{(m,n)}+\frac{1}{\rho}\left[N_{x}\left(\frac{\pi\,m}{ a}\right)^2+N_{y}\left(\frac{\pi\,n}{b}\right)^2\right]~,
\label{e:analytical_sol_rect_nonlinear_modal}
\eqe
where $N_{x}$ and $N_{y}$ are the normal stress components along the $x$ and $y$ directions \citep{leissa1969Vibration_of_Plates}. For the material model of \citet{Kumar2014_01} under pure dilatation, $N_{x}$ and $N_{y}$ can be written as
\eqb{lll}
N_{x}=N_{y}=\ds \varepsilon\,\hat{\alpha}^2\,\ln(J)\,e^{-(1+\hat{\alpha})\,\ln(J)}~.
\eqe}
\subsection{Circular plate}
In this subsection, the characteristic equations of circular plates with clamped and simply supported boundary conditions are obtained.
\subsubsection{Circular clamped}
The boundary conditions of a clamped circular plate are
\eqb{l}
\ds w=0;~w_{,r}=0~\text{at}~ r=a~,\\[3mm]
\ds w~\&~w_{,r}~\rightarrow~\text{finite}~\text{at}~r=0~,\\
\eqe
where $a$ is the radius of the plate. The solution of Eq.~(\ref{e:equib_cart_w}) can be decomposed as
\eqb{l}
\ds w(r, \varphi, t) = R(r) e^{im\varphi}e^{-i\hat{\omega} t}~.
\eqe
The final solution is \citep{leissa1969Vibration_of_Plates,ugural2009_Stresses_in_Beams_Plates}
\eqb{l}
\ds w(r, \varphi, t) = \sum\limits_{m,n=0}^{\infty} {\left[D_{(m,n)}\cos(m\varphi)+ E_{(m,n)}\sin(m\varphi)\right]\,R_{(m,n)}(r)\,e^{-i\hat{\omega}_{(m,n)}t}}~,
\eqe
where $R_{(m,n)}$ and $\hat{\omega}_{(m,n)}$ are
\eqb{l}
\ds R_{(m,n)} = I_{m}\left(\gamma_{(m,n)}\right)\,J_{m}\left(\frac{\gamma_{(m,n)}r}{a}\right) -J_{m}\left(\gamma_{(m,n)}\right)\,I_{m}\left(\frac{\gamma_{(m,n)}r}{a}\right)~,
\eqe
\eqb{l}
\ds \hat{\omega}_{(m,n)} = \frac{\gamma_{(m,n)}^2}{a^2}\sqrt{\frac{c}{\rho}}~,
\label{e:freq_lin_circ_clamp}
\eqe
where $J_{m}$ and $I_{m}$ are the Bessel and modified Bessel functions of the first kind of order m. Furthermore, $\gamma_{(m,n)}$ are the solution of
\eqb{l}
\ds \frac{J_{m+1}(\gamma)}{J_{m}(\gamma)}+\frac{I_{m+1}(\gamma)}{I_{m}(\gamma)}=0;~~\text{with}~m=0,1,2,3,...~.
\eqe
$D_{(m,n)}$ and $E_{(m,n)}$ can be computed by Fourier transformation.
\subsubsection{Circular simply supported}
The boundary conditions of a simply supported circular plate are
\eqb{l}
\ds w=0;~\&~M_{rr}=-c\,w_{,rr}=0~\text{at}~r=a~,\\[3mm]
\ds w~\&~w_{,r}~\rightarrow~\text{finite}~\text{at}~r=0~,\\
\eqe
where $a$ is the radius of the plate. The final solution is \citep{leissa1969Vibration_of_Plates,ugural2009_Stresses_in_Beams_Plates}
\eqb{l}
\ds w(r, \varphi, t) = \sum\limits_{m,n=0}^{\infty} {\left[D_{(m,n)}\,\cos(m\varphi)+ E_{(m,n)}\,\sin(m\varphi)\right]\,R_{(m,n)}(r)\,e^{-i\hat{\omega}_{(m,n)}t}}~,
\eqe
where $R_{(m,n)}$ and $\hat{\omega}_{(m,n)}$ are
\eqb{l}
\ds R_{(m,n)} = I_{m}\left(\gamma_{(m,n)}\right)\,J_{m}\left(\frac{\gamma_{(m,n)}r}{a}\right)
-J_{m}\left(\gamma_{(m,n)}\right)\,I_{m}\left(\frac{\gamma_{(m,n)}r}{a}\right)~,
\eqe
\eqb{l}
\ds \hat{\omega}_{(m,n)} = \frac{\gamma_{(m,n)}^2}{a^2}\,\sqrt{\frac{c}{\rho}}~,
\label{e:freq_lin_circ_simp}
\eqe
where $\gamma_{(m,n)}$ are solution of
\eqb{l}
\ds \frac{J_{m+1}(\gamma)}{J_{m}(\gamma)}+\frac{I_{m+1}(\gamma)}{I_{m}(\gamma)}=2\gamma;~~\text{with}~~m=0,1,2,3,...~.
\eqe
This relation is different from the one which can be obtained for the Koiter shell theory \citep{leissa1969Vibration_of_Plates}.
\section{Stiffness and mass matrix}\label{s:stiffness_mass_matrix}
In this section, the finite element mass and stiffness matrices are given. They are defined based on NURBS shape functions $N_i$ and their parametric ``,'' and covariant ``;'' derivatives. The matrix form of NURBS shape functions and their derivatives can be written as \citep{Duong2016_01}
\eqb{lll}
\mN \is [N_1\,\boldsymbol{1}, N_2\,\boldsymbol{1}, ..., N_{n_e}\,\boldsymbol{1}]~,\\[1mm]
\mN_{,\alpha} \is [N_{1,\alpha}\,\boldsymbol{1}, N_{2,\alpha}\,\boldsymbol{1}, ..., N_{n_e,\alpha}\,\boldsymbol{1}]~, \\[1mm]
\mN_{,\alpha\beta} \is [N_{1,\alpha\beta}\,\boldsymbol{1}, N_{2,\alpha\beta}\,\boldsymbol{1}, ..., N_{n_e,\alpha\beta}\,\boldsymbol{1}]~,\\ [1mm]
\tilde{\mN}_{;\alpha\beta} \is \mN_{,\alpha\beta}-\Gamma_{\alpha\beta}^{\gamma}\,\mN_{,\gamma}~,
\eqe
where $n_e$ is the number of control points per element.
The mass matrix is independent of the deformation and can therefore be precomputed at the beginning of the simulation. It is
\eqb{lll}
\mM \is \ds \int\limits_{\mathcal{S}_{0}}{\rho_{0}\,\mN^{T}\,\mN\,\dif A}~.
\eqe
The stiffness matrix can be written as
\eqb{lll}
\mK \is \mk_{\text{mat}}+\mk_{\text{geo}}+\mk_{\mathrm{c}}+\mk_{\mathrm{p}}~.
\eqe
The material stiffness matrix $\mk_{\text{mat}}$ can be written as
\eqb{lll}
\mk_{\text{mat}} \is \mk_{\tau\tau}+\mk_{\tau M}+\mk_{M\tau}+\mk_{MM},
\eqe
with
\eqb{lll}
\mk_{\tau\tau} \dis \ds \int\limits_{\mathcal{S}_{0} }{\cabgd\,\mN_{,\alpha}^{\mathrm{T}}\,(\ba_{\beta}\otimes\ba_{\gamma})\,\mN_{,\delta}~\dif A}~,\\
\mk_{\tau M} \dis \ds \int\limits_{\mathcal{S}_{0} }{\dabgd\,\mN_{,\alpha}^{\mathrm{T}}\,(\ba_{\beta}\otimes\bn)\,\tilde{\mN}_{;\gamma\delta}~\dif A}~,\\
\mk_{M\tau} \dis \ds \int\limits_{\mathcal{S}_{0} }{\eabgd\,\tilde{\mN}_{;\alpha\beta}^{\mathrm{T}}\,(\bn\otimes\ba_{\gamma})\,\mN_{,\delta}~\dif A}~,\\
\mk_{MM} \dis \ds \int\limits_{\mathcal{S}_{0} }{\fabgd\,\tilde{\mN}_{;\alpha\beta}^{\mathrm{T}}\,(\bn\otimes\bn)\,\tilde{\mN}_{;\gamma\delta}~\dif A}~.
\eqe
Furthermore, the geometrical stiffness matrix $\mk_{\text{geo}}$ is defined as
\eqb{lll}
\mk_{\text{geo}} \dis \mk_{\tau}+\mk_{M}~,
\eqe
with
\eqb{lll}
\mk_{\tau} \dis \ds \int\limits_{\mathcal{S}_{0} }{\mN_{,\alpha}\,\tau^{\alpha\beta}\,\mN_{,\beta}~\dif A}~,
\eqe
\eqb{lll}
\mk_{M} \dis \ds \mk_{M1}+\mk_{M2}+\mk_{M2}^{\mathrm{T}}~,
\eqe
with
\eqb{lll}
\mk_{M1} \dis \ds -\int\limits_{\mathcal{S}_{0} }{\buab\,\Mab_0\,\agd\, \mN_{,\gamma}^{\mathrm{T}}\,(\bn\otimes\bn)\,\mN_{,\delta}~\dif A}~,\\
\mk_{M2} \dis \ds -\int\limits_{\mathcal{S}_{0} }{\Mab_0\,\mN_{,\gamma}^{\mathrm{T}}\,(\bn\otimes\,a^{\gamma})\,\tilde{\mN}_{;\alpha\,\beta}~\dif A}~.
\eqe
The elasticity tensors of $\cabgd$ , $\dabgd$ , $\eabgd$ and $\fabgd$ are given in \citet{Ghaffari2017_01} for graphene based on the definition in \citet{Sauer2017_01}. The contact stiffness matrix $\mk_{\mathrm{c}}$ is given in \citet{Ghaffari2017_01}. Clamped boundary conditions are applied with a penalty parameter and its stiffness matrix $\mk_{\mathrm{p}}$ is given in \citet{Duong2016_01}.

\FloatBarrier

\bibliographystyle{model1-num-names}
\bibliography{bibliography}

\end{document}